\documentclass[12pt,a4paper]{article} 
\usepackage{etex}
\pdfoutput=1
\usepackage{jheppub}
\usepackage{epstopdf}
\usepackage{graphicx}
\usepackage{epsfig}
\usepackage{dcolumn}  
\usepackage{bm}    
\usepackage{amssymb} 
\usepackage{amsmath,bm}
\usepackage{amsfonts}    
\usepackage{slashed}  
\usepackage{youngtab}
\usepackage[mathscr]{euscript}
\usepackage{epsfig}
\usepackage[utf8]{inputenc}
\usepackage{tikz}
\usetikzlibrary{tikzmark,calc,,arrows,shapes,decorations.pathreplacing}
\usepackage{xcolor}
\usepackage{mathrsfs}
\usepackage{verbatim}
\usepackage{cases}
\usepackage{bbding}
\usepackage{pifont}
\usepackage{ulem}
\usepackage{dcolumn}
\usepackage{graphicx}
\usepackage{pgf}
\usepackage{pstricks}
\usepackage{pst-node}
\usepackage[all]{xy}

\usepackage{young}
\usepackage{graphicx,rotating,booktabs}

\newdimen\tableauside\tableauside=1.0ex
\newdimen\tableaurule\tableaurule=0.4pt
\newdimen\tableaustep
\def\phantomhrule#1{\hbox{\vbox to0pt{\hrule height\tableaurule width#1\vss}}}
\def\phantomvrule#1{\vbox{\hbox to0pt{\vrule width\tableaurule height#1\hss}}}
\def\sqr{\vbox{%
		\phantomhrule\tableaustep
		\hbox{\phantomvrule\tableaustep\kern\tableaustep\phantomvrule\tableaustep}%
		\hbox{\vbox{\phantomhrule\tableauside}\kern-\tableaurule}}}
\def\squares#1{\hbox{\count0=#1\noindent\loop\sqr
		\advance\count0 by-1 \ifnum\count0>0\repeat}}
\def\tableau#1{\vcenter{\offinterlineskip
		\tableaustep=\tableauside\advance\tableaustep by-\tableaurule
		\kern\normallineskip\hbox
		{\kern\normallineskip\vbox
			{\gettableau#1 0 }%
			\kern\normallineskip\kern\tableaurule}%
		\kern\normallineskip\kern\tableaurule}}
\def\gettableau#1 {\ifnum#1=0\let\next=\null\else
	\squares{#1}\let\next=\gettableau\fi\next}

\newcommand{\includeCroppedPdf}[2][]{%
    \IfFileExists{./#2-crop.pdf}{}{%
        \immediate\write18{pdfcrop #2 #2-crop.pdf}}%
    \includegraphics[#1]{#2-crop.pdf}}

\newcommand{\be}{ \begin{equation}}
\newcommand{\ee}{\end{equation}}
\newcommand{\bea}[1]{\begin{eqnarray}\label{#1} }
\newcommand{\eea}{\end{eqnarray}}

\def\ZZZ{{\hskip-3pt\hbox{ Z\kern-1.6mm Z}}}
\def\zzz{{\hskip-3pt\hbox{ z\kern-1mm z}}}

\def\s{\sigma}
\def\bal#1\eal{\begin{align}#1\end{align}}

\def\one{{\hbox{ 1\kern-.8mm l}}}
\def\zero{{\hbox{ 0\kern-1.5mm 0}}}

\def\>{\rangle}
\def\<{\langle}

\title{Twin-plane-partitions and $\mathcal{N}=2$ affine Yangian}

\author{Matthias R.\ Gaberdiel$^a$,  Wei Li$^b$ and  Cheng Peng$^c$} 
\affiliation{$^a$ Institut f\"ur Theoretische Physik, ETH Zurich, \\
\hspace*{0.3cm}CH-8093 Z\"urich, Switzerland}
\affiliation{$^b$ Institute of Theoretical Physics, Chinese Academy of Sciences\\
\hspace*{0.3cm}100190 Beijing, P.R.\ China}
\affiliation{$^c$ Department of Physics, Brown University, \\
\hspace*{0.3cm}182 Hope Street, Providence  , RI 02912, USA}

\emailAdd{gaberdiel@itp.phys.ethz.ch, weili@itp.ac.cn, cheng$\underline{~}$peng@brown.edu}

\abstract{The universal enveloping algebra of ${\cal W}_{1+\infty}$ is isomorphic to the affine Yangian of $\mathfrak{gl}_1$. 
We study the ${\cal N}=2$ supersymmetric version of this correspondence, and identify the full set of defining relations of the supersymmetric affine Yangian. 
These relations can be deduced by demanding that the algebra has a representation on twin-plane-partitions, which we construct by gluing pairs of plane partitions. 
We define the action of the algebra on these twin-plane-partitions explicitly. }

\begin{document}

\maketitle

\makeatletter
\g@addto@macro\bfseries{\boldmath}
\makeatother

\section{Introduction}

During the last few years it has become clear that the CFT duals of higher spin theories on ${\rm AdS}_3$ \cite{Vasiliev:2003ev} are characterized by an ${\cal W}_\infty$ symmetry algebra \cite{Henneaux:2010xg,Campoleoni:2010zq,Gaberdiel:2010pz}, see e.g.\ \cite{Gaberdiel:2012uj} for a review. 
The higher spin theories are expected to arise in the tensionless limit of string theory on ${\rm AdS}_3$ \cite{Gaberdiel:2014cha,Gaberdiel:2017oqg,Ferreira:2017pgt,Gaberdiel:2018rqv,Giribet:2018ada}, see also \cite{Sundborg:2000wp,Witten,Mikhailov:2002bp} for earlier work outlining the general philosophy. 
More interestingly, the higher spin symmetry is only a subalgebra of the hidden stringy symmetry that can manifest itself at the tensionless point. 
%In particular, at the tensionless point of string theory in ${\rm AdS}_3$, the stringy symmetry takes the form of the so-called Higher Spin Square (HSS) \cite{Gaberdiel:2015mra}, and is an extension of the ${\cal W}_\infty$ algebra.

Another type of symmetry structure that appears in string theory is integrability, see e.g.\ \cite{Beisert:2010jr} for a review, which is typically associated with the emergence of a Yangian symmetry.  
Understanding the relation and interplay between these two different types of symmetry structures, higher spin symmetry and Yangian symmetry, might shed light on the nature of stringy symmetries.

For  AdS$_3$, it was shown in \cite{Prochazka:2015deb,Gaberdiel:2017dbk} that the higher spin symmetry is actually equivalent to a certain Yangian symmetry. 
%Intriguingly, the higher spin symmetry for AdS$_3$ is equivalent to a certain Yangian symmetry. 
%At the tensionless point of string theory on ${\rm AdS}_3$, 
Furthermore, at the tensionless point the ${\cal W}_\infty$ algebra characterizing the higher spin theory is extended to the stringy symmetry algebra that takes the form of the so-called Higher Spin Square (HSS) \cite{Gaberdiel:2015mra}.
%extension of the ${\cal W}_\infty$ algebra.
A promising way towards  characterizing the HSS algebra is therefore from this Yangian perspective. 

For the bosonic case originally studied in \cite{Gaberdiel:2010pz}, the relevant Yangian symmetry is that of the affine Yangian  of $\mathfrak{gl}_1$ \cite{SV,MO,Tsymbaliuk,Tsymbaliuk14}, and the corresponding isomorphism was studied in detail in \cite{Prochazka:2015deb,Gaberdiel:2017dbk}.\footnote{The $q$-deformed version of this isomorphism, i.e.\ between the quantum-deformed $\mathcal{W}_{1+\infty}$ algebra and the quantum toroidal algebra of $\mathfrak{gl}_1$ was studied earlier in \cite{Miki,FFJMM1,FFJMM2,FJMM1}.}
 One important aspect of this relation is that both the affine Yangian  of $\mathfrak{gl}_1$ and ${\cal W}_{1+\infty}$ have (faithful) irreducible representations on plane partitions:
\begin{equation}\label{bosonictriangle}
\xymatrix@C=1pc@R=2.2pc{
& \textrm{  affine Yangian of } \mathfrak{gl}_1
\ar[dl]_{\textrm{``iso"}} %\ar[dr]
&  \\
\mathcal{W}_{1+\infty}[\mu]  \ar[ur]      &&   \ar[ll]_{\textrm{irreps}}  \ar[ul]^{\qquad\qquad \qquad \qquad\qquad\qquad  \qquad \qquad \textrm{irreps}}        \textrm{plane partitions}
} 
\end{equation}
This correspondence is not only interesting from a conceptual viewpoint, e.g.\ explaining the triality symmetry of the algebra \cite{Gaberdiel:2012ku}, it also allows one to compute
%, for example, 
the characters of the $\mathcal{W}_{1+\infty}[\mu]$ algebra quite efficiently, see e.g.\ \cite{Datta:2016cmw}. 
\smallskip

Given the usefulness of these relations, it is natural to try to generalize them for other $\mathcal{W}_\infty$ algebras, in particular, for the case with supersymmetry. 
Unfortunately, the supersymmetric analogue of the affine Yangian of $\mathfrak{gl}_1$ is, to our knowledge, not yet known,\footnote{A first attempt at an ${\cal N}=1$ supersymmetric generalization of the construction was undertaken in \cite{Zhu:2015nha}.} and thus it also needs to be defined in the process. 
This program was initiated in \cite{GLPZ} where we identified the general structure of the affine Yangian that is isomorphic to $\mathfrak{u}(1)\oplus\mathcal{W}^{\mathcal{N}=2}_{\infty}[\mu]$,\footnote{As will become clear later, the introduction of $ \mathfrak{u}(1)$ here is merely to make the construction more symmetric and is not essential since the $\mathfrak{u}(1)$ can be decoupled. We should also mention that this is the simplest supersymmetric generalization since the  $\mathcal{N}=1$  $\mathcal{W}_{\infty}[\mu]$ only exists for the special value of $\mu=\frac{1}{2}$, and is therefore less generic and interesting.} for related work see also \cite{Gaiotto:2017euk,PR}. 

The basic idea of \cite{GLPZ} was to use the (conjectural) observation that the algebra
$\mathfrak{u}(1)\oplus\mathcal{W}^{\mathcal{N}=2}_{\infty}[\mu]$ has two commuting bosonic $\mathcal{W}_{1+\infty}$ subalgebras, see also \cite{Romans:1991wi,Datta:2012km,Gaiotto:2017euk}. 
Furthermore, the fermionic generators of $\mathcal{W}^{\mathcal{N}=2}_{\infty}[\mu]$ transform as $(\lambda, {\lambda^{\star}})$ w.r.t.\ these two bosonic subalgebras, where $\lambda$ runs over a specific set of representations of $\mathcal{W}_{1+\infty}$, and $\lambda^\star$ is the conjugate of the transpose representation. 
Since each $\mathcal{W}_{1+\infty}$ maps to an affine Yangian of $\mathfrak{gl}_1$, this suggests that the $\mathcal{N}=2$ affine Yangian can be built up from two bosonic affine Yangian  algebras by adding suitable fermionic generators, see \cite{Gaiotto:2017euk} for the general strategy. 
In order to identify the relations that are satisfied by the fermionic generators, the free field realization of $\mathcal{W}^{\mathcal{N}=2}_{\infty}[\mu=0]$ was used to find the relevant identities at $\mu=0$. 
Using the representation theory of the two bosonic affine Yangians, natural conjectures for the deformation of these relations away from $\mu=0$ could then be proposed \cite{GLPZ}. 
%
%inspired the construction of  in \cite{GLPZ} by taking two copies of the bosonic $\widehat{\mathcal{Y}(\mathfrak{gl}_1)}$ and introduce new fermionic generators and use them to glue the two bosonic subalgebras. 
%The $\mathcal{W}^{\mathcal{N}=2}_{\infty}[\lambda]$ algebra has a free field realization at $\lambda=0$. 
%At this free point, we defined explicitly the $\mathcal{N}=2$ affine Yangian. 
%Away from the free point, based on the expected $\mathcal{W}$ charges for the basic building blocks, we proposed the set of defining relations. 
\smallskip

While this general strategy was largely successful, it was not quite strong enough to fix all the defining relations. 
It is the aim of this paper to fill this gap. 
%In particular, the relations between the bosonic generators and fermionic ones was left unfixed (although we made some conjectures on their form.) 
%In this paper, we would like to fill this gap. 
Our main inspiration comes from the fact that the bosonic isomorphism could be best understood in terms of the irreducible representation on plane partition configurations, see (\ref{bosonictriangle}). 
This suggests that the ${\cal N}=2$ supersymmetric generalization of the affine Yangian (and the isomorphism to the ${\cal N}=2$ ${\cal W}_\infty$ algebra) may be constructed in terms of its representation on pairs of plane partitions. %, where the two plane partitions correspond to the two bosonic affine Yangian subalgebras. 
The fermionic generators generate infinite rows of boxes connecting the two plane partitions, and hence ``glue" them together along one of the three legs; we shall call the resulting configurations \textit{twin-plane-partitions} in the following.

%We will show that the plane partition type representations for $\mathcal{W}^{\mathcal{N}=2}_{\infty}[\lambda]$ are pairs of plane partitions glued along one of the three legs (called twin-plane-partition from now on).

\begin{equation}\label{susictriangle}
\xymatrix@C=1pc@R=2.2pc{
& \mathcal{N}=2 \,\,\textrm{affine Yangian of } \mathfrak{gl}_1
\ar[dl]_{\textrm{``iso"}}
&  \\
\mathcal{N}=2 \,\,\, \mathcal{W}_{\infty}[\lambda]  \ar[ur]      &&   \ar[ll]^{\textrm{irreps}}  \ar[ul]_{\textrm{irreps}}        \textrm{twin-plane-partitions}
} 
\end{equation}
\smallskip

More specifically, we shall show in this paper how to define the action of the ${\cal N}=2$ affine Yangian generators on these twin-plane-partitions. 
As we shall see, this action is largely determined by the representation theory of the two bosonic affine Yangians. 
This allows us to fix the remaining freedom in the defining relations of the $\mathcal{N}=2$ affine Yangian. 
The existence of a consistent action on twin-plane-partitions also shows that our relations are self-consistent. 
\medskip

The paper is organized as follows. 
We start by reviewing in section~\ref{sec:review} the salient features of the  bosonic triangle and our construction of the $\mathcal{N}=2$ affine Yangian from \cite{GLPZ}. 
As was already explained there, conjugate representations (whose plane partition description was hitherto not known) play a crucial role for the construction, and we explain some of their features in more detail in section~\ref{sec:conj}. 
% we explain some crucial feature of the construction in detail, namely the emergence of the conjugate representations, which is crucial in the gluing of two plane partitions to form a twin-plane partition.
In section~\ref{sec:4}, we define twin-plane-partitions and show how to compute their eigenfunctions with respect to the Cartan generators of the two bosonic affine Yangians. 
This allows us to determine the action of the Yangian generators on at least some twin-plane-partition configurations. 
With these results at hand, we can then fix in section~\ref{sec:5} all but a few parameters in the defining relations of the supersymmetric affine Yangian. 
The remaining freedom is finally determined in sections~\ref{sec:xaction} and \ref{sec:annih} where we define the action of the remaining Yangian generators on generic configurations of twin-plane-partitions. 
Among other things, this shows that our set of defining relations is non-trivial and consistent. 
Finally, we close in section~\ref{sec:sum} with a summary and discussion of future directions. 
%
%Along the way, we fix the remaining parameters. In section 7 we discuss the direct dictionary to $\mathcal{W}^{\mathcal{N}=2}_{\infty}[\lambda]$. Finally section 8 contains our summary and discussions.  

%\break
%
%
%
%
%To perturb away from the two free points, one then use the twin plane partition picture, and fix the eigenfunction for the fermionic building blocks. 
%This actually allowed us to determined most of the relations. 
%
%However, to proceed further, we need to use 
%
%In \cite{GLPZ}, we adopted a route that is inspired by the bosonic triangle.
%
%
%proposed a construction for the $\mathcal{N}=2$ version of the affine Yangian of $\mathfrak{gl}_1$.
%
%mention construction of N=2 Yangian via gluing
%
%
%Emphasize that the relations of the $\mathcal{N}=2$ affine Yangian are not yet fixed.
%
%motivate the set of twin plane partition as representation
%
%fix algebra by demanding it having TPL as faithful
%
%
%\break

\section{Review}\label{sec:review}

In this section we review the three relations in the bosonic triangle (\ref{bosonictriangle}) and summarize the construction of the $\mathcal{N}=2$ affine Yangian of $\mathfrak{gl}_1$ from \cite{GLPZ}.

\subsection{Affine Yangian of $\mathfrak{gl}_1$}

Let us begin by reviewing the structure of the bosonic affine Yangian of $\mathfrak{gl}_1$; more details can be found in \cite{Prochazka:2015deb,Gaberdiel:2017dbk}.

\subsubsection{Defining relations of affine Yangian of $\mathfrak{gl}_1$}

The defining relations of the affine Yangian are most conveniently expressed in terms of the fields 
\begin{equation}\label{generating}
e(z) = \sum_{j=0}^{\infty} \, \frac{e_j}{z^{j+1}} \ , \qquad 
f(z) = \sum_{j=0}^{\infty} \, \frac{f_j}{z^{j+1}} \ , \qquad 
\psi(z)  = 1 + \sigma_3 \, \sum_{j=0}^{\infty} \frac{\psi_j}{z^{j+1}} \ .
\end{equation}
In this language, the defining relations can be written as \cite{Prochazka:2015deb,Gaberdiel:2017dbk} 
\begin{equation}\label{bosonicdef}
\begin{aligned}
\psi(z)\, e(w) \, \ \sim \ \ & \varphi_3(\Delta)\, e(w)\, \psi(z) \\
\psi(z)\, f(w) \, \ \sim \ \ & \varphi_3^{-1}(\Delta)\, f(w)\, \psi(z) \\
%\end{aligned}
%\end{equation}
%and
%\begin{equation}
%\begin{aligned}
e(z)\, e(w) \, \ \sim \ \ & \varphi_3(\Delta)\, e(w)\, e(z) \\
f(z)\, f(w) \, \ \sim \ \ & \varphi_3^{-1}(\Delta)\, f(w)\, f(z)  \\
[e(z)\,, f(w)]  \ \sim \ \ & - \frac{1}{\sigma_3}\, \frac{\psi(z) - \psi(w)}{z-w} \ , 
\end{aligned}
\end{equation}
where from now on $\Delta$ is defined as  
\begin{equation}
\Delta\equiv z-w \ ,
\end{equation}
and ``$\sim$" means equality up to terms that are regular at $z=0$ or $w=0$, see the discussion around eq.~(5.15) in \cite{Gaberdiel:2017dbk}.
 The function $\varphi_3(z)$ is defined as 
\begin{equation}\label{varphidef}
\varphi_3(z) \equiv \frac{(z+h_1) (z+h_2) (z+h_3)}{(z-h_1) (z-h_2) (z-h_3)}   \ ,
%= \frac{z^3 + \sigma_2 z + \sigma_3}{z^3 + \sigma_2 z - \sigma_3}\ .
\end{equation}
and the parameters $h_1,h_2,h_3$ satisfy
\begin{equation}\label{sumh}
h_1+h_2+h_3=0 \ .
\end{equation}
Finally, the above relations need to be supplemented by the Serre relations
\begin{equation}\label{Serre}
\begin{aligned}
&\sum_{\pi \in \mathcal{S}_3}\, \bigl(z_{\pi(1)} - 2 z_{\pi(2)} + z_{\pi(3)} \bigr)\, e(z_{\pi(1)})\, e(z_{\pi(2)})\, e(z_{\pi(3)}) \sim 0 \\
&\sum_{\pi \in 
\mathcal{S}_3}\, \bigl(z_{\pi(1)} - 2 z_{\pi(2)} + z_{\pi(3)} \bigr)\, f(z_{\pi(1)})\, f(z_{\pi(2)})\, f(z_{\pi(3)}) \sim 0 \ .
\end{aligned}
\end{equation}
Note that the defining relations of the affine Yangian are manifestly invariant under the permutation group $\mathcal{S}_3$ acting on the triplet $(h_1,h_2,h_3)$. 
This feature of the algebra plays a significant role in the derivation of the higher spin AdS$_3$/CFT$_2$ holography \cite{Gaberdiel:2012ku}.

The defining relations of the affine Yangian can also be written in terms of modes $(e_j,f_j,\psi_j)$. To see how this can be deduced from eq.~(\ref{bosonicdef}), we multiply both sides of each identity with the denominator of the rational function on the right-hand-side. We then require that the equality holds up to terms that are regular at $z=0$ or $w=0$ --- this then gives the corresponding relation in terms of modes. For example, the first identity in (\ref{bosonicdef}) really means that 
%\begin{equation}\label{psie}
%\prod_{j=1}^{3} (z-w-h_i)\, \psi(z)\, e(w) \sim e(w) \, \psi(z) \, \prod_{j=1}^{3} (z-w+h_i)\ ,
%\end{equation}
\begin{equation}\label{psie}
\Bigl( (z-w)^3 + \sigma_2 (z-w) - \sigma_3 \Bigr)  \psi(z)\, e(w) \sim \Bigl( (z-w)^3 + \sigma_2 (z-w) + \sigma_3 \Bigr) \, e(w) \, \psi(z) \ , 
\end{equation}
where we have used (\ref{sumh}) and defined the $\mathcal{S}_3$ invariant expressions 
\begin{equation}\label{sigma23}
\sigma_2 \equiv h_1 h_2 + h_2 h_3 + h_3 h_1 \ , \qquad \sigma_3 \equiv h_1 h_2 h_3 \ . 
\end{equation}
Expanding $\psi(z)$ and $e(z)$ then gives 
\begin{equation}
\begin{aligned}
\sigma_3 \{e_j,e_k \}   &=  [e_{j+3},e_k] - 3  [e_{j+2},e_{k+1}] + 3  [e_{j+1},e_{k+2}]  - [e_{j},e_{k+3}]   \nonumber \\
& \ + \sigma_2 [e_{j+1},e_{k}] - \sigma_2  [e_{j},e_{k+1}] 
\end{aligned}
\end{equation}
and similarly in the other cases.\footnote{For a complete list of the defining relations in terms of modes, see e.g.\ \cite{Prochazka:2015deb,Gaberdiel:2017dbk,GLPZ}.}

The relations in terms of modes need to be supplemented by  ``initial relations"
\begin{equation}\label{initial}
\begin{aligned}
&[\psi_0,e_r] = 0 \ , \qquad [\psi_1,e_r] = 0 \ , \qquad [\psi_2,e_r] = 2\, e_r \ , \\
&[\psi_0,f_r] = 0 \ , \qquad [\psi_1,f_r] = 0 \ , \qquad [\psi_2,f_r] = - 2\, f_r \ .
\end{aligned}
\end{equation}
They actually follow from (\ref{psie}) and the corresponding $\psi (z) f(w)$ relation if we demand that they do not just hold up to regular terms, but are also true for the terms of the form $z^n w^{-r}$, with $n=0,1,2,3$ and $r>0$.
\begin{comment} 
In order to see this, one uses the expansions (\ref{generating}), as well as the property of the $h_i$ parameters to satisfy $h_1+h_2+h_3=0$, with 
\begin{equation}\label{sigma23}
\sigma_2 = h_1 h_2 + h_2 h_3 + h_1 h_3 \ , \qquad \sigma_3 = h_1 h_2 h_3 \ . 
\end{equation}
Thus (\ref{psie}) becomes 
\begin{equation}
\Bigl( (z-w)^3 + \sigma_2 (z-w) - \sigma_3 \Bigr)  \psi(z)\, e(w) \sim \Bigl( (z-w)^3 + \sigma_2 (z-w) + \sigma_3 \Bigr) \, e(w) \, \psi(z) \ , 
\end{equation}
and the above statements follow directly. The same analysis also applies to the $\psi(z)\, f(w)$ OPE, leading to the initial conditions 
\begin{equation}
{}[\psi_0,f_r] = 0 \ , \qquad [\psi_1,f_r] = 0 \ , \qquad [\psi_2,f_r] = - 2\, f_r \ . 
\end{equation}
\end{comment}
We note that the natural analogue of these initial conditions for the $e(z)\, e(w)$ OPE  is the Serre relation (\ref{Serre}).  
%\be
%\sum_{\pi \in S_3}\, \bigl(z_{\pi(1)} - 2 z_{\pi(2)} + z_{\pi(3)} \bigr)\, e(z_{\pi(1)})\, e(z_{\pi(2)})\, e(z_{\pi(3)}) \sim 0 \ .
%\ee
Indeed, the Serre relation is almost a consequence of the $e(z)\, e(w)$ OPE since\footnote{The analysis for the $f(z)\, f(w)$ OPE and Serre relation is essentially identical.}
\bal
0\sim \ & \sum_{\pi \in S_3}\, \bigl( p(z_{\pi(1)}-z_{\pi(2)}) - p(z_{\pi(2)}-z_{\pi(3)}) \bigr) 
\, e(z_{\pi(1)})\, e(z_{\pi(2)})\, e(z_{\pi(3)}) \\ 
%+p(z_2-z_1)e(z_2)e(z_1)e(z_3) \nonumber \\
%&+p(z-u)e(z)e(u)e(w)+p(u-z)e(u)e(z)e(w) \nonumber \\
%&+p(w-u)e(w)e(u)e(z)+p(u-w)e(u)e(w)e(z) \nonumber \\
%& - \sum_{\pi \in S_3}\, p(z_{\pi(2)}-z_{\pi(3)})\, e(z_{\pi(1)})\, e(z_{\pi(2)})\, e(z_{\pi(3)}) \\
%& -p(w-u)e(z)e(w)e(u)-p(u-w)e(z)e(u)e(w)\nonumber \\
%&-p(z-u)e(w)e(z)e(u)-p(u-z)e(w)e(u)e(z)\nonumber \\
%&-p(z-w)e(u)e(z)e(w)-p(w-z)e(u)e(w)e(z)\nonumber \\[4pt]
\sim\ & ( z_1^2 + z_2^2 + z_3^2 - z_1 z_2-z_1 z_3-z_2 z_3 + \sigma_2 )\label{prefactor} \\
& \times \sum_{\pi \in S_3}\, \bigl(z_{\pi(1)} - 2 z_{\pi(2)} + z_{\pi(3)} \bigr)\, e(z_{\pi(1)})\, e(z_{\pi(2)})\, e(z_{\pi(3)})\ , \label{Serre1}
\eal
where we have used the short-hand expression $p(\Delta) =  ( \Delta^3 + \sigma_2 \Delta - \sigma_3 )$, see eq.~(\ref{psie}). Note that (\ref{Serre1}) is precisely the Serre relation, but we cannot deduce it from this analysis because of the prefactor in (\ref{prefactor}). (All of these relations are only true up to regular terms, and hence we cannot divide by the prefactor.)

\subsubsection{Isomorphism between affine Yangian of $\mathfrak{gl}_1$ and $\mathcal{W}_{1+\infty}$}

It was shown in \cite{Prochazka:2015deb,Gaberdiel:2017dbk} that the affine Yangian of  $\mathfrak{gl}_1$ is isomorphic to the universal enveloping algebra of $\mathcal{W}_{1+\infty}$. In terms of the conformal field theory language, the $h_i$ parameters and $\psi_0$ can be expressed as
\begin{equation}\label{h123}
h_1 =  -\sqrt{\frac{N+k+1}{N+k}} \ , \quad h_2 =  \sqrt{\frac{N+k}{N+k+1}} \ , \quad h_3 = \frac{1}{\sqrt{(N+k)(N+k+1)}} \ , 
\end{equation}
and
\begin{equation}
\psi_0 = N  \ , 
\end{equation}
see eqs.~(3.51)  and (3.52) of \cite{Gaberdiel:2017dbk}. Furthermore, we have 
\begin{equation}
W^{(s)}_{-1} \sim e_{s-1} \qquad \qquad W^{(s)}_{0} \sim \psi_{s} \qquad  \qquad W^{(s)}_{1} \sim f_{s-1} \ , 
\end{equation}
up to ``sub-leading" correction terms. For the first few spins we can be quite explicit. For example, at spin $s=2$, the conformal scaling operator is identified with 
\begin{equation}
L_0 =\frac{1}{2}\psi_2  \ ,
\end{equation}
while at spin $s=3$ we have
\begin{equation}
W^{(3)}_{0}=  -\frac{1}{3}\psi_3-\frac{\s_3 \psi_0}{6}\psi_2+\s_3 \Bigl[\frac{1}{6}\psi_1\psi_1 +\frac{1}{2}\sum_{\ell} |\ell| :J_{-\ell}J_{\ell}: \Bigr] \ . 
\end{equation}
Note that only for the ground states, on which the last term in the above expression vanishes, the natural twin-plane-partition states are eigenstates of $W^{(3)}_0$,  with eigenvalue 
\begin{equation}\label{W30}
W^{(3)}_{0}=  -\frac{1}{3}\psi_3-\frac{\s_3 \psi_0}{6}\psi_2+\frac{\s_3}{6}\psi_1\psi_1 \ . 
\end{equation}
However, already at the first excited level, the eigenstates of $W^{(3)}_{0}$ are not the individual plane partition configurations. Indeed, using that 
\begin{equation}\label{u1gen}
J_1=-f_0\ , \qquad J_{-1}=e_0\ , 
\end{equation}
we have
\begin{equation}\label{2.16}
W^{(3)}_{0}=  -\frac{1}{3}\psi_3-\frac{\s_3 \psi_0}{6}\psi_2+\frac{\s_3}{6}(\psi_1\psi_1 -e_0 f_0) \ , 
\end{equation}
and the $e_0 f_0$ term does not act diagonally on plane partition configurations. 
The situation is similar for spins $s\geq 4$. 

\subsection{Plane partitions}

The relation between $\mathcal{W}_{1+\infty}$ and the affine Yangian is useful because the latter has an elegant
%comes from the fact that for the latter the representations can be described very geometrically. 
representation theory in terms of plane partitions. 
By the isomorphism reviewed in the previous subsection, the set of plane partitions therefore also furnishes a representation for the $\mathcal{W}_\infty$ algebra. This is very useful for understanding representations of the $\mathcal{W}_\infty$ algebra, both conceptually (such as in manifestly seeing the triality symmetry of \cite{Gaberdiel:2012ku}, which is crucial to understanding the bosonic higher spin holography for AdS$_3$/CFT$_2$) and computationally (as in computing the characters of the $\mathcal{W}_{\infty}$ algebra via box counting combinatorics), see e.g.\ \cite{Datta:2016cmw}.
\smallskip

The set of plane partitions (with given asymptotics $(\lambda_1, \lambda_2, \lambda_3)$ along the three directions) furnishes a representation for the affine Yangian of $\mathfrak{gl}_1$, where the actions of $(\psi, e, f)$ on a plane partition configuration $\Lambda$ is given by (for details see \cite{Prochazka:2015deb,Gaberdiel:2017dbk})
\begin{equation}\label{ppart}
\begin{aligned}
\psi(z)|\Lambda \rangle & = \psi_{\Lambda}(z)|\Lambda \rangle  \ ,\\
e(z) | \Lambda \rangle & =  \sum_{ {\tiny \yng(1)} \in {\rm Add}(\Lambda)}\frac{\bigl[ -  \frac{1}{\sigma_3} {\rm Res}_{w = h({\tiny \yng(1)})} \psi_{\Lambda}(w) \bigr]^{\frac{1}{2}}}{ z - h({\tiny \yng(1)}) } 
| \Lambda + {\tiny \yng(1)} \rangle \ , \\
f(z) | \Lambda \rangle & =  \sum_{ {\tiny \yng(1)} \in {\rm Rem}(\Lambda)}\frac{\bigl[ -  \frac{1}{\sigma_3} {\rm Res}_{w = h({\tiny \yng(1)})} \psi_{\Lambda}(w) \bigr]^{\frac{1}{2}}}{ z - h({\tiny \yng(1)}) } | \Lambda - {\tiny \yng(1)} \rangle \ . 
\end{aligned}
\end{equation}
Here ``Res" denotes the residue, and $\psi(z)$ acts diagonally on $\Lambda$ with eigenvalue $\psi_\Lambda(z)$ defined by 
\be\label{psieig}
\psi_\Lambda(z) =\psi_0(z) \, \prod_{ {\tiny \yng(1)} \in (\Lambda)} \varphi_3(z - h({\tiny \yng(1)}) ) \ , 
\ee
where 
\begin{equation}\label{psi0def}
\psi_0(z)\equiv 1 + \frac{\psi_0 \sigma_3}{z} 
\end{equation}
is the vacuum factor and 
\be\label{hbox}
h({\tiny \yng(1)}) \equiv h_1 x_1({\tiny \yng(1)}) + h_2 x_2({\tiny \yng(1)}) +  h_3 x_3({\tiny \yng(1)}) 
\ee
with $x_i({\tiny \yng(1)})$ the $x_i$-coordinate of the box. $e(z)$ adds one box to $\Lambda$ at all possible positions, and $f(z)$ removes one box from $\Lambda$ at all possible positions.
One can check that under the action (\ref{ppart}), the set of all plane partitions $\Lambda$  indeed forms a faithful representation of the affine Yangian algebra (\ref{bosonicdef}). 

\begin{figure}[h!]
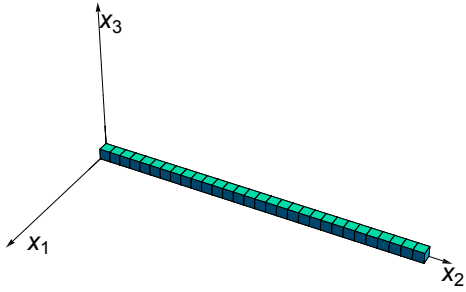

	\centering
	\includeCroppedPdf[width=0.4\textwidth]{"xleft"}
	\caption{The minimal representation corresponding to an infinite row in the $x_2$ direction.}
	\label{figxleftmin}
\end{figure}

The different irreducible representations of the affine Yangian are parametrized by the asymptotic Young diagrams $(\lambda_1, \lambda_2, \lambda_3)$ along the three directions;\footnote{It is clear from eq.~(\ref{ppart}) that the action of the affine Yangian generators does not modify the asymptotics.}  for example, Fig.~\ref{figxleftmin} shows the ground state of the (``minimal") representation $(0,{\tiny{\yng(1)}},0)$. The character of each such representation can be easily computed by box counting.
%\footnote{This is an easier and more transparent method of computing $\mathcal{W}_\infty$ algebra characters than the Kac-Peterson formula; for an application in string theory, see \cite{Datta:2016cmw}. }
For example, the vacuum character of the affine Yangian of $\mathfrak{gl}_1$ equals the generating function of plane partitions with trivial asymptotics, i.e.\ the MacMahon function (see e.g.\ \cite{Prochazka:2015deb,Gaberdiel:2017dbk})
\begin{equation}\label{MacMahon}
\chi_{\rm pp} = \prod_{n=1}^{\infty} \frac{1}{(1-q^n)^n} \ .
\end{equation}
This agrees precisely with the vacuum character of the ${\cal W}_{1+\infty}[\mu]$ algebra. Similarly, for the minimal representation $(0,{\tiny{\yng(1)}},0)$, the character equals 
\be
\chi_{\rm min} = \chi_{\rm pp} \cdot \chi^{({\rm wedge})}_{(0,{\tiny{\yng(1)}},0)} \ , \qquad \hbox{with} \qquad
\chi^{({\rm wedge})}_{(0,{\tiny{\yng(1)}},0)} = \frac{q^h}{(1-q)} \ .
\ee

The different plane partition configurations $\Lambda$ are in one-to-one correspondence with their eigenvalue functions $\psi_{\Lambda}(u)$, defined in (\ref{psieig}). For example, the 
%encodes its charges $\{\psi_{j}\}$ with $j\in\mathbb{N}_0$.
vacuum state (which we will denote by $|\emptyset\rangle$) has charge  function
\begin{equation}
 \psi_0(u) =  1+ \frac{\sigma_3\psi_0}{u}  \ . 
\end{equation}
The various descendant states are obtained by repeated action of the $e$ generators. 
We denote the first descendant by
\begin{equation}
 |{\tiny\yng(1)}\rangle \equiv e_0  |\emptyset\rangle \ , \qquad \hbox{where} \qquad 
e(z) |\emptyset\rangle \sim \frac{1}{z} |{\tiny\yng(1)}\rangle \ . 
\end{equation}
The $\psi(u)$ eigenvalue of  $|{\tiny\yng(1)}\rangle$ is then 
\begin{equation}
|{\tiny\yng(1)}\rangle: \qquad 
\psi(u)=\psi_0(u) \cdot \varphi_3(u)\ . 
 % \qquad \textrm{and}\qquad
\end{equation}
%Similarly
%\begin{equation}
% |\hat{e}\rangle \equiv e_0  |\emptyset\rangle
%\end{equation}
%The $(\psi(u), \hat{\psi}(u))$ eigenvalues of  $|\hat{{\tiny\yng(1)}}\rangle$ are
%\begin{equation}
%\hat{e}(z) |\emptyset\rangle \sim \frac{1}{z} |\hat{e}\rangle
%\end{equation}
%\begin{equation}
%|\hat{{\tiny\yng(1)}}\rangle: \qquad 
%\begin{cases}
%\begin{aligned} 
%\psi(u)&=\psi_0(u) \\
%%\qquad \textrm{and}\qquad
% \hat{\psi}(u)&=\hat{\psi}_0(u)  \cdot \varphi_3(u)
%\end{aligned}
% \end{cases}
%\end{equation}
Similarly, we have bosonic annihilators defined by 
\begin{equation}
f(z) |{\tiny\yng(1)}\rangle \sim \frac{1}{z} |\emptyset\rangle \ ,
\end{equation}
where 
\begin{equation}
f_0 |{\tiny\yng(1)}\rangle = - \psi_0\,  |\emptyset\rangle  \ , 
\end{equation}
as follows from the defining relations of the affine Yangian (together with the fact that $f_r |\emptyset\rangle = 0$). 
\medskip

For non-trivial asymptotics, the charges of the corresponding states are still given by (\ref{psieig}), except that now the infinite product (over the infinitely many boxes defining the asymptotic configuration) must be suitably regularized. For example, for the plane partition representation described by Fig.~\ref{figxleftmin}, the ground state has the charge function
\begin{align}
\psi_{\blacksquare}(u)  & =   \psi_0(u) \, \prod_{n=0}^{\infty} \varphi_3(u - n h_2) %\nonumber %\\ 
%& =:  \bigl( 1 + \frac{\psi_0 \sigma_3}{u} \bigr) \, \varphi_2(u) 
%\frac{u (u+h_2)} {(u-h_1) (u-h_3)} 
\ . %\label{psiudef}
\end{align}
Evaluating the infinite product we obtain a crucial identity \cite{Prochazka:2015deb}
\begin{equation}\label{psiudef0}
\psi_{\blacksquare}(u)= \psi_0(u) \, \varphi_2(u) \qquad \textrm{with}\qquad \varphi_2(u) \equiv \frac{u(u +h_2)}{(u -h_1)(u -h_3)} \ . 
\end{equation}
%where  $\varphi_2(u)$ is defined as
%\be\label{varphi2}
%\varphi_2(u) = \frac{u(u +h_2)}{(u -h_1)(u -h_3)} \ , 
%\ee
This is one of the main ingredients in the construction of the $\mathcal{N}=2$ Yangian in \cite{GLPZ}, as will be reviewed in the next section.

\begin{figure}[h!]
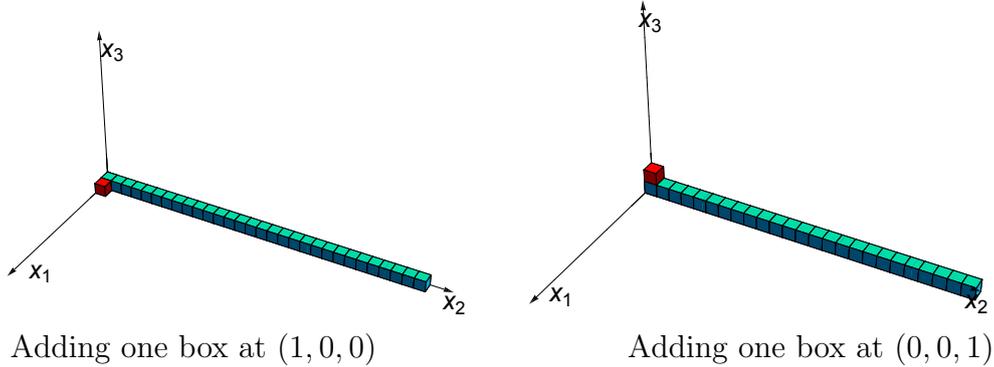

	\centering
	\begin{tabular}{c}
	\includeCroppedPdf[width=.4\textwidth]{"xleftexcited1"}\qquad
		\includeCroppedPdf[width=.4\textwidth
		]{"xleftexcited3"}
  \\
		\vspace{-1.0cm} Adding one box at $(1,0,0)$\qquad \qquad \qquad \qquad Adding one box  at $(0,0,1)$\vspace{1.5cm}
	\end{tabular}	
	\caption{The first descendant states of the minimal representation.}
	\label{figxleftex}
\end{figure}

This ``minimal" representation has two single box descendants, shown in Fig.~\ref{figxleftex}.
If we denote the ground state of the minimal representation by $|\blacksquare\rangle$, then this means that 
\be
e(z)\, |\blacksquare\rangle = \frac{1}{z-h_1} |\blacksquare+{\tiny\yng(1)_{\ 1}}\rangle +\frac{1}{z-h_3} |\blacksquare+{\tiny\yng(1)_{\ 3}}\rangle \ ,
\ee
where the relevant states have the charges
\begin{equation}\label{xecharges}
|\blacksquare+{\tiny\yng(1)_{\ i}}\rangle  : \qquad \qquad  \psi(u) = \psi_0(u) \, \varphi_2(u) \, \varphi_3(u-h_i) \ , 
%|\blacksquare+{\tiny\yng(1)_{\ 3}}\rangle & : \qquad & \psi(u) = \psi_0(u) \, \varphi_2(u) \, \varphi_3(u-h_3) \ . 
\end{equation}
see Fig.~\ref{figxleftex}.

\subsection{${\cal N}=2$ affine Yangian}

In \cite{GLPZ} the construction of the ${\cal N}=2$ version of the affine Yangian was initiated. 
The basic idea of the construction relies on the observation (going back already to \cite{Romans:1991wi,Datta:2012km,Gaiotto:2017euk}) that the corresponding ${\cal N}=2$ ${\cal W}_\infty$ algebra contains two bosonic ${\cal W}_\infty$ algebras as commuting subalgebras
\be
{\cal W}^{({\cal N}=2)}_{N,k} \, \supset \,  {\cal W}_{N,k} \oplus {\cal W}_{k,N} \ . 
\ee
It turns out to be convenient to add to the  ${\cal W}_{\infty}^{({\cal N}=2)}$ algebra a single free boson; the vacuum character of the resulting algebra
\begin{equation}\label{N2W}
\mathfrak{u}(1)\oplus{\cal W}_{\infty}^{({\cal N}=2)}
\end{equation}
is then 
\begin{equation}\label{chi0}
\chi^{\textrm{Full}}_0(q,y) = \prod_{n=1}^{\infty} \frac{(1+yq^{n+\frac{1}{2}})^{n}(1+\frac{1}{y}q^{n+\frac{1}{2}})^{n}}{(1-q^n)^{2n}} \ .
\end{equation}
This combined system (\ref{N2W}) is the starting point of our $\mathcal{N}=2$ affine Yangian construction. 

We can organize this character in terms of representations of the two bosonic subalgebras 
\begin{equation}
{\cal W}_{1+\infty}[\lambda] \oplus {\cal W}_{1+\infty}[1-\lambda] \ . 
\end{equation}
In particular, the denominator of $\chi_0(q)$ in (\ref{chi0}) corresponds to the vacuum characters of the two bosonic ${\cal W}_{1+\infty}$ algebras, while the numerator of  (\ref{chi0}) accounts for the fermionic excitations. 
It was noted in \cite{GLPZ} that 
\begin{equation}\label{charid}
\prod_{n=1}^{\infty} (1+y\, q^{n+\frac{1}{2}})^{n} = \sum_{R} y^{|R|}\chi_{R}^{({\rm wedge})\,  [\lambda]}(q) \cdot \chi_{R^\star}^{({\rm wedge})\, [1-\lambda]}(q) \ ,
\end{equation}
where  $R$ runs over all $2d$ Young diagrams, labelling the asymptotic behaviour of the first plane partition in the $x_2$-direction, and 
\be\label{stardef}
R^{\star} \equiv \overline{R^T} 
\ee
is the conjugate of the representation corresponding to the transpose Young diagram. 
Furthermore $|{R}|$ denotes the number of boxes in $R$. 
Similarly, the conjugate factor can be written as 
\begin{equation}\label{charidconj}
\prod_{n=1}^{\infty} (1+\frac{1}{y}\, q^{n+\frac{1}{2}})^{n} = \sum_{S}  \frac{1}{y^{|S|}}\, \chi_{S^\star}^{({\rm wedge})\,  [\lambda]}(q) \cdot \chi_{S}^{({\rm wedge})\, [1-\lambda]}(q) \ ,
\end{equation}
where  $S$ runs over all $2d$ Young diagrams, now labelling the asymptotics along the $\hat{x}_2$-direction from the perspective of the second plane partition. Combining (\ref{MacMahon}), (\ref{charid}), and (\ref{charidconj}), the full vacuum character of the combined system $\mathfrak{u}(1)\oplus \mathcal{W}^{\mathcal{N}=2}[\mu]$ is 
\begin{equation}
\begin{aligned}
\chi^{\textrm{Full}}_0(q,y)&=\chi_{\rm pp}(q) \left(\sum_{R} y^{|R|}\, \chi_{R}^{({\rm wedge})\,  [\lambda]}(q) \cdot \chi_{R^\star}^{({\rm wedge})\, [1-\lambda]}(q)\right)\\
&\qquad \qquad \cdot \left(\sum_{{S}}\frac{1}{ y^{|{S}|}}\chi_{S^\star}^{({\rm wedge})\,  [\lambda]}(q) \cdot \chi_{S}^{({\rm wedge})\, [1-\lambda]}(q)\right)\chi_{\rm pp}(q) \\
&=1+ \sum_{R} y^{|{R}|}\chi_{R}^{ [\lambda]}(q) \cdot \chi_{R^\star}^{ [1-\lambda]}(q)+ \sum_{S} \frac{1}{y^{|S|}} \, \chi_{S^\star}^{ [\lambda]}(q) \cdot \chi_{R}^{ [1-\lambda]}(q)+\cdots \ ,
\end{aligned}
\end{equation}
where in the last line we have used the fact that for each representation $U$ of ${\cal W}_{1+\infty}$, the full character is the product of the vacuum character and its wedge character,
\begin{equation}
\chi_{U}(q)= \chi_{\rm pp}(q) \cdot \chi_{U}^{({\rm wedge})}(q) \ .
\end{equation}
Furthermore, the wedge character of the representation associated to $(R,S^\star)$ is the product of the corresponding wedge characters.

We see from this character analysis that the combined system $\mathfrak{u}(1)\oplus{\cal W}_{\infty}^{({\cal N}=2)}$ can be decomposed w.r.t\ the bosonic subalgebra ${\cal W}_{1+\infty}[\lambda] \oplus {\cal W}_{1+\infty}[1-\lambda]$, and that all the states organize themselves into representations of these two algebras in such a way that the representation with respect to the second factor is the conjugate transpose of the one with respect to the first factor. 
Furthermore, all the representations appear in tensor powers of the two ``bi-minimal" building blocks that transform as 
\begin{itemize}
\item $x$: minimal w.r.t.\ the first ${\cal W}_{1+\infty}$ and anti-minimal w.r.t\ the second one; 
\item $\bar{x}$:  anti-minimal w.r.t.\ the first ${\cal W}_{1+\infty}$ and minimal w.r.t.\ the second one. 
\end{itemize}
Here $x$ and $\bar{x}$ to label the creation operators of the two bi-minimals (with $x$ adding a box to $R$, and $\bar{x}$ adding a box to $S$). 
Together with their annihilators, $y$ for $x$ and  $\bar{y}$ for $\bar{x}$, see Table~\ref{tab1}, they constitute the building blocks for the fermionic generators of the ${\cal N}=2$ algebra. 
%Now we describe these representations from the viewpoint of the affine Yangian. 
%
%\begin{equation}\label{iden}
%x_{1/2} = \frac{1}{\sqrt{2}} \,  G^+_{-3/2} \ , \qquad  \bar{x}_{1/2} = \frac{1}{\sqrt{2}} \, G^-_{-3/2} \ .
%\end{equation}
%\begin{equation}\label{iden}
%y_{1/2} = \frac{1}{\sqrt{2}} \,  G^+_{+3/2} \ , \qquad  \bar{y}_{1/2} = \frac{1}{\sqrt{2}} \, G^-_{3/2} \ .
%\end{equation}
%
%Thus the $\mathcal{N}=2$ affine Yangian will be generated by the basic players of Fig.~\ref{figOPEeverybody1}. The goal of this paper is to find the relations between them. Some of them were already found in \cite{GLPZ}. 

\begin{table}
\begin{center}
\begin{tabular}{|c|c|c|}
\hline \\[-15pt]
\hbox{generator} & \hbox{unhatted algebra ${\cal Y}$ } & \hbox{hatted algebra $\hat{{\cal Y}}$} \\ \hline
&& \\[-12pt]
$x$ & \hbox{minimal} & \hbox{conj. minimal} \\
$\bar{x}$ &  \hbox{conj.\ minimal} & \hbox{minimal} \\
$y$ & \hbox{conj.\ minimal} & \hbox{minimal} \\
$\bar{y}$ &  \hbox{minimal} & \hbox{conj.\ minimal} \\ \hline
\end{tabular}
\end{center}
\caption{The representation properties of the fermionic generators.}\label{tab1}
\end{table}

%\begin{figure}[h!]
%	\centering
%	\includegraphics[trim=2cm 12cm 0cm 4cm, width=0.5\textwidth]{"OPEeverybody1"}
%	\caption{
%	}
%	\label{figOPEeverybody1}
%\end{figure}

\begin{figure}[h!]
	\centering
	\includegraphics[trim=2cm 12cm 0cm 4cm, width=.8\textwidth]{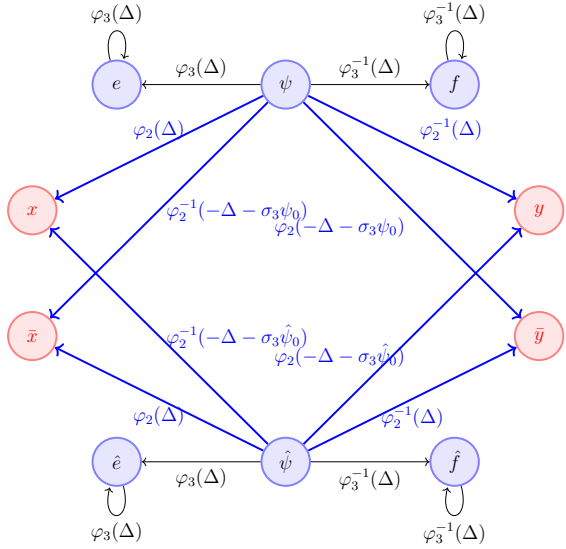}
	\caption{The OPEs with the Cartan generators $\psi(u)$ and $\hat{\psi}(u)$.}
	\label{figOPEeverybody2}
	
\end{figure}

\subsection{Relations between bosonic and fermionic generators}\label{sec:somerelations}

As explained above, the supersymmetric affine Yangian contains two bosonic affine Yangian subalgebras that commute with one another. 
We shall use the convention of \cite{GLPZ} that the generators of the first affine Yangian are denoted by $e_r$, $f_r$ and $\psi_r$, while those of the second one are $\hat{e}_r$, $\hat{f}_r$ and $\hat{\psi}_r$. 
The left affine Yangian $\mathcal{Y}$ corresponds to $\mathcal{W}_{1+\infty}[\lambda]$ and has parameters given by (\ref{h123}) and $\psi_0=N$. Since the right affine Yangian corresponds to $\mathcal{W}_{1+\infty}[1-\lambda]$, it has the same parameter as (\ref{h123}) but
\begin{equation}
\hat{\psi}_0=k \ .
\end{equation}

It follows from the charge assignments above that the commutation relations of the $x$ and $\bar{x}$ generators with $\psi$ and $\hat{\psi}$ are \cite{GLPZ},
\begin{equation}\label{psiFBx0} 
%\begin{cases}
\begin{aligned}
\psi(z) \, x(w)  &\sim  \varphi_2(\Delta) \, x(w) \,\psi(z)   \\
\hat{\psi}(z) \, x(w)  &\sim  \varphi^{-1}_2(-\Delta-\sigma_3\hat{\psi}_0) \, x(w) \,\hat{\psi}(z) 
\end{aligned}
%\end{cases}
\end{equation}
and
\begin{equation}\label{psiFBxbar0} 
%\begin{cases}
\begin{aligned}
\psi(z) \, \bar{x}(w)  &\sim  \varphi^{-1}_2(-\Delta-\sigma_3\psi_0) \, \bar{x}(w)\, \psi(z)   \\
\hat{\psi}(z) \, \bar{x}(w)  &\sim  \varphi_2(\Delta) \, \bar{x}(w)\, \hat{\psi}(z)  \ , 
\end{aligned}
%\end{cases}
\end{equation}
see Fig.~\ref{figOPEeverybody2}.\footnote{Note that in all figures the positions of $y$ and $\bar{y}$ have been interchanged relative to those in \cite{GLPZ}.}
Here $\varphi_2(u)$ was defined in (\ref{psiudef0}), and the tilde indicates, as before, that these relations  are only true up to terms that are regular at either $z=0$ or $w=0$.
(We shall come back to a more detailed analysis of the charges of the conjugate representations in Section~\ref{sec:conj}.)

For the relations between $x$, and $e,f$ we made the ansatz in \cite{GLPZ} that 
\begin{eqnarray}
e(z) \, x(w)  &\sim  &G(\Delta)\, x(w) \, e(z)  \label{g1}\\
f(z) \, x(w) & \sim  &H(\Delta)\, x(w) \, f(z) \ .  \label{g2}
\end{eqnarray}
Since the OPE type relation between the charge functions $\psi(z)$ (and $\hat{\psi}(z)$) and each of these three players are already fixed, $G$ and $H$ are not independent. 
From the consistency with the existing OPE relations, in particular the last one in (\ref{bosonicdef})  and  (\ref{psiFBx0}), we derived a relation between $G$ and $H$ \cite{GLPZ}
\begin{equation}\label{keyiden}
G(\Delta) \, H(\Delta) = \varphi_2(\Delta) \ . 
\end{equation}
However, we were not able to fix $G(\Delta)$ and $H(\Delta)$ separately. 
Note that the free field limit (from which our construction in \cite{GLPZ} originated) leads to some constraints on $G(\Delta)$ and $H(\Delta)$, but these were not sufficient, see \cite{GLPZ}.

\begin{figure}[h!]
	\centering
%\includeCroppedPdf[width=.7\textwidth]{"OPEbosonicxy-conj"}
%\includegraphics[trim=0cm 10cm 0cm 5cm, width=1\textwidth]{"OPEbosonicxy-conj"}
\includegraphics[trim=2cm 12cm 0cm 4cm, width=.8\textwidth]{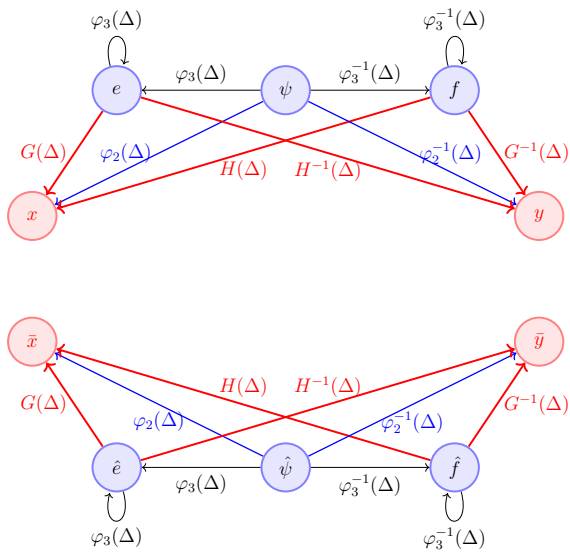}
%\includeCroppedPdf[width=.4\textwidth]{"OPEbosonicxy-conj"}
\caption{The OPE relations of the unhatted fields with $x$ and $y$, and those of the hatted fields with $\bar{x}$ and $\bar{y}$. 
	}
	\label{figOPEbosonic-xy}
	
\end{figure}

\begin{figure}[h!]
	\centering
%\includeCroppedPdf[width=.7\textwidth]{"OPEbosonicxy-conj"}
%\includegraphics[trim=0cm 10cm 0cm 5cm, width=1\textwidth]{"OPEbosonicxy-conj"}
\includegraphics[trim=2cm 12cm 0cm 4cm, width=.8\textwidth]{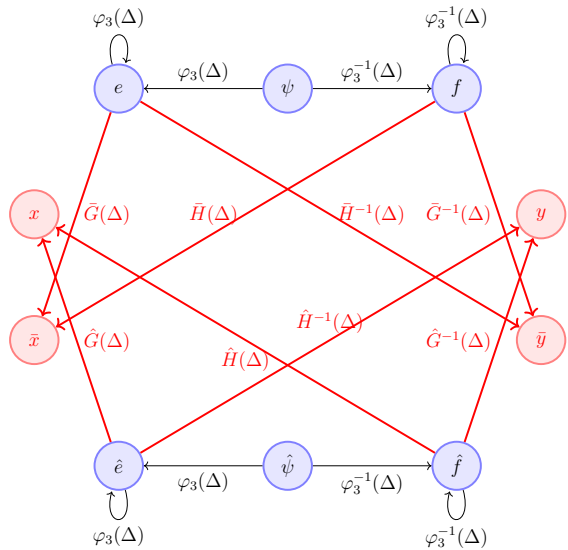}
%\includeCroppedPdf[width=.4\textwidth]{"OPEbosonicxy-conj"}
\caption{The OPE relations of the unhatted fields with $\bar{x}$ and $\bar{y}$, and those of the hatted fields with ${x}$ and ${y}$. 
	}
	\label{figOPEbosonicfull}
	
\end{figure}

Similar relations can also be found for the corresponding annihilation generator $y$, and similarly for $\bar{x}$ and $\bar{y}$ with respect to the hatted fields, see Fig.~\ref{figOPEbosonic-xy}. 
Finally, the structure of the remaining OPEs is sketched in Fig.~\ref{figOPEbosonicfull}. 
%the whole structure can be summarized in Fig.~\ref{figOPEbosonic-xy0}. Furthermore, the analysis is completely analogous for $\bar{x}$ and $\bar{y}$ with respect to the hatted fields, and their structure can thus be similarly realized, see Fig.~\ref{figOPEbosonicxy-conj0}. 

%\begin{figure}[h!]
%	\centering
%		\includegraphics[trim=2cm 12cm 0cm 4cm, width=.45\textwidth]{"OPEbosonicfull"}
%		\includegraphics[trim=2cm 12cm 0cm 4cm, width=.45\textwidth]{"OPEbosonicfull-conj"}
%	\caption{The additional relations (in red and blue) together with some of the relations from Fig.~\ref{figOPEbosonic-xy} for comparison. 
%	}
%	\label{figOPEbosonicfull}
%	\end{figure}
%

%\begin{figure}[h!]
%	\centering
%	\includegraphics[trim=0cm 15cm 0cm 4cm, width=.5\textwidth]{"OPEbosonic-xy"}
%	\caption{The OPE relations involving $x$ and $y$. 
%	}
%	\label{figOPEbosonic-xy0}
%	
%\end{figure}
%
%
%\begin{figure}[h!]
%	\centering
%	\includegraphics[trim=0cm 15cm 0cm 4cm, width=.5\textwidth]{"OPEbosonicxy-conj"}
%	\caption{The OPE relations involving $\bar{x}$ and $\bar{y}$.
%	}
%	\label{figOPEbosonicxy-conj0}
%\end{figure}

\subsection{Initial relations}\label{sec:N=2initial}

In \cite{GLPZ} we also imposed the initial relations 
\be\label{inigen}
\begin{array}{rclrcl}
{} [\psi_0,x_s] & = & 0 \qquad \qquad & [\psi_{1},{x}_s]&=& - h_2^{-1} \, {x}_s  \\
{}[\hat\psi_{0},{x}_s]&=& 0  \qquad \qquad & [\hat\psi_{1},{x}_s]&=& h_2^{-1} x_s\\
{} [\psi_0,\bar{x}_s] & = & 0 \qquad \qquad & [\psi_{1},\bar{x}_s]&=&  h_2^{-1} \, \bar{x}_s  \\
{}[\hat\psi_{0},\bar{x}_s]&=& 0  \qquad \qquad & [\hat\psi_{1},\bar{x}_s]&=&- h_2^{-1} \bar{x}_s \ ,
%[\hat\psi_{2},{x}_s]&=& \frac{2}{h_2} x_{s+1} - x_s - h_1 h_3 \hat{\psi}_0 x_s  \ .
\end{array}
\ee
as well as similar relations for the fermionic annihilation generators. As in the bosonic case reviewed above, they arise from the OPE relations (\ref{psiFBx0}) and (\ref{psiFBxbar0}) upon demanding that they also hold for 
the terms of the form $z^n w^{-r}$ with $n=0,1,2$ and $r>0$.

\section{The conjugate representation}\label{sec:conj}

For the following it will be important to understand in more detail the conjugate representation in terms of plane partitions. The basic idea was already described in \cite{GLPZ}, but here we will be more explicit. 
To be specific, we shall concentrate on the conjugate representation of the minimal representation, whose asymptotic plane partition configuration was described in Fig.~\ref{figxleftmin}; this is the representation that appears as the building block of our construction. 

\begin{figure}[h!]
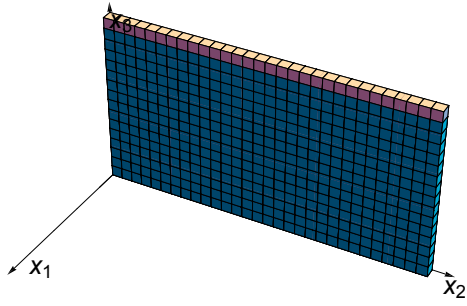

	\centering
	\includeCroppedPdf[width=0.4\textwidth]{"xbarleft"}
	\caption{The conjugate minimal representation. The yellow row of boxes on top of the wall are meant to indicate that only one row of boxes can be added.}
	\label{figxright}
\end{figure}

It was proposed in \cite{GLPZ} that the charge of the ground state of the conjugate representation of the minimal representation (denoted by $|\overline{\blacksquare}\rangle$) is given by 
\be\label{psiconjmin}
\psi_{\overline{\blacksquare}}(u) =\psi_0(u) \, \varphi_2^{-1}(-u-\sigma_3\psi_0 ) =
\frac{(u+\sigma_3\psi_0 + h_1) (u+\sigma_3\psi_0 + h_3)}{u (u+\sigma_3\psi_0 - h_2)} \ ,
\ee
where we have used the explicit formulae for $\psi_0(u)$ and $\varphi_2(u)$, see eqs.~(\ref{psi0def}) and (\ref{psiudef0}), respectively. 
In terms of plane partitions, this conjugate representation corresponds to the asymptotic Young diagram depicted in Fig.~\ref{figxright}, which consists of a high wall on which only a single row of boxes can be added. 
(This is indicated by the yellow row of boxes on top of the wall.) 
%We shall denote the ground state of the conjugate minimal representation by $|\overline{\blacksquare}\rangle$.

Since the poles of $\psi_{\overline{\blacksquare}}(u)$ are at $u=0$ and $u=h_2 - \sigma_3\psi_0$, the two excited states correspond to adding a box at either of these positions; thus the resulting charge functions are 
\begin{eqnarray}
\psi_{\overline{\blacksquare}+{\tiny{\yng(1)}}_{\, 0}}(u) & = & \psi_{\overline{\blacksquare}}(u)  \, \varphi_3(u) \nonumber \\
& = & \frac{(u+\sigma_3\psi_0 + h_1) (u+\sigma_3\psi_0 + h_3) (u+h_1) (u+h_2) (u+h_3)}{u (u+\sigma_3\psi_0 - h_2)(u-h_1) (u-h_2) (u-h_3)}
\end{eqnarray}
and 
\begin{eqnarray}
\psi_{\overline{\blacksquare}+{\tiny\yng(1)_{\ \textrm{top}}}}(u) & = & \psi_{\overline{\blacksquare}}(u)  \, \varphi_3(u+\sigma_3\psi_0 - h_2)  \nonumber \\
& = & 
\frac{(u+\sigma_3\psi_0 -h_2 + h_1) (u+\sigma_3\psi_0 -h_2 + h_3) (u+\sigma_3\psi_0 )}{u (u+\sigma_3\psi_0 - h_2) (u+\sigma_3\psi_0 - 2 h_2)} \label{mbar+2psi}
\ . 
\end{eqnarray}
In terms of operators, this then amounts to the identity 
\begin{equation}
e(z) |\overline{\blacksquare}\rangle \sim \frac{1}{z} \, |\overline{\blacksquare}+{\tiny\yng(1)_{\ 0}}\rangle +\frac{1}{(z+\sigma_3 \psi_0-h_2)}\, |\overline{\blacksquare}+{\tiny\yng(1)_{\ \textrm{top}}}\rangle \ .
\end{equation}

\begin{figure}[h!]
	\centering
	\begin{tabular}{cc}
	\includeCroppedPdf[width=.4\textwidth]{"xbarleftexcited0"}&
		\includeCroppedPdf[width=.4\textwidth
		]{"xbarleftexcitedtop"}
  \\
		 $e$ acts by adding a box at $(0,0,0)$\qquad  \qquad&$e$ acts by adding a box on top. \\
	\end{tabular}	
	\caption{The first descendant states of the conjugate minimal representation.}
	\label{figconjdes}
\end{figure}

Note that the two descendants of the conjugate representation look less symmetric than for the minimal case. 
The two descendants of the minimal representation both have a charge function that is fractional quartic:
\begin{equation}
\begin{aligned}
|\blacksquare+{\tiny\yng(1)_{\ 1}}\rangle:\qquad \psi(u) &= \frac{(u+\psi_0 \sigma_3) \, u (u+h_2-h_1) (u+h_3 - h_1)}{(u-h_1) (u-h_3) (u-2 h_1) (u+h_3)} \\
|\blacksquare+{\tiny\yng(1)_{\ 3}}\rangle:\qquad \psi(u) 
%&= \bigl( 1 + \frac{\psi_0 \sigma_3}{u} \bigr) \, \varphi_2(u) \, \varphi_3(u-h_3)\\
&= \frac{(u+\psi_0 \sigma_3) \, u (u+h_2-h_3) (u+h_1 - h_3)}{(u-h_1) (u-h_3) (u-2 h_3) (u+h_1)} \  ,
\end{aligned}
\end{equation}
whereas in the conjugate case one is fractional cubic and one fractional quintic,
\begin{equation}\label{mbar+2}
\begin{aligned}
|\overline{\blacksquare}+{\tiny\yng(1)_{\ 0}}\rangle: \qquad \psi(u) 
%& = & \psi_{\bar{\rm m}}(u)  \, \varphi_3(u) \nonumber \\
& =  \frac{(u+\sigma_3\psi_0 + h_1) (u+\sigma_3 \psi_0+ h_3) (u+h_1) (u+h_2) (u+h_3)}{u (u+\sigma_3\psi_0 - h_2)(u-h_1) (u-h_2) (u-h_3)}\\
 |\overline{\blacksquare}+{\tiny\yng(1)_{\ \textrm{top}}}\rangle: \qquad \psi(u) 
%& = & \psi_{\bar{\rm m}}(u)  \, \varphi_3(u+\sigma_3\psi_0 - h_2)  \nonumber \\
& =  
\frac{(u+\sigma_3\psi_0 -h_2 + h_1) (u+\sigma_3\psi_0 -h_2 + h_3) (u+\sigma_3\psi_0 )}{u (u+\sigma_3\psi_0 - h_2) (u+\sigma_3\psi_0 - 2 h_2)}  \ .
\end{aligned}
\end{equation}

This can be understood from the plane partition picture. 
The pole at $u=0$ corresponds to the first box that can be added on the floor, see Fig.~\ref{figconjdes}, while the pole at $u=h_2 - \sigma_3\psi_0$ describes the box that can be added on top of the wall. 
The fact that their charge functions have a different structure then just reflects that there are more descendants of the former state than of the latter. 

Given this apparent asymmetry between the two descriptions, one may worry whether the proposal for the conjugate representation is indeed correct. 
In order to check this, we have evaluated the eigenvalues of $W^{(3)}_0$ (see eq.~(\ref{W30})) and $W^{(4)}_0$ on the ground states of the minimal and the conjugate minimal representation, and confirmed that their spin-$3$ charges are opposite while the spin-$4$ charges agree.
%y agree (up to a sign for the case of $s=3$). 

%The details are as follows. 
More specifically, the eigenvalue of the ground state of the minimal representation is 
%we find the eigenvalue 
\begin{eqnarray}
W^{(3)}_0 & = & \frac{1}{6}\left(h_2-h_3+\frac{h_1^2}{h_2}-\sigma_3 \psi_0(3-h_1h_3\psi_0)\right) \nonumber \\
& = & \frac{(\lambda+1)(\lambda+2)}{3!}+\frac{1}{N} \left(-\frac{\lambda^2(\lambda+3)}{12}\right) \ . \label{W3min}
\end{eqnarray}
This expression is exact in $\frac{1}{N}$, and it agrees with the prediction 
\begin{equation}
{W}^{(3)}_0=\frac{h}{3}\left(\frac{-5N_3 (16 h^2+2 \tilde{c}\, h+\tilde{c} -10 h)}{2 \tilde{c}\, h - 3 \tilde{c}-2 h}\right)^{\frac{1}{2}}
\end{equation}
of \cite[eq.~(B.13)]{Gaberdiel:2012ku}, using that the conformal dimension of the minimal representation equals 
\be
h = \frac{1}{2}(1-h_1 h_3 \psi_0) \ , 
\ee
the central charge $\tilde{c}$ of the corresponding ${\cal W}_\infty$ algebra is
\be
\tilde{c} =-\psi_0\left[\s_2+(\s_3)^2(\psi_0)^2\right]-1 \  ,
\ee
and the $N_3$ parameter of \cite{Gaberdiel:2012ku} is identified with
\be
N_3 = \frac{1}{5}\frac{\left(-(\s_3)^2(\psi_0)^3-4\s_2\psi_0-8\right) }{\psi_0 } \ , \label{N3}
\ee
see \cite[eq.~(3.43)]{Gaberdiel:2017dbk}.\footnote{Incidentally, there is a misprint in the last formula of this equation: the numerator should be $(-\sigma_3^2\psi_0^3-4\sigma_2 \psi_0 -8)$.}
The analysis for the conjugate representation works similarly, and the corresponding eigenvalue is 
\begin{eqnarray}
\bar{W}^{(3)}_0 & = & \frac{1}{6}\left(-2h_2+\frac{h_1h_3}{h_2}+\sigma_3 {\psi}_0(3-h_1h_3 {\psi}_0)\right) \nonumber \\
& = & -\frac{(1+\lambda)(2+\lambda)}{3!}+\frac{1}{N} \left(\frac{\lambda^2(\lambda+3)}{12}\right) \ ,
\end{eqnarray}
which is indeed the negative of (\ref{W3min}). For spin $s=4$ we find similarly
\begin{equation}
W^{(4)}_0=\frac{(\lambda+1)(\lambda+2)(\lambda+3)}{4!}+\frac{1}{N} \left(-\frac{\lambda^2(\lambda+5)(\lambda+1)}{4!}\right) = \bar{W}^{(4)}_0 \ . 
\end{equation}
We have also checked this correspondence for the first excited states, using eq.~(\ref{2.16}). As explained there, the plane partition states are in general not eigenstates of $W^{(3)}_0$ and $W^{(4)}_0$, and hence we need to diagonalise the action of $W^{(3)}_0$ and $W^{(4)}_0$ on the two descendants. After this is done, we find for the 
$W^{(3)}_0$ eigenvalues of the first descendants of the minimal representation 
\begin{equation}
W^{(3)}_0: \qquad 
\begin{cases}
\begin{aligned}
&\frac{(1+\lambda)(2+\lambda)}{6}+\frac{1}{N}\left(-\frac{2\lambda}{2+\lambda}-\frac{1}{12}\lambda^2(\lambda+3)\right)\\
&\frac{(2+\lambda)(7+\lambda)}{6}+\frac{1}{N}\left(-\frac{\lambda^2(\lambda+5)(\lambda+6)}{12(\lambda+2)}\right) \ ,
\end{aligned}
\end{cases}
\end{equation}
while for the conjugate minimal representation we obtain the same eigenvalues with the opposite sign. We have also checked that the spin $4$ eigenvalues agree between the first descendants of the minimal and the conjugate minimal representation.

\section{Twin-plane-partitions}\label{sec:4}

For the bosonic affine Yangian, the set of plane partitions 
%with given asymptotics $(\lambda_1,\lambda_2,\lambda_3)$ 
furnishes a (faithful) representation \cite{Tsymbaliuk14,Prochazka:2015deb,Gaberdiel:2017dbk}, 
%In particular, every plane partition configuration $\Lambda$ is an eigenstate of $\psi(u)$ and is characterized uniquely by by its charge function $\psi_{\Lambda}(u)$, de fined in (\ref{psieig}).
%The generators $e(u)$ and $f(u)$ act by adding and removing boxes from $\Lambda$, respectively, according to (\ref{ppart}).
and this allows one to deduce the algebra relations (\ref{bosonicdef}) directly from the action (\ref{ppart}). % and the fact that the set of $\Lambda$ with fixed asymptotics $(\lambda_1,\lambda_2,\lambda_3)$ forms a faithful representation. 
We want to imitate this idea for the ${\cal N}=2$ affine Yangian. 

In this section, we shall explain the construction of the twin-plane-partitions that form a natural representation of the ${\cal N}=2$ affine Yangian from \cite{GLPZ}. 
%, a natural representation is defined by the set of twin-plane-partitions, i.e.\ by a pair of plane partitions glued together along one leg. 
We shall then show how to efficiently compute the charges of these configurations, and how to use these charge functions to fix the pole structure of the action of the Yangian generators on them. 
%, that plays the role of faithful representation. 
%Then we will explain (1) the construction of twin plane partitions, (2) how to compute its charge functions, and (3) how to use these charge functions to fix the pole structure of the action of the Yangian generators. 

\subsection{Set of twin-plane-partitions}

As reviewed in section 2, the $\mathfrak{u}(1)\oplus\mathcal{W}^{\mathcal{N}=2}_{\infty}$ algebra has two commuting bosonic $\mathcal{W}_{1+\infty}$ algebras, and the fermionic generators transform in representations $(R\otimes S^{\star}, R^{\,\star}\otimes S)$  w.r.t.\ these two bosonic subalgebras.\footnote{Here by $R\otimes S^{\star}$ we mean the representation that involves both `boxes' and `anti-boxes', with the box-part being described by $R$, and the anti-box-part by $S^T$.} Here both $R$ and $S$ are described in terms of Young diagrams, and $R^{\,\star}$ is the conjugate of the transpose of $R$, see eq.~(\ref{stardef}), and similarly for $S$. 
By the isomorphism between the bosonic $\mathcal{W}_{1+\infty}$ algebra and the affine Yangian,  the ${\cal N}=2$ affine Yangian contains two commuting bosonic affine Yangian subalgebras (denoted as $\mathcal{Y}$ and $\widehat{\mathcal{Y}}$), and the fermionic generators transform as $(\lambda\otimes \hat{\rho}^{\star}, {\lambda}^{\star}\otimes\hat{\rho})$ w.r.t.\ these two bosonic subalgebras.

\subsubsection{Gluing rules}\label{sec:gluing}

In the following we shall mainly consider the vacuum module of the ${\cal N}=2$ affine Yangian. It consists of pairs of plane partitions glued together along the common $x_2$ and $\hat{x}_2$ directions, as we shall now explain.
The ground state is just the empty configuration. 
The bosonic raising operators add boxes in the left and right corners. 
The fermionic raising operators $x(u)$ create infinitely long rows connecting the left and right plane partitions. In particular, they add a box to $\lambda$, describing the asymptotic behaviour along the $x_2$ direction, and hence simultaneously affect the asymptotic behaviour $\lambda^\star$ along the $\hat{x}_2$ direction, where $\lambda^\star$ is the conjugate transpose of $\lambda$, see eq.~(\ref{stardef}). 
Similarly, the fermionic raising operators $\bar{x}(u)$ create infinitely long rows along the $\hat{x}_2$ direction
from the perspective of $\hat{\rho}$, and thus simultaneously affect the asymptotic behaviour ${\hat{\rho}}^{\,\star}$  with respect to the unhatted modes.  %with projection $\lambda$ onto the $x$-$z$ plane and $\overline{\lambda^t}$ onto the $\hat{x}$-$\hat{z}$ plane, where $\overline{\lambda^t}$ denotes the conjugate of the transpose of the Young diagram $\lambda$. The fermionic raising operator $\bar{x}(u)$ create infinitely long rows with $\lambda$

As will become apparent below, it is natural to 
%show later, to match the $\mathcal{W}^{\mathcal{N}=2}_{\infty}$ expectation and satisfy some basic consistency conditions, we can 
think of the conjugate representation in the internal leg to grow along the negative $x_1$ and $x_3$ (or $\hat{x}_1$ and $\hat{x}_3$) directions. 
In particular, the asymptotics associated to the conjugate (anti-box) representations can coexist with those of the regular (box) representations: the former describe the asymptotics in the quadrant with $x_1, x_3<0$, while the latter characterize the asymptotics in the quadrant with $x_1,x_3>0$. 
%Namely, one can take the tensor products of internal leg created by $x$ and those created by $\bar{x}$. 

Finally, each configuration in the vacuum module can be viewed as a pair of plane partitions with the following asymptotics
\begin{equation}
(0, \lambda\otimes {\hat{\rho}}^{\, \star} , 0) \qquad \textrm{and} \qquad (0, {\lambda}^{\star} \otimes \hat{\rho},0) \ .
\end{equation}
More generic representations are labeled by the four asymptotics $(\mu_1,\mu_3,\hat{\mu}_1,\hat{\mu}_3)$, and each state can be viewed as a pair of plane partitions with asymptotics
\begin{equation}
(\mu_1,\lambda\otimes {\hat{\rho}}^{\, \star}, \mu_3) \qquad \textrm{and} \qquad (\hat{\mu}_1, {\lambda}^{\star}\otimes \hat{\rho},\hat{\mu}_3) \ .
\end{equation}
%where $\lambda$ runs over all Young diagrams.
We shall call these configurations twin-plane-partitions. In the rest of the paper we shall mainly concentrate on the twin-plane-partition configurations with trivial asymptotics along $x_1$, $x_3$, $\hat{x}_1$ and $\hat{x}_3$, i.e.\  the vacuum module. 
\medskip

%As a consequence, the states of the ${\cal N}=2$ affine Yangian are described by pairs of plane partitions \textcolor{red}{glued along one leg}, or twin plane partitions. 

\begin{figure}[h!]
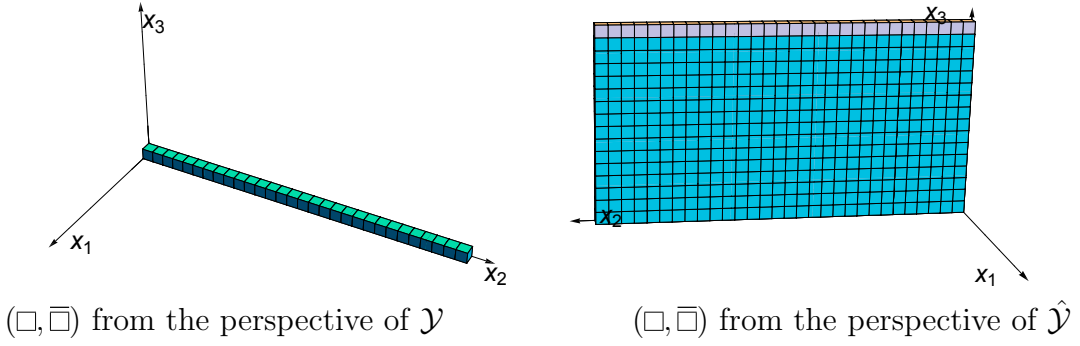

	\centering
	\begin{tabular}{c}
	\includeCroppedPdf[width=.4\textwidth]{"xleft"}\qquad
		\includeCroppedPdf[width=.4\textwidth
		]{"xright"}
  \\
		\vspace{-1.0cm} $({\tiny{\yng(1)}},\overline{{\tiny{\yng(1)}}})$ from the perspective of $\mathcal{Y}$\qquad \qquad  \qquad $({\tiny{\yng(1)}},\overline{{\tiny{\yng(1)}}})$ from the perspective of $\hat{\mathcal{Y}}$  \vspace{1.5cm}
	\end{tabular}	
	\caption{The state $|\blacksquare\rangle  $ in terms of twin plane partition. }
	\label{figxleftright}
\end{figure}

For example, the twin-plane-partition configuration in the vacuum module with bi-fundamental $({\tiny{\yng(1)}},\overline{{\tiny{\yng(1)}}})$ as internal leg is given by Fig. \ref{figxleftright}.
The one with bi-rep $({\tiny{\yng(2)}},\overline{{\tiny{\yng(1,1)}}})$ as internal leg is shown in Fig.~\ref{figxleftadd1} and the one with bi-rep $({\tiny{\yng(1,1)}},\overline{{\tiny{\yng(2)}}})$ in Fig.~\ref{figxleftadd3}. The conjugate of $({\tiny{\yng(1)}},\overline{{\tiny{\yng(1)}}})$ in Fig.~\ref{figxleftright} is $(\overline{{\tiny{\yng(1)}}},  {\tiny{\yng(1)}})$, and the corresponding twin-plane-partition is the mirror image of Fig.~\ref{figxleftright}, see Fig.~\ref{figxbarleftright}.

\begin{figure}[h!]
\centering 
\begin{tabular}{cc}
\includeCroppedPdf[width=.4\textwidth]{"xx2h1"}&
\includeCroppedPdf[width=.4\textwidth]{"xx2h1right"}\\
$({\tiny{\yng(2)}},\overline{{\tiny{\yng(1,1)}}})$ from the perspective of ${\cal Y}$ &
$({\tiny{\yng(2)}},\overline{{\tiny{\yng(1,1)}}})$ from the perspective of $\widehat{\cal Y}$
\end{tabular}	
\caption{The state $|\blacksquare\blacksquare_{1}\rangle$ in terms of twin plane partition. }
\label{figxleftadd1}
\end{figure}
\begin{figure}[h!]
\centering
\begin{tabular}{cc}
\includeCroppedPdf[width=.4\textwidth]{"xx2h3"}&
\includeCroppedPdf[width=.4\textwidth]{"xx2h3right"}\\
$({\tiny{\yng(1,1)}},\overline{{\tiny{\yng(2)}}})$ from the perspective of ${\cal Y}$ &
$({\tiny{\yng(1,1)}},\overline{{\tiny{\yng(2)}}})$ from the perspective of $\widehat{\cal Y}$
\end{tabular}	
\caption{The state $|\blacksquare\blacksquare_{3}\rangle$ in terms of twin plane partition.}
\label{figxleftadd3}
\end{figure}
\begin{figure}[h!]
	\centering
	\begin{tabular}{c}
	\includeCroppedPdf[width=.4\textwidth]{"xbarleft"}\qquad
		\includeCroppedPdf[width=.4\textwidth]{"xbarright"}
  \\
		\vspace{-1.0cm} $(\overline{{\tiny{\yng(1)}}},  {\tiny{\yng(1)}})$ from the perspective of $\mathcal{Y}$\qquad \qquad   $(\overline{{\tiny{\yng(1)}}},  {\tiny{\yng(1)}})$ from the perspective of $\widehat{\mathcal{Y}}$  \vspace{1.5cm}
	\end{tabular}	
	\caption{The state $|\overline{\blacksquare}\rangle$ in terms of twin plane partition.}
	\label{figxbarleftright}
\end{figure}

%\subsubsection{Gluing}
\subsubsection{Coordinate system}
%Now we explain the convention for the coordinates system for the twin plane partitions. 

Recall that for a plane partition $\Lambda$, a box in $\Lambda$ is labelled by its coordinates $x_i({\tiny{\yng(1)}})$, with $i=1,2,3$ and $x_i({\tiny{\yng(1)}})\in \mathbb{N}_0$, i.e.\ the box at the origin has coordinate $(0,0,0)$. 
We now generalize this to a coordinate system for the twin-plane-partitions. 

First, we use coordinates $x_i$ to label  the boxes in the left corner (denoted by ${\tiny{\yng(1)}}$), and $\hat{x}_i$ for hatted boxes  $\hat{{\tiny{\yng(1)}}}$ in the right corner. 
The coordinate for the box sitting at the bottom is the same as in the bosonic case:
\begin{equation}
\begin{aligned}
{\tiny\yng(1)}\textrm{ at bottom}:\qquad &x_1({\tiny\yng(1)})\,, x_2({\tiny\yng(1)})\,,x_3({\tiny\yng(1)})\,=0,1,2,3\dots\\
\widehat{{\tiny\yng(1)}}\textrm{ at bottom}:\qquad &\hat{x}_1(\widehat{{\tiny\yng(1)}})\,, \hat{x}_2(\widehat{{\tiny\yng(1)}})\,,\hat{x}_3(\widehat{{\tiny\yng(1)}})\,=0,1,2,3\dots
\end{aligned}
\end{equation}
Namely, for both left and right, the first box in the corner has coordinate $(0,0,0)$.

For boxes sitting on top of the conjugate representations, the natural coordinate system is determined by the pole structure of its first descendent in (\ref{mbar+2}),
\begin{equation}\label{coordtop}
{\tiny\yng(1)} \textrm{ on top}:\qquad \begin{cases} x_1({\tiny\yng(1)})&=0,-1,-2,\dots\qquad \\ x_2({\tiny\yng(1)})&=0,1,2,\dots\qquad\\
 x_3 ({\tiny\yng(1)})\,&=0,-1,-2,\dots \ , \qquad
\end{cases}
\end{equation}
and similarly for $\widehat{{\tiny\yng(1)}}$ sitting on top of the right window-sill,
\begin{equation}\label{coordhattop}
\widehat{{\tiny\yng(1)}} \textrm{ on top}:\qquad \begin{cases} \hat{x}_1(\widehat{{\tiny\yng(1)}})&=0,-1,-2,\dots\qquad \\
\hat{x}_2(\widehat{{\tiny\yng(1)}})&=0,1,2,\dots\qquad\\
\hat{x}_3(\widehat{{\tiny\yng(1)}})\,&=0,-1,-2,\dots\ . \qquad 
\end{cases}
\end{equation}
In either case we define 
\begin{align}
h({\tiny\yng(1)}) & \equiv x_1({\tiny\yng(1)})\,h_1+x_2({\tiny\yng(1)})\,h_2+x_3({\tiny\yng(1)})\,h_3  + \delta_{{\rm top}} (h_2-\sigma_3  \psi_0) \\ 
\hat{h}(\widehat{{\tiny\yng(1)}}) & \equiv \hat{x}_1(\widehat{{\tiny\yng(1)}})\,h_1+\hat{x}_2(\widehat{{\tiny\yng(1)}})\,h_2+\hat{x}_3(\widehat{{\tiny\yng(1)}})\,h_3 + \delta_{{\rm top}}  (h_2-\sigma_3 \hat{\psi}_0)  \ ,
\end{align}
where $\delta_{{\rm top}}=1$ if the box sits on top of the wall, and $\delta_{{\rm top}}=0$ otherwise; explicitly 
\begin{equation}\label{hbottop}
h({\tiny\yng(1)})=\begin{cases} 
\begin{array}{ll}
m h_1 +n h_3 + \ell h_2 \qquad \qquad   \qquad & \qquad {\tiny\yng(1)}\in \textrm{bottom}\\
-m h_3 - n h_1 + \ell h_2 +(h_2-\sigma_3 \psi_0)  & \qquad {\tiny\yng(1)}\in \textrm{top}
\end{array}
\end{cases}
\end{equation}
and 
\begin{equation}\label{hhatbottop}
\hat{h}(\widehat{{\tiny\yng(1)}})=\begin{cases} 
\begin{array}{ll}
\hat{m} h_1 + \hat{n} h_3 + \ell h_2 \qquad \qquad   \qquad & \qquad \widehat{{\tiny\yng(1)}}\in \textrm{bottom}\\
- \hat{m} h_3 - \hat{n} h_1 + \ell h_2 +(h_2-\sigma_3 \hat{\psi}_0)  & \qquad \widehat{{\tiny\yng(1)}}\in \textrm{top} \ ,
\end{array}
\end{cases} 
\end{equation}
\smallskip

\noindent where $m, n, \hat{m}, \hat{n}, \ell \in \mathbb{N}_{0}$.
Thus $h({\tiny\yng(1)})$ and $\hat{h}(\widehat{{\tiny\yng(1)}})$ describe correctly the poles of the corresponding descendants. Note that  $x_1,x_3<0$ and $\hat{x}_1,\hat{x}_3<0$ in eqs.~(\ref{coordtop}) and (\ref{coordhattop}); this is the reason why we may think of the conjugate representations in terms of ``high walls" that are located in the quadrant with $x_1,x_3<0$ and $\hat{x}_1,\hat{x}_3<0$, see the comments in Section~\ref{sec:gluing}.
\smallskip

We also need to introduce labels for the individual (infinitely long) rows in $(\lambda, {\lambda}^{\star})$ and those in $(\hat{\rho}^{\,\star},\hat{\rho})$.  
For each internal leg, it is enough to focus on the representation $\lambda$ and $\hat{\rho}$ (as opposed to ${\lambda}^{\star}$ and $\hat{\rho}^{\, \star}$). 
For $(\lambda, {\lambda}^{\star})$, we label each box in the Young diagram $\lambda$ by $\blacksquare$, and choose its coordinate
\begin{equation}
\blacksquare :\qquad x_1(\blacksquare)\,, x_3(\blacksquare)=0,1,2,3 \ldots \ .
\end{equation}
Since a $\blacksquare$ is visible from both sides, it has two coordinate functions, defined as
\begin{eqnarray}g(\blacksquare)&\equiv &x_1(\blacksquare) h_1+ x_3(\blacksquare) h_3  \label{gdef}\\
\hat{g}(\blacksquare)&\equiv &-x_3(\blacksquare) h_1- x_1(\blacksquare) h_3+h_2-\sigma_3\hat{\psi}_0\label{ghatdefx} \ ,
\end{eqnarray}
which reflect the fact that $\blacksquare$ is inside the Young diagram $\lambda$ on the unhatted side, and the conjugate transpose ${\lambda}^{\star}$ on the hatted side. Note the parallel between the definition of the coordinate function $\hat{h}(\hat{\tiny\yng(1)})$ when $\hat{\tiny\yng(1)}$ is on top and $\hat{g}({\blacksquare})$.
\smallskip

Similarly,  for $(\hat{\rho}^{\,\star}, \hat\rho)$, we label each box in the Young diagram $\hat{\rho}$ by $\overline{\blacksquare}$, and choose its coordinate
\begin{equation}
\overline{\blacksquare} :\qquad \hat{x}_1(\overline{\blacksquare})\,, \hat{x}_3(\overline{\blacksquare})=0,1,2,3  \ldots \ , 
\end{equation}
with coordinate function
\begin{eqnarray}
g(\overline\blacksquare)&\equiv&-\hat{x}_3(\overline{\blacksquare}) \, h_1- \hat{x}_1(\overline{\blacksquare}) \, h_3+ h_2-\sigma_3\psi_0 \label{gdefxbar}\\
\hat{g}(\overline\blacksquare)&\equiv&\hat{x}_1(\overline{\blacksquare}) \, h_1+ \hat{x}_3(\overline{\blacksquare}) \, h_3 \ . \label{ghatdef}
\end{eqnarray}
Note the parallel between the definition of the coordinate function $h({\tiny\yng(1)})$ when ${\tiny\yng(1)}$ is on top and $g(\overline{\blacksquare})$.

For the following it will also be convenient to define another coordinate function for these boxes, 
%which is more related to the pole structure 
\begin{align}
h(\blacksquare) & \equiv  x_1(\blacksquare) h_1 + x_3(\blacksquare) h_3 + \bigl(x_1(\blacksquare)+ x_3(\blacksquare)  \bigr)h_2 \\
\hat{h}(\overline{\blacksquare}) & \equiv\hat{x}_1(\overline{\blacksquare}) h_1 + \hat{x}_3(\overline{\blacksquare}) h_3 + \bigl(\hat{x}_1(\overline{\blacksquare})+ \hat{x}_3(\overline{\blacksquare})  \bigr)h_2  \ .
\end{align}
Note that because of $h_1+h_2+h_3=0$, these can also be rewritten as\footnote{As will be explained in Section~\ref{sec:4.5},  $h(\blacksquare)$ is directly related to the pole corresponding to adding a $\blacksquare$, whereas $\hat{g}(\blacksquare)$ appears in the conjugate charge function of  $\blacksquare$. The situation is similar for $\hat{h}(\overline{\blacksquare})$ and $g(\overline{\blacksquare})$, respectively %Therefore although $h(\blacksquare)$ and $\hat{g}(\blacksquare)$ share the same function form due to $h_1+h_2+h_3=0$, we distinguish between them in the derivation of formulae; the identity between the two is crucial for many results. 
}
\begin{align}
h(\blacksquare) & =  - x_3(\blacksquare) h_1 - x_1(\blacksquare) h_3 =\hat{g}(\blacksquare) + \sigma_3\hat{\psi}_0-h_2\label{hblack}\\
\hat{h}(\overline{\blacksquare}) & = - \hat{x}_3(\overline{\blacksquare}) h_1 - \hat{x}_1(\overline{\blacksquare}) h_3 =g(\overline{\blacksquare}) + \sigma_3\psi_0-h_2\label{hbarblack} \ . 
\end{align}

%\begin{equation}
%h(\blacksquare)=\begin{cases} \begin{aligned}&
%x_1(\blacksquare) h_1 + x_3(\blacksquare) h_3 + (x_1(\blacksquare)+ x_3(\blacksquare)  )h_2 \qquad   \qquad &\blacksquare\in \textrm{bottom}\\
%&-m h_3 - n h_1  +(h_2-\sigma_3 \psi_0)  &\blacksquare\in \textrm{top}
%\end{aligned}
%\end{cases} \qquad m, n \in \mathbb{Z}_{+}
%\end{equation}
%\begin{equation}
%h(\overline{\blacksquare})=\begin{cases} \begin{aligned}&
%m h_1 +n h_3 + (m+n)h_2 \qquad   \qquad &\overline{\blacksquare}\in \textrm{bottom}\\
%&-m h_3 - n h_1  +(h_2-\sigma_3 \hat{\psi}_0)  &\overline{\blacksquare}\in \textrm{top}
%\end{aligned}
%\end{cases} \qquad m, n\in \mathbb{Z}_{+}
%\end{equation}

%\qquad \hat{g}(\overline\blacksquare)=\hat{x}(\blacksquare) h_1+ \hat{z}(\blacksquare) h_3
%\end{equation}
%\begin{equation}
%\begin{cases}
%\psi_{\blacksquare}(u)=\varphi_2(u-g(\blacksquare))\\
%\hat{\psi}_{\blacksquare}(u)=\varphi^{-1}_2(-(u-g(\blacksquare))-\sigma_3\hat{\psi}_0)
%\end{cases}
%\end{equation}
\subsection{Eigenvalues of a twin-plane-partition}

Just as a plane partition configuration is uniquely characterized by its eigenvalue function $\psi(u)$, a twin-plane-partition configuration is uniquely characterized by its eigenvalues $\psi(u)$ and $\hat{\psi}(u)$ with respect to the two bosonic affine Yangians.
% --- we shall denote the generators of the second affine Yangian by a hat. 
%Since there is a one-to-one map from these $(\psi(u), \hat{\psi}(u))$eigenvalues to $W^{(s)}_0$ charges, one can use this to determine twin-plane partition configurations that are known to have certain higher spin charges. 

A twin-plane-partition configuration (with trivial boundary conditions along $x_1$, $x_3$, $\hat{x}_1$ and $\hat{x}_3$) consists of four types of contributions: 
%\begin{comment}
%\begin{enumerate}
%\item  A bi-representation $(\lambda,\bar{\lambda}^{T})$  generated by $x$.
%\item A bi-representation $(\bar{\hat{\lambda}^{t}}, \hat{\lambda})$, generated by $\bar{x}$.
%\item Collection $\mathcal{E}$ of individual boxes on the left corner, generated by $e$.
%\item Collection $\hat{\mathcal{E}}$ of individual boxes on the left corner, generated by $\hat{e}$
%\end{enumerate}
%\end{comment}
\begin{enumerate}
\item A bi-representation $(\lambda, {\lambda}^{\star})$  that is recursively generated by $x$.

It is enough to focus on $\lambda$. Let us label each box in the Young diagram $\lambda$ by $\blacksquare$. Then, using the OPEs (\ref{psiFBx0}) and (\ref{psiFBxbar0}), the contribution of $(\lambda,{\lambda}^{\star})$ to the $(\psi(u),\hat{\psi}(u))$ eigenfunctions is 
\begin{equation}\label{xforpsi}
\begin{aligned}
&\begin{cases}
\psi_{\lambda}(u)=\prod_{\blacksquare\in\lambda}\psi_{\,\blacksquare}(u)\\
\hat{\psi}_{\lambda}(u)=\prod_{\blacksquare\in \lambda}\hat{\psi}_{\,\blacksquare}(u)
\end{cases}  \\ 
&\,\, \textrm{with}  \qquad\qquad \begin{cases}
\psi_{\,\blacksquare}(u)\equiv\varphi_2(u-g(\blacksquare))\\
\hat{\psi}_{\,\blacksquare}(u)\equiv\varphi^{-1}_2(u-\hat{g}(\blacksquare))=\varphi^{-1}_2(-u+h(\blacksquare)-\sigma_3\hat{\psi}_0)  
\end{cases}
\end{aligned}
\end{equation}
where in deriving the contribution to $\hat{\psi}(u)$ we have used $\varphi_2(-u)=\varphi_2(u-h_2)$.

\item A bi-representation $({\hat{\rho}}^{\,\star}, \hat{\rho})$, generated recursively by $\bar{x}$.\\
%This is the infinitely row with projection $\hat{\lambda}$ onto the $\hat{x}$-$\hat{z}$ plane and $\overline{\hat{\lambda}^{t}}$ onto the $x$-$z$ plane. 
Each box in the Young diagram $\hat{\rho}$ is labelled by $\overline{\blacksquare}$, and the contribution of $({\hat{\rho}}^{\,\star}, \hat{\rho})$ to the $(\psi(u),\hat{\psi}(u))$ eigenfunctions is 
\begin{equation}\label{xbarforpsi}
\begin{aligned}
&\begin{cases}
\psi_{\hat{\rho}}(u)=\prod_{\overline{\blacksquare}\in \hat{\rho}}\psi_{\, \overline{\blacksquare}}(u)\\
\hat{\psi}_{\hat{\rho}}(u)=\prod_{\overline{\blacksquare}\in \hat{\rho}}\hat{\psi}_{\, \overline{\blacksquare}}(u)
\end{cases}  \\
&\,\, \textrm{with} \qquad \qquad\begin{cases}
\psi_{\, \overline{\blacksquare}}(u)\equiv \varphi^{-1}_2(u-g(\overline{\blacksquare}))=\varphi^{-1}_2(-u+\hat{h}(\overline{\blacksquare}) -\sigma_3\psi_0) \\
\hat{\psi}_{\, \overline{\blacksquare}}(u)\equiv\varphi_2(u-\hat{g}(\overline{\blacksquare})) \ , 
\end{cases}
\end{aligned}
\end{equation}
which mirrors (\ref{xforpsi}).

\item Collection $\mathcal{E}$ of individual boxes in the left corner, generated by $e$.\\
Each individual box  is labeled by ${\tiny{\yng(1)}}$, and irrespective of whether it sits at the bottom or on top, every ${\tiny\yng(1)}$ contributes to the $\psi(u)$ eigenfunction as
\begin{equation}\label{eforpsi}
\psi_{\mathcal{E}}(u)=\prod_{{\tiny\yng(1)}\in\mathcal{E}}\psi_{\,{\tiny\yng(1)}}(u)
 \qquad \textrm{with} \qquad
\psi_{\,{\tiny\yng(1)}}(u)\equiv\varphi_3(u-h({\tiny\yng(1)})) \ . 
\end{equation}
It does not contribute to the $\hat{\psi}(u)$ eigenfunction.

\item Collection $\hat{\mathcal{E}}$ of individual boxes in the right corner, generated by $\hat{e}$.\\
Each individual box  is labeled by $\widehat{{\tiny{\yng(1)}}}$, and irrespective of whether it sits at the bottom or on top, a $\widehat{{\tiny\yng(1)}}$ contributes to the $\hat{\psi}(u)$ eigenfunction as
\begin{equation}\label{hateforpsi}
\hat{\psi}_{\hat{\mathcal{E}}}(u)=\prod_{\widehat{{\tiny\yng(1)}}\in\mathcal{E}}\hat{\psi}_{\,\widehat{{\tiny\yng(1)}}}(u)
 \qquad \textrm{with} \qquad
\hat{\psi}_{\,\widehat{{\tiny\yng(1)}}}(u)\equiv\varphi_3(u-\hat{h}(\widehat{{\tiny\yng(1)}}))\ . 
\end{equation}
It does not contribute to the $\psi(u)$ eigenfunction.
\end{enumerate}
In summary, a twin-plane-partition $\Lambda$ can thus be labeled by the quartet $(\lambda, \hat{\rho},\mathcal{E},\hat{\mathcal{E}})$.  
Its eigenfunctions $(\psi(u),\hat{\psi}(u))$ are then
\begin{equation}\label{chargefunctions}
\begin{cases}
\psi_{\Lambda}(u)=\psi_0(u) \cdot \psi_{\lambda}(u)\cdot \psi_{\hat{\rho}}(u)\cdot \psi_{\mathcal{E}}(u)\\
\hat{\psi}_{\Lambda}(u)=\hat{\psi}_0(u) \cdot \hat{\psi}_{\lambda}(u)\cdot \hat{\psi}_{\hat{\rho}}(u)\cdot\hat{\psi}_{\hat{\mathcal{E}}}(u) \ , 
\end{cases}
\end{equation}
where  the vacuum factors $\psi_0(u)$ and $\hat{\psi}_0(u)$ are defined as 
\begin{equation}
\psi_0(u)\equiv1+ \frac{\sigma_3\psi_0}{u} \qquad\textrm{and} \qquad \hat{\psi}_0(u)\equiv1+\frac{\sigma_3  \hat{\psi}_0}{u} \ .
\end{equation}
\begin{comment}
For simple configurations such as states in the vacuum module, one can compute their $\psi(u)$ and $\hat{\psi}(u)$ eigenvalues directly from the definition (\ref{psieig}) and its analogue for  $\hat{\psi}(u)$. 
Unfortunately, this is unfeasible for generic twin-plane-partitions. In particular, for the case of a non-trivial module, the number of boxes is infinite and need to be regulated.\footnote{ For example, in deriving the $\psi(u)$ eigenvalue of the minimal representation (\ref{psiudef0}) we have implicitly chosen a regularization scheme.} 
However, it is not clear whether one can choose a consistent regularization scheme for all cases; in particular, it is not obvious how to obtain the charge function of the conjugate representation (\ref{psiconjmin}) in this manner. 
\textbf{WL: rewrite above}
\end{comment}
%Now we explain these four types of contributions to  (\ref{chargefunctions}) in turn.
%contribution to the charge function 

\subsection{Using eigenvalues to fix poles of generators}
Next we explain how to use the twin-plane-partitions to determine the algebra relations. We will first focus on the pole structures, and only look at the detailed coefficients in Section~\ref{sec:xaction}.

The main idea behind this approach is to demand that the algebra acts within the set of twin-plane-partitions.
In the previous subsections we defined the allowed set of twin-plane-partition configurations, and identified their charge functions $(\psi(u),\hat{\psi(u)})$. These charge functions characterize the allowed plane partition configurations uni\-quely. Thus, given a pair of putative charge functions $(\psi(u),\hat{\psi}(u))$, we can determine whether they correspond to an allowed twin-plane-partition configuration or not. This constraint is sufficient to determine the pole structure of the algebra  action, as we shall now explain.
\smallskip

Recall that the creation operators of the ${\cal N}=2$ affine Yangian are $e$, $\hat{e}$, $x$ and $\bar{x}$, and that their OPE relations with $\psi$ and $\hat{\psi}$ are
\begin{equation}\label{psiFBe}
e:\qquad \qquad
\begin{cases}
\begin{aligned}
\psi(z) \, e(w)  &\sim  \varphi_3(\Delta) \, e(w)\, \psi(z)   \\
\hat{\psi}(z) \, e(w)  &\sim   e(w) \, \psi(z)   
\end{aligned}
\end{cases}
\end{equation}
\begin{equation}\label{psiFBehat}
\hat{e}:\qquad \qquad
\begin{cases}
\begin{aligned}
\psi(z) \, \hat{e}(w)  &\sim \hat{e}(w)\, \psi(z)   \\
\hat{\psi}(z) \, \hat{e}(w)  &\sim \varphi_3(\Delta) \,   \hat{e}(w) \, \psi(z)   
\end{aligned}
\end{cases}
\end{equation}
\begin{equation}\label{psiFBx} 
x:\qquad \qquad
\begin{cases}
\begin{aligned}
\psi(z) \, x(w)  &\sim  \varphi_2(\Delta) \, x(w) \,\psi(z)   \\
\hat{\psi}(z) \, x(w)  &\sim  \varphi^{-1}_2(-\Delta-\sigma_3\hat{\psi}_0) \, x(w) \,\hat{\psi}(z) 
\end{aligned}
\end{cases}
\end{equation}
and
\begin{equation}\label{psiFBxbar} 
\bar{x}:\qquad \qquad
\begin{cases}
\begin{aligned}
\psi(z) \, \bar{x}(w)  &\sim  \varphi^{-1}_2(-\Delta-\sigma_3\psi_0) \, \bar{x}(w)\, \psi(z)   \\
\hat{\psi}(z) \, \bar{x}(w)  &\sim  \varphi_2(\Delta) \, \bar{x}(w)\, \hat{\psi}(z) \ .
\end{aligned}
\end{cases}
\end{equation}
In order to define the successive action of these generators on the vacuum, we need to understand their pole structure when acting on a twin-plane-partition. 

%The first piece of information we need to fix is the pole structure when successively applying the  generators $\{e,\hat{e}, x, \bar{x}\}$ on the vacuum. 
There is a simple method to evaluate  the $\psi(u)$ and $\hat{\psi}(u)$ eigenvalues of any state obtained by the action of these generators, using the OPE relations (\ref{psiFBe}) -- (\ref{psiFBxbar}). 
To see this, let us  
%Since we know the OPEs (\ref{psiFBe}) -- (\ref{psiFBxbar}), we can use the following method to evaluate the $\psi(u)$ and $\hat{\psi}(u)$ eigenvalues of any twin plane partition generated by a series of these generators. Let us 
consider a state that is generated by applying operators from the set $\{e,\hat{e}, x, \bar{x}\}$ on the vacuum 
\begin{equation}\label{genericstate}
g_1(z_1) g_2 (z_2) \dots g_n (z_n) |\emptyset\rangle \qquad \textrm{with}\qquad g_i \in \{e,\hat{e}, x, \bar{x}\} \ . 
\end{equation}
The resulting state is then a linear combination of the form
\begin{equation}
g_1(z_1) g_2 (z_2) \dots g_n (z_n) |\emptyset\rangle =\sum_{\{z^{*}_i\}}\, \prod_{i=1}^{n} \frac{\#}{z_i-z^{*}_i}\,  |\Phi(\{z^{*}_i\})\rangle \ , 
\end{equation}
where the sum runs over a finite number of ordered sets $\{z^{*}_1,\ldots, z^{*}_n\}$, labelling the poles of the different functions. A priori, we do not know the positions of all of these poles (and hence the set of poles we have to sum over), but we can fix them by demanding that the $(\psi(u),\hat{\psi}(u))$ eigenvalues of $ |\Phi(\{z^{*}_i\})\rangle$ correspond to an allowed twin-plane-partition. 

More specifically, we can compute these eigenvalues by passing $\psi(u)$ and $\hat{\psi}(u)$ through (\ref{genericstate}), using the OPE relations (\ref{psiFBe}) -- (\ref{psiFBxbar}), as well as the fact that all of these relations only hold up to regular terms. This leads to the simple formula
\begin{equation}\label{generalcharge}
 |\Phi(\{z^{*}_i\})\rangle: \qquad 
\begin{cases}
\begin{aligned} 
\psi(u)&=\psi_0(u)\prod_{i=1}^{n} \phi[g_i](u-z^{*}_i) \\
%\qquad \textrm{and}\qquad
 \hat{\psi}(u)&=\hat{\psi}_0(u)\prod_{i=1}^{n}  \hat{\phi}[g_i]( u-z^{*}_i) \ ,
\end{aligned}
 \end{cases}
\end{equation}
where 
\be\label{charge1}
\begin{array}{ll}
\phi[e](u)\equiv \varphi_3(u)\qquad \qquad  \qquad \qquad &\phi[\hat{e}](u)\equiv 1 \\
\phi[x](u)\equiv \varphi_2(u) \qquad\qquad \qquad \qquad &\phi[\bar{x}](u)\equiv \varphi^{-1}_2(-u-\sigma_3 \psi_0)
\end{array}
\ee
and
\begin{equation}\label{charge2}
\begin{array}{ll}
\hat{\phi}[e](u)\equiv 1 \qquad \qquad & \hat{\phi}[\hat{e}](u)\equiv \varphi_3(u)\\
\hat{\phi}[x](u)\equiv \varphi_2^{-1}(-u-\sigma_3\hat{\psi}_0) \qquad \qquad & \hat{\phi}[\bar{x}](u)\equiv \varphi_2(u) \ .
\end{array}
\end{equation}

Let us illustrate this method first with some examples where we already know the answer from first principles, and then apply it to cases where the resulting box configurations can be determined in this manner. 

\subsection{Some illustrative examples}

For the vacuum state $|\emptyset\rangle$ the charge functions equal the vacuum factors, 
%\begin{equation}
% \psi_0(u)\equiv 1+\sigma_3 \frac{\psi_0}{u} \qquad \textrm{and}\qquad \hat{\psi}_0(u)\equiv 1+\sigma_3 \frac{\hat{\psi}_0}{u} 
%\end{equation}
%The $(\psi(u), \hat{\psi}(u))$ eigenvalues of the vacuum $|\emptyset\rangle$ are
\begin{equation}
|\emptyset\rangle: \qquad 
\begin{cases}
\begin{aligned} 
\psi(u)&=\psi_0(u)\equiv 1+ \frac{\sigma_3\psi_0}{u} \\
%\qquad \textrm{and}\qquad
 \hat{\psi}(u)&= \hat{\psi}_0(u)\equiv 1+ \frac{\sigma_3\hat{\psi}_0}{u} \ .
\end{aligned}
 \end{cases}
\end{equation}

\subsubsection{$e$ and $\hat{e}$ descendants of vacuum}

Let us denote, as before, the state with one single box in the left corner (the one corresponding to the unhatted affine Yangian generators) by $|{\tiny\yng(1)}\rangle$, i.e.\ 
\begin{equation}
 |{\tiny\yng(1)}\rangle \equiv e_0  |\emptyset\rangle  \ , \qquad 
e(z) |\emptyset\rangle \sim \frac{1}{z} |{\tiny\yng(1)}\rangle \ .
\end{equation}
The $(\psi(u), \hat{\psi}(u))$ eigenvalues of  $|{\tiny\yng(1)}\rangle$ are then 
\begin{equation}
|{\tiny\yng(1)}\rangle: \qquad 
\begin{cases}
\begin{aligned} 
\psi(u)&=\psi_0(u) \cdot \varphi_3(u)\\
 % \qquad \textrm{and}\qquad
\hat{\psi}(u)&=\hat{\psi}_0(u) \ , 
\end{aligned}
 \end{cases}
\end{equation}
in agreement with (\ref{generalcharge}). 
The situation is analogous for the case of a single box in the right corner (the one corresponding to the hatted affine Yangian generators), whose state we denote by $|\hat{{\tiny\yng(1)}}\rangle$, where
\begin{equation}
 |\hat{{\tiny\yng(1)}}\rangle \equiv \hat{e}_0  |\emptyset\rangle \ , \qquad \hat{e}(z) |\emptyset\rangle \sim \frac{1}{z} |\hat{{\tiny\yng(1)}}\rangle \ . 
\end{equation}
The $(\psi(u), \hat{\psi}(u))$ eigenvalues of  $|\hat{{\tiny\yng(1)}}\rangle$ are then 
\begin{equation}
|\hat{{\tiny\yng(1)}}\rangle: \qquad 
\begin{cases}
\begin{aligned} 
\psi(u)&=\psi_0(u) \\
%\qquad \textrm{and}\qquad
 \hat{\psi}(u)&=\hat{\psi}_0(u)  \cdot \varphi_3(u) \ .
\end{aligned}
 \end{cases}
\end{equation}

\subsubsection{$x$ and $\bar{x}$ descendants of vacuum}

For the modes of $x_r$ and $\bar{x}$ we define the corresponding generating functions as 
\be
x(z) = \sum_{r=1/2}^{\infty} \frac{x_r}{z^{r+1/2}} \ , \qquad 
\bar{x}(z) = \sum_{r=1/2}^{\infty} \frac{\bar{x}_r}{z^{r+1/2}} \ .
\ee
The condition that the vacuum is annihilated by $x_s$ with $s\geq \frac{3}{2}$, implies that 
\begin{equation}
x(z) |\emptyset\rangle \sim \frac{1}{z} |\blacksquare\rangle \ , \qquad \hbox{where} \qquad 
|\blacksquare\rangle \equiv x_{\frac{1}{2}}  |\emptyset\rangle \ . 
\end{equation}
Since $|\blacksquare\rangle$ transforms in the minimal (conjugate minimal) representation with respect to the unhatted  (hatted) modes, we have 
\begin{equation}\label{eigenx}
|\blacksquare\rangle: \qquad 
\begin{cases}\begin{aligned}
 \psi(u)=&\psi_0(u) \cdot \varphi_2(u) \\%\qquad \textrm{and}\qquad 
 \hat{\psi}(u)=&\hat{\psi}_0(u)\cdot \varphi^{-1}_2(-u-\sigma_3 \hat{\psi}_0) \ , 
 \end{aligned}
 \end{cases}
\end{equation}
again in agreement with eq.~(\ref{generalcharge}). In terms of twin-plane-partitions, the relevant state is therefore described by Fig.~\ref{figxleftright}. 
For the conjugate generator, we have similarly
\begin{equation}
\bar{x}(z) |\emptyset\rangle \sim \frac{1}{z} |\overline{\blacksquare}\rangle \ , \qquad \hbox{where} \qquad 
|\overline{\blacksquare}\rangle \equiv \bar{x}_{\frac{1}{2}}  |\emptyset\rangle \ ,
\end{equation}
and the corresponding charges are now 
\begin{equation}
|\overline{\blacksquare}\rangle: \qquad \left\{
%\begin{cases}
\begin{aligned} 
\psi(u)&=\psi_0(u) \cdot \varphi^{-1}_2(-u-\sigma_3\psi_0) \\
%\qquad \textrm{and}\qquad 
\hat{\psi}(u)&=\hat{\psi}_0(u)\cdot \varphi_2(u) \ .
\end{aligned} \right.
% \end{cases}
\end{equation}
In terms of twin-plane-partitions we now have the situation depicted in Fig.~\ref{figxbarleftright}.

\subsubsection{$e$/$\hat{e}$ descendents of $|\blacksquare\rangle$ and $|\overline{\blacksquare}\rangle$}

As we have seen above, the charge function of the ground state of the minimal representation along the $x_2$ direction is given by (\ref{eigenx}). 
Next we want to study its descendants created by the action of  $e(z)$. 
This is straightforward since the action of $e(z)$ on $|\blacksquare\rangle$ only affects the unhatted algebra, and since we know the structure of the minimal representation, following (\ref{ppart}). In fact, we simply find 
\begin{equation}\label{eonxfinal}
e(z)x(w) |\emptyset\rangle \sim  \frac{1}{w} \, \Bigl[\frac{1}{z-h_1}\,  |\blacksquare+{\tiny\yng(1)_{\, 1}}\rangle +\frac{1}{z-h_3} \, |\blacksquare+{\tiny\yng(1)_{\, 3}}\rangle\Bigr] \ , 
\end{equation}
where 
\begin{equation}\label{m+1}
|\blacksquare+{\tiny\yng(1)_{\, j}}\rangle:\qquad 
\begin{cases}
\begin{aligned}
\psi(u) &= \psi_0(u) \, \varphi_2(u) \, \varphi_3(u-h_j) \\
\hat{\psi}(u) &= \psi_0(u) \, \varphi_2(u)  \ ,
%&= \frac{(u+\psi_0 \sigma_3) \, u (u+h_2-h_1) (u+h_3 - h_1)}{(u-h_1) (u-h_3) (u-2 h_1) (u+h_3)} \ , 
\end{aligned}
\end{cases}
\end{equation}
%and
%\begin{equation}\label{m+3}
%|\blacksquare+{\tiny\yng(1)_{\, 3}}\rangle:\qquad
%\begin{cases} \begin{aligned}
%\psi(u) &= \psi_0(u) \, \varphi_2(u) \, \varphi_3(u-h_3) \\
%\hat{\psi}(u) &= \psi_0(u) \, \varphi_2(u)  \ , \\
%%&= \frac{(u+\psi_0 \sigma_3) \, u (u+h_2-h_1) (u+h_3 - h_1)}{(u-h_1) (u-h_3) (u-2 h_1) (u+h_3)} \ , 
%\end{aligned}
%\end{cases}
%\end{equation}
and $j=1,3$, see eq.~(\ref{xecharges}). The analysis for $e(z)$ on $|\overline{\blacksquare}\rangle$ works similarly, now using instead 
eq.~(\ref{mbar+2}); this leads to 
\begin{equation}\label{4.23}
e(z) \bar{x}(w)   |\emptyset\rangle \sim  \frac{1}{w} \, \Bigl[
\frac{1}{z} \, |\overline{\blacksquare}+{\tiny\yng(1)_{\, 0}}\rangle +\frac{1}{(z+\sigma_3 \psi_0-h_2)} \, |\overline{\blacksquare}+{\tiny\yng(1)_{\, \textrm{top}}}\rangle \Bigr] \ , 
\end{equation}
where 
\begin{equation}
|\overline{\blacksquare}+{\tiny\yng(1)_{\, 0}}\rangle:\qquad 
\begin{cases}
\begin{aligned}
\psi(u) &= \psi_0(u)\cdot \varphi^{-1}_2(-u-\sigma_3 \psi_0)\cdot \varphi_3(u) \\
\hat{\psi}(u) &=\hat{\psi}_0(u)\cdot \varphi_2(u)
\end{aligned}
\end{cases}
\end{equation}
and
\begin{equation}\label{xtopeigenvalue}
 |\overline{\blacksquare}+{\tiny\yng(1)_{\, \textrm{top}}}\rangle:\qquad 
\begin{cases}
\begin{aligned}
\psi(u) &= \psi_0(u)\cdot \varphi^{-1}_2(-u-\sigma_3 \psi_0) \cdot \varphi_3(u+\sigma_3\psi_0-h_2)\\
\hat{\psi}(u) &=\hat{\psi}_0(u)\cdot \varphi_2(u) \ . 
\end{aligned}
\end{cases}
\end{equation}
All of the resulting charge functions are of the form of (\ref{generalcharge}).
\smallskip
%\subsection{Applying $\hat{e}$ on $|\blacksquare\rangle$}

The situation is also analogous for the action of $\hat{e}$. 
For example, applying $\hat{e}$ on $|\blacksquare\rangle$, we have 
\begin{equation}\label{ehatonx}
\hat{e}(z) |\blacksquare\rangle \sim \frac{1}{z} \, |\blacksquare+\hat{{\tiny\yng(1)}}_{\, 0}\rangle +\frac{1}{(z+\sigma_3 \hat{\psi}_0-h_2)} \, |\blacksquare+\hat{{\tiny\yng(1)}}_{\, \textrm{top}}\rangle \ , 
\end{equation}
where 
\begin{equation}\label{m+hat0}
|\blacksquare+\hat{{\tiny\yng(1)}}_{\, 0}\rangle:\qquad 
\begin{cases}
\begin{aligned}
\psi(u) &= \psi_0(u)\cdot \varphi_2(u) \\
\hat{\psi}(u) &=\hat{\psi}_0(u)\cdot \varphi^{-1}_2(-u-\sigma_3 \hat{\psi}_0)\cdot \varphi_3(u)
\end{aligned}
\end{cases}
\end{equation}
and
\begin{equation}\label{xtophateigenvalue}
|\blacksquare+\hat{{\tiny\yng(1)}}_{\, \textrm{top}}\rangle:\qquad 
\begin{cases}
\begin{aligned}
\psi(u) &= \psi_0(u)\cdot \varphi_2(u)\\
\hat{\psi}(u) &=\hat{\psi}_0(u)\cdot \varphi^{-1}_2(-u-\sigma_3 \hat{\psi}_0)\cdot \varphi_3(u+\sigma_3\hat{\psi}_0-h_2)  \ .
\end{aligned}
\end{cases}
\end{equation}

\begin{figure}[h!]
	\centering
	\begin{tabular}{cc}
	\includeCroppedPdf[width=.4\textwidth]{"xbarexcited0"}&
		\includeCroppedPdf[width=.4\textwidth
		]{"xbarexcitedtop"}
  \\
 \hspace*{-0.5cm} $\hat{e}$ acts on $|\blacksquare\rangle$ by adding a box at $(0,0,0)$\qquad    &$\hat{e}$ acts on $|\blacksquare\rangle$ by adding a box on top \\
	\end{tabular}	
	\caption{The first hatted descendants of $|\blacksquare\rangle$.}
	\label{figxleft}
\end{figure}

\subsection{Fermionic raising operators on existing boxes}\label{sec:4.3}

Next we want to apply the technique illustrated above to situations where the answer is not obvious. We begin by considering 
\begin{equation}
%\begin{aligned}
x(w) e(z) |\emptyset\rangle \sim \frac{1}{z}\, x(w) |{\tiny\yng(1)}\rangle \sim\frac{1}{z} \sum_{i} \frac{1}{w-w^{*}_i} |\Phi^{xe}\rangle \ ,
%\end{aligned}
\end{equation}
where $w^{*}_i$ are all possible poles for which the state $|\Phi^{xe}_{i}\rangle$ is a legitimate twin-plane-partition configuration. 
It follows from our general formula, eq.~(\ref{generalcharge}), that the resulting charges are 
\begin{equation}
|\Phi^{xe}_{i}\rangle:\qquad 
\begin{cases}
\begin{aligned}
\psi(u) &= \psi_0(u)\cdot \varphi_2(u-w^{*}_i)\cdot \varphi_3(u) \\
\hat{\psi}(u) &=\hat{\psi}_0(u)\cdot \varphi^{-1}_2(-(u-w^{*}_i)-\sigma_3\hat{\psi}_0) \ . 
\end{aligned}
\end{cases}
\end{equation}
Since the resulting state involves a single $x$ generator, these states should be $e$ or $\hat{e}$ descendants of $|\blacksquare\rangle$. 
Furthermore, since the total conformal dimension operator is simply 
\be\label{Ltotal}
L_0^{\rm total} = L_0 + \hat{L}_0 \ , \qquad L_0 = \frac{1}{2}\psi_2 \ , \qquad \hat{L}_0 = \frac{1}{2} \hat{\psi}_2 \ , 
\ee
we find that the total conformal dimension of the state $|\Phi^{xe}_{i}\rangle$ is 
\begin{equation}
L_0^{\rm total} =\frac{5}{2} \ ,
\end{equation}
independent of the position of the pole $w^*_i$.
Namely, the state $|\Phi^{xe}_{i}\rangle$ involves only a single box excitation of $|\blacksquare\rangle$. 
The possible candidates are therefore (\ref{m+1}), (\ref{m+hat0}), and (\ref{xtophateigenvalue}). 
Exploring the different possibilities, one can show that there is only one consistent value for $w^{*}$, namely 
\begin{equation}\label{wpole}
w^{*}=h_2 \ . 
\end{equation}
Indeed, using 
\begin{equation}\label{magic}
\varphi_2(u-h_2) \, \varphi_3(u)  =  \varphi_2(u) \ ,
\end{equation}
we find that for $w^*=h_2$ 
\begin{equation}
|\Phi^{xe}\rangle:\qquad 
\begin{cases}
\begin{aligned}
\psi(u) &= \psi_0(u)\cdot \varphi_2(u)\\
\hat{\psi}(u) &=\hat{\psi}_0(u)\cdot \varphi^{-1}_2(-u-\sigma_3\hat{\psi}_0)\cdot\varphi_3(u+\sigma_3\hat{\psi}_0-h_2) \ , 
\end{aligned}
\end{cases}
\end{equation}
and hence 
\begin{equation}\label{xonefinal}
\boxed{ x(w) |{\tiny\yng(1)}\rangle \sim \frac{1}{w-h_2} \, |\blacksquare+\hat{{\tiny\yng(1)}}_{\, \textrm{top}}\rangle \ . }
\end{equation}

This result has a simple geometric interpretation: since the starting configuration is a single box in the left corner, the only natural way in which one can add an infinite row is to start this infinite row at the position $(x_1,x_2,x_3) = (0,1,0)$, thus leading to the pole (\ref{wpole}). 
Furthermore, it is suggestive that one extra box sticks out at the right corner, and thus leads to an $\hat{e}$ descendant of $|\blacksquare\rangle$. 
\smallskip

The analysis for determining the action of $x$ on $|\hat{{\tiny\yng(1)}}\rangle$ works similarly. 
Now we make the ansatz 
\begin{equation}
\begin{aligned}
x(w) \hat{e}(z) |\emptyset\rangle& \sim \frac{1}{z}\, x(w) |\hat{{\tiny\yng(1)}}\rangle \sim\frac{1}{z} \, \sum_{i} \frac{1}{w-w^{*}_i} \, |\Phi^{x\hat{e}}_i\rangle \ ,
\end{aligned}
\end{equation}
where the charges of $|\Phi^{x\hat{e}}_{i}\rangle$ depend on $w^*_i$ via
\begin{equation}
|\Phi^{x\hat{e}}_{i}\rangle:\qquad 
\begin{cases}
\begin{aligned}
\psi(u) &= \psi_0(u)\cdot \varphi_2(u-w^{*}_i) \\
\hat{\psi}(u) &=\hat{\psi}_0(u)\cdot \varphi^{-1}_2(-(u-w^{*}_i)-\sigma_3\hat{\psi}_0) \cdot \varphi_3(u) \ . 
\end{aligned}
\end{cases}
\end{equation}
By the same arguments as above, these charges must agree with one of (\ref{m+1}), (\ref{m+hat0}) or (\ref{xtophateigenvalue}), and one finds that the only possible solution is 
\begin{equation}\label{xonehatfinal}
\boxed{x(w) |\hat{{\tiny\yng(1)}}\rangle \sim \frac{1}{w} \, |\blacksquare+\hat{{\tiny\yng(1)}}_{\, \textrm{0}}\rangle}  \ ,
\end{equation}
corresponding to $w^*=0$. 
Again, this fits together with the previous geometrical intuition since now there is no box in the left corner, and hence the infinite row should start at position $(x_1,x_2,x_3) = (0,0,0)$.
Finally, for the conjugate cases we find similarly
\begin{equation}\label{xbaronefinal}
\boxed{ \bar{x}(w) |{\tiny\yng(1)}\rangle \sim \frac{1}{w} \, |\overline{\blacksquare}+{{\tiny\yng(1)}}_{\, 0}\rangle }  \ ,
\end{equation}
and
\begin{equation}\label{barxonehatfinal}
\boxed{\bar{x}(w) |\hat{{\tiny\yng(1)}}\rangle \sim \frac{1}{w-h_2} \, |\overline{\blacksquare}+{{\tiny\yng(1)}}_{\, \textrm{top}}\rangle } \ .
\end{equation}

\subsection{$x$ on generic states}\label{sec:interactf}

%\subsubsection{Low level examples}
We can proceed in this manner and consider more complicated descendants, e.g., those involving two $x$ generators. 
The simplest case is 
\be
\begin{aligned}
x(z) \cdot x(w) \, |\emptyset\rangle & \sim \frac{1}{w} \, x(z) \, |\blacksquare\rangle
& \sim \frac{1}{w} \, \sum_{i} \frac{1}{z-z^{*}_i} \, |\Phi^{xx}_{i}\rangle \ ,
\end{aligned}
\end{equation}
whose charges are then 
\begin{equation}\label{psixx}
|\Phi^{xx}_{i}\rangle:\qquad 
\begin{cases}
\begin{aligned}
\psi(u) &= \psi_0(u)\cdot \varphi_2(u-z^{*}_i) \cdot \varphi_2(u) \\
\hat{\psi}(u) &=\hat{\psi}_0(u)\cdot \varphi^{-1}_2(-(u-z^{*}_i)-\sigma_3\hat{\psi}_0)\cdot \varphi^{-1}_2(-u-\sigma_3\hat{\psi}_0)  \ . 
\end{aligned}
\end{cases}
\end{equation}
One might expect that the resulting states should match 
one of the two configurations corresponding to two infinite rows of boxes of Fig.~\ref{figxleftadd1} or Fig.~\ref{figxleftadd3}, remembering that the Young diagrams characterizing the two representations are transposes of one another, see eq.~(\ref{charid}).
The charge functions of these two possible ground states equal 
\begin{equation}
|\blacksquare\blacksquare_{1}\rangle:\qquad 
\begin{cases}
\begin{aligned}
\psi(u) &= \psi_0(u)\cdot \varphi_2(u) \cdot \varphi_2(u-h_1) \\
\hat{\psi}(u) &=\hat{\psi}_0(u)\cdot \varphi^{-1}_2(-u-\sigma_3\hat{\psi}_0) \cdot \varphi^{-1}_2(-u-h_3-\sigma_3\hat{\psi}_0) \ ,
\end{aligned}
\end{cases}
\end{equation}
and
\begin{equation}
|\blacksquare\blacksquare_{3}\rangle:\qquad  
\begin{cases}
\begin{aligned}
\psi(u) &= \psi_0(u)\cdot \varphi_2(u) \cdot \varphi_2(u-h_3) \\
\hat{\psi}(u) &=\hat{\psi}_0(u)\cdot \varphi^{-1}_2(-u-\sigma_3\hat{\psi}_0) \cdot \varphi^{-1}_2(-u-h_1-\sigma_3\hat{\psi}_0) \ ,
\end{aligned}
\end{cases}
\end{equation}
see eq. (\ref{xforpsi}).
However, it is clear that no value of $z^{*}_i$ can give rise to these charge functions, hence leading to the conclusion that 
\be\label{xxnull}
x(z) \cdot x(w) \, |\emptyset\rangle = 0 \ . 
\ee
This is compatible with the fact that we expect from the CFT perspective that 
\be\label{xiden}
x_r \sim W^{(r+\frac{1}{2})+}_{-3/2} \ ,
\ee
where we have used the conventions of \cite{Candu:2012tr} to denote the generators of the ${\cal N}=2$ ${\cal W}_\infty$ algebra. All of these generators anti-commute with one another, and all, except for $x_{1/2}$, annihilate the vacuum. This then implies that $x_r x_s \, |\emptyset\rangle = 0$, in agreement with (\ref{xxnull}). 
\smallskip

In order to generate a non-trivial state, we need to apply an $e$ generator in between, i.e.\ we consider instead 
\begin{equation}
\begin{aligned}
x(z)\cdot e(w)\cdot x(v) |\emptyset\rangle &\sim  \frac{1}{v} \, x(z)\cdot e(w) |\blacksquare\rangle  \\
&\sim  \frac{1}{v} \, x(z) \left[\frac{1}{w-h_1}|\blacksquare+{\tiny\yng(1)_{\, 1}}\rangle+\frac{1}{w-h_3} |\blacksquare+{\tiny\yng(1)_{\, 3}}\rangle\right]   \ ,
\end{aligned}
\end{equation}
and then make the ansatz that 
\begin{equation}
x(z) |\blacksquare+{\tiny\yng(1)_{\, i}}\rangle \sim\sum_{j} \frac{1}{z-z^{*}_{i,j}} \, |\Phi^{xex}_{i,j}\rangle \ , \qquad \hbox{where $i=1,3$.}
\end{equation}
The corresponding charges are then
\begin{equation}
|\Phi^{xex}_{i,j}\rangle:\qquad 
\begin{cases}
\begin{aligned}
\psi(u) &= \psi_0(u)\cdot \varphi_2(u-z^{*}_{ i,j})\cdot\varphi_3(u-h_i)\cdot \varphi_2(u) \\
\hat{\psi}(u) &=\hat{\psi}_0(u)\cdot \varphi^{-1}_2(-(u-z^{*}_{i,j})-\sigma_3\hat{\psi}_0)\cdot \varphi^{-1}_2(-u-\sigma_3\hat{\psi}_0) \ , 
\end{aligned}
\end{cases}
\end{equation}
 for $i=1,3$.
We found that for each $i$, there is only pole\begin{equation}
z^{*}_{ i}=h_i +h_2  \ , \qquad \hbox{where $i=1,3$ ,}
\end{equation}
that leads to a consistent twin-plane-partition. 
Thus we find 
%\begin{equation}
%|\Phi^{x\texttt{ex}}_i\rangle:\qquad 
%\begin{cases}
%\begin{aligned}
%\psi(u) &= \psi_0(u)\cdot \varphi_2(u-h_i)\cdot \varphi_2(u) \\
%\hat{\psi}(u) &=\hat{\psi}_0(u)\cdot \varphi^{-1}_2(-(u-h_i)-\sigma_3\hat{\psi}_0)\cdot \varphi^{-1}_2(-u-\sigma_3\hat{\psi}_0)
%\end{aligned}
%\end{cases}
%\end{equation}
%
%
\begin{equation}\label{xexfinal}
\begin{aligned}
x(z)\cdot e(w)\cdot x(v) \, |\emptyset\rangle &\sim  \frac{1}{v} \, x(z)\cdot e(w)\,  |\blacksquare\rangle  \\
&\sim  \frac{1}{v}  \cdot\left[\frac{1}{z+h_3}\cdot\frac{1}{w-h_1}\, |\blacksquare\blacksquare_1\rangle+\frac{1}{z+h_1}\cdot\frac{1}{w-h_3} \, |\blacksquare\blacksquare_3\rangle\right]    \ .
\end{aligned}
\end{equation}
Note that this has also a nice CFT interpretation: the presence of the single box introduces a bosonic $-1$ mode, whose commutator with $W^{(r+1/2)+}_{-3/2}$ gives rise to generators of the form $W^{(r'+1/2)+}_{-5/2}$. These generators do not annihilate the vacuum for $r'=1/2$ and $r'=3/2$, and hence there are two possible descendant states, corresponding to $|\blacksquare\blacksquare_1\rangle$ and $|\blacksquare\blacksquare_3\rangle$.

\subsubsection{General formula}\label{sec:4.5}

It is not too difficult to extract from these considerations the general rule for how the $x$-action works. 
In order for $x$ to be allowed to add an infinite row along the $x_2$ direction at $(x_1,x_3)=(m,n)$, it must be possible to add a new box $\blacksquare$ at that position to the $2$-dimensional Young diagram $\lambda$. 
In addition, there must be at least a bud of $m+n$ boxes extending in the $x_2$ direction at that position, i.e.\ the box configuration must already contain boxes at\footnote{Note that this generalizes the discussion in Section~\ref{sec:interactf}.}
\be\label{stub0}
(x_1,x_2,x_3) = (m,0,n), (m,1,n), (m,2,n), \ldots , (m,m+n-1,n) \ . 
\ee
Before we continue we should mention that this also has a nice CFT interpretation: the presence of $m+n=p$ boxes means that we can now produce (upon commuting the fermionic generators to the right) fermionic generators of the form $W^{(r+1/2)+}_{-3/2-p}$. There are then $p+1$ different such generators that do not annihilate the vacuum (namely those with $r=1/2,\ldots, p+1/2$), and they correspond to the $p+1$ different possible positions of a $\blacksquare$ with $m+n=p$.
\smallskip

In this minimal configuration, i.e. when there are exactly $m+n$ boxes along the $x_2$ direction at $(x_1, x_3)=(m+n)$,  the $x(z)$ operator has a pole at 
\begin{equation}\label{pole}
z^\ast = - (n h_1 + m h_3) = h(\blacksquare)=\hat{g}(\blacksquare)+\sigma_3\hat{\psi}_0-h_2\ , 
\end{equation}
see eq.~(\ref{hblack}).
Thus the $x(u)$ action with this pole contributes to the two charge functions by
\begin{eqnarray}
\psi(u): &\qquad \qquad &\varphi_2(z-h(\blacksquare))\label{xchargeleft}\\
\hat{\psi}(u): &\qquad \qquad &\varphi^{-1}_2(z-\hat{g}(\blacksquare))\ , \label{xchargeright}
\end{eqnarray}
where in deriving the  contribution to $\hat{\psi}(u)$ we have used 
\be\label{varphi2m}
\varphi_2(-u)=\varphi_2(u-h_2) \ . 
\ee

On the hatted side, the contribution (\ref{xchargeright}) already describes the hatted charge function of $\blacksquare$, corresponding to a wall at the position $(\hat{x}_1,\hat{x}_3) = (-n,-m)$ from the hatted viewpoint, see eq.~(\ref{xforpsi}).
On the unhatted side, the contribution (\ref{xchargeleft}) combines with the charge functions of the $m+n$ existing boxes in the minimal-length bud to give 
\begin{equation}
\varphi_2(u -  h(\blacksquare)) \prod_{j=1}^{m+n} \varphi_3\bigl(u - ( h(\blacksquare) - j h_2 )\bigr) 
= \varphi_2\bigl(u-g(\blacksquare) \bigr) \ , 
\end{equation}
where we have used recursively the identity eq.~(\ref{magic}). (Here the product over $j$ describes the charge contribution coming from the boxes in (\ref{stub0}), with the $j$'th term describing the box at position
$(m,m+n-j,n)$.) 
Therefore the $x$ action with the pole (\ref{pole}) creates a $\blacksquare$ at $(x_1, x_3)=(m,n)$, with the correct charge functions (\ref{xforpsi}).

\smallskip

We should also mention that if the bud in (\ref{stub0}) is longer, i.e.\ if there are $\ell$ additional boxes, at position 
\be\label{stub1}
(x_1,x_2,x_3) = (m,m+n,n), (m,m+n+1,n), \ldots , (m,m+n+\ell-1,n) \ ,
\ee
then the pole $z^\ast$ of eq.~(\ref{pole}) gets shifted  to 
\be\label{pplus}
z^\ast = - (n h_1 + m h_3)  + \ell h_2  = h(\blacksquare) + \ell h_2 =\hat{g}(\blacksquare)+\ell h_2+(\sigma_3\hat{\psi}_0-h_2)\ .
\ee
On the unhatted side we again just get $\varphi_2\bigl(u- g(\blacksquare))$, corresponding to an $x_2$ row at  position $(x_1,x_3)=(m,n)$, while the hatted charge function is  now
\begin{equation}
\begin{aligned}
 &\varphi_2^{-1} (u - \hat{g}(\blacksquare)- \ell h_2 ) \\
   &\qquad =  \varphi_2^{-1} (u - \hat{g}(\blacksquare) )
 \prod_{j=0}^{\ell-1} \varphi_3(u - (\hat{g}(\blacksquare) + j\, h_2)) \ ,  \label{4.80}
\end{aligned}
\end{equation}
which correspond to $\ell$ additional hatted boxes lined up along the newly created wall at the position $(\hat{x}_1,\hat{x}_3) = (-n,-m)$. (This is always an allowed box configuration.)

There is one final complication: suppose that $x(u)$ acts on a state which already contains some non-trivial $(\lambda,{\lambda}^{\star})$ background, and that furthermore there are already some hatted boxes on top of the ${\lambda}^{\star}$ configuration. 
Then we need to ensure that, after the addition of the additional $\blacksquare$ to $\lambda$ and ${\lambda}^{\star}$, the configuration of hatted boxes on top of $\lambda^{\star}$ remains allowed. 
In general, this will not be automatic, but will depend on the number of boxes on top of the adjacent walls, i.e.\ the walls corresponding to $(\hat{x}_1,\hat{x}_3) = - (n-1,m)$ and $- (n,m-1)$ --- these walls are already present since otherwise adding $\blacksquare$ to $\lambda$ would not be  allowed. 
Given that the new wall is created closer to the origin, the condition is simply that the bud must have at least length $m+n+k$, where $k$ is the larger of the number of (hatted) boxes on top of the two walls $(\hat{x}_1,\hat{x}_3) =-(n-1,m)$ and $-(n,m-1)$. 
Because then, using eq.~(\ref{4.80}), at least $k$ boxes will appear on top of the newly created hatted wall, and thus make the resulting hatted box configuration consistent.

\subsection{Conjugate fermions}\label{sec:conjf}

The analysis above works similarly for the $\bar{x}(u)$ generators, but an interesting situation arises if we consider both $x$ and $\bar{x}$ excitations. 
More specifically, let us consider the action of $x(u)$ on $ |\overline{\blacksquare}\rangle$ 
\begin{equation}
x(z)\bar{x}(w) |\emptyset\rangle \sim  \frac{1}{w}\, x(z) |\overline{\blacksquare}\rangle \ \sim \ 
\frac{1}{w} \cdot\frac{1}{z-z^{*}_{x\bar{x}}} |\Phi^{x\bar{x}}\rangle \ , 
\end{equation}
for which the charge functions are  
\begin{equation}\label{xxbareigenvaluepole}
|\Phi^{x\bar{x}}\rangle:\qquad 
\begin{cases}
\begin{aligned}
\psi(u) &= \psi_0(u)\cdot \varphi_2(u-z^{*}_{x\bar{x}}) \cdot \varphi^{-1}_2(-u-\sigma_3 \psi_0)  \\
\hat{\psi}(u) &=\hat{ \psi}_0(u)\cdot  \varphi^{-1}_2(-(u-z^{*}_{x\bar{x}})-\sigma_3 \hat{\psi}_0) \cdot \varphi_2(u) \ . 
\end{aligned}
\end{cases}
\end{equation}
Expanding out the power series, we can read off the $\psi_2$ and $\hat{\psi}_2$ eigenvalues, and we find 
\begin{equation}
\psi_2 =2- 2 h_1h_3\psi_0-\frac{2 z^{*}_{x\bar{x}}}{h_2} \qquad \qquad \hat{\psi}_2=2-2h_1h_3\hat{\psi}_0+\frac{2z^{*}_{x\bar{x}}}{h_2} \ . 
\end{equation}
Therefore the total $L_0^{\rm total}$ eigenvalue, see eq.~(\ref{Ltotal}), equals 
\begin{equation}
L^{\textrm{total}}_0= \frac{1}{2} \Bigl( \psi_2 + \hat{\psi}_2 \Bigr) = 3 \ , 
\end{equation}
independent of the position of the pole $z^{*}_{x\bar{x}}$, where we have used
\begin{equation}\label{3.30}
\psi_0+\hat{\psi}_0=-\frac{1}{h_1h_3} \ ,
\end{equation}
see eq.~(3.30) of \cite{GLPZ}.
Thus the resulting state should either have three boxes, or be the ground state of the representation that is, with respect to both bosonic algebras, minimal and conjugate-minimal, i.e.\ for which both $\lambda$ and $\hat{\rho}$ contain one box. For the case of $\mathfrak{su}(N)$ this is just the adjoint representation of $\mathfrak{su}(N)$, and we shall therefore also refer to it as the adjoint here.
\smallskip

Knowing what the state $|\Phi^{x\bar{x}}\rangle$ can be, allows us to fix the pole. 
First, the adjoint representation arises when 
\begin{equation}
z^{*}_{x\bar{x}}=0 \ ,
\end{equation} 
for which the charge function is 
\begin{equation}\label{ansatz}
|\Phi^{x\bar{x}}\rangle:\qquad 
\begin{cases}
\begin{aligned}
\psi(u) &= \psi_0(u)\cdot \varphi_2(u) \cdot \varphi^{-1}_2(-u-\sigma_3 \psi_0)   \\
\hat{\psi}(u) &=\hat{ \psi}_0(u)\cdot   \varphi_2(u) \cdot \varphi^{-1}_2(-u-\sigma_3 \hat{\psi}_0) \ .
\end{aligned}
\end{cases}
\end{equation}
Indeed, this agrees with the charges of the adjoint representation (with respect to both bosonic affine algebras); for example, the conformal dimension with respect to the unhatted algebra equals 
\be
\frac{1}{2} \, \psi_2 =  1-  h_1h_3\psi_0 =  1 + \frac{N}{N+k} \ , 
\ee
which agrees with $h({\rm adj},0)$ in the coset language, see e.g.\ \cite[eq.~(2.15)]{Gaberdiel:2010pz}.\footnote{Since this is a real representation, there is no need to subtract out the $\mathfrak{u}(1)$ contribution.}

As regards the other possible state, the one consisting of three boxes, we note that there are 
18 such configurations.\footnote{First, there are four ways to distribute the three boxes among the two corners: $(3,0)$, $(2,1)$, $(1,2)$, and $(0,3)$. Since there are $6$ configurations for each $3$-box corner and $3$ for each $2$-box one, we have in total $18$ possibilities.} It is straightforward to write down their $(\psi(u),\hat{\psi}(u))$ eigenvalues and compare them with (\ref{ansatz}), and we find that the only match happens when
\begin{equation}
z^{*}_{x\bar{x}}=h_2-\sigma_3 \psi_0 \ , 
\end{equation}
leading to 
\begin{equation}
|\Phi^{x\bar{x}}\rangle:\qquad 
\begin{cases}
\begin{aligned}
\psi(u) &= \psi_0(u)  \\
\hat{\psi}(u) &=\hat{ \psi}_0(u) \cdot \varphi_3(u)  \cdot  \varphi_3(u-h_2)\cdot\varphi_3(u-2h_2) \ . 
\end{aligned}
\end{cases}
\end{equation}
Geometrically, this describes the configuration of three hatted boxes, lined up along the $x_2$ direction, see Fig.~\ref{figxonxbar}.
\begin{figure}[h!]
	\centering
	\includeCroppedPdf[width=0.5\textwidth]{"xonxbar"}
	\caption{}
	\label{figxonxbar}
\end{figure}
Thus altogether we find 
\begin{equation}\label{xonxbar}
\boxed{x(z) |\overline{\blacksquare}\rangle \sim \frac{1}{(z+\sigma_3 \psi_0-h_2)}\, |\widehat{{\tiny\yng(3)}}_{\,2}\rangle 
+ \frac{1}{z}\,  |({\rm adj},{\rm adj})\rangle } \ .
 \end{equation}
\smallskip

We can similarly study the action of $\bar{x}(u)$ on $|\blacksquare\rangle$, 
\begin{equation}
\bar{x}(z)x(w) |\emptyset\rangle \sim  \frac{1}{w}\, \bar{x}(z) |\blacksquare\rangle  \ \sim \ 
\frac{1}{w} \cdot\frac{1}{z-z^{*}_{\bar{x}x}} |\Phi^{\bar{x}x}\rangle \ , 
\end{equation}
for which the resulting state has the charges
\begin{equation}\label{ansatz1}
|\Phi^{\bar{x}x}\rangle:\qquad 
\begin{cases}
\begin{aligned}
\psi(u) &= \psi_0(u)\cdot\varphi^{-1}_2(-(u-z^{*}_{\bar{x}x})-\sigma_3 \psi_0) \cdot \varphi_2(u)  \\
\hat{\psi}(u) &=\hat{ \psi}_0(u)\cdot   \varphi_2(u-z^{*}_{\bar{x}x}) \cdot \varphi^{-1}_2(-u-\sigma_3 \hat{\psi}_0) \ .
\end{aligned}
\end{cases}
\end{equation}
%\begin{equation}
%|\Phi_{x\bar{x}}\rangle:\qquad 
%\begin{cases}
%\begin{aligned}
%\frac{1}{2}\psi_2 &=(1-h_1 h_3 \psi_0)+\frac{1}{h_2} z^{*}_{\bar{x}x}\\
%\frac{1}{2}\hat{\psi}_2&=(1-h_1 h_3 \hat{\psi}_0)-\frac{1}{h_2} z^{*}_{\bar{x}x}
%\end{aligned}
%\end{cases}
%\end{equation}
Similar to the  previous case of $x(u)$ on $|\overline{\blacksquare}\rangle$, the pole $z^{*}_{\bar{x}x}=0$ gives rise to the $|({\rm adj},{\rm adj})\rangle$ state. 
The other alternative, i.e.\ the three box configurations, corresponds to 
\begin{equation}
z^{*}_{\bar{x}x}=h_2-\sigma_3 \hat{\psi}_0
\end{equation}
leading to 
\begin{equation}
|\Phi^{\bar{x}x}\rangle:\qquad 
\begin{cases}
\begin{aligned}
\psi(u) &= \psi_0(u) \cdot \varphi_3(u)  \cdot  \varphi_3(u-h_2)\cdot\varphi_3(u-2h_2) \\
\hat{\psi}(u) &=\hat{ \psi}_0(u) \ .
\end{aligned}
\end{cases}
\end{equation}
This is precisely the charge function for the configuration of three unhatted boxes lined up along the $x_2$ direction, see Fig.~\ref{figxbaronx}. 
 \begin{figure}[h!]
	\centering
	\includeCroppedPdf[width=0.5\textwidth]{"xbaronx"}
	\caption{}
	\label{figxbaronx}
\end{figure}
Thus, altogether we find 
\begin{equation} \label{barxonxfinal}
\boxed{\bar{x}(z) |\blacksquare\rangle \sim \frac{1}{(z+\sigma_3 \hat{\psi}_0-h_2)} \, |{\tiny\yng(3)}_{\,2}\rangle 
+ \frac{1}{z}\, |({\rm adj},{\rm adj})\rangle}    \ . 
\end{equation}
%
%\noindent In the free field limit, the position of the pole is now at $z^{*}_{x\bar{x}}=1$, which again agrees with the free field computation.
%
%
%\begin{equation}\label{xxbareigenvalue}
%|\Phi_{x\bar{x}}\rangle:\qquad 
%\begin{cases}
%\begin{aligned}
%\psi(u) &= \psi_0(u) \\
%\hat{\psi}(u) &=\hat{ \psi}_0(u)\cdot \varphi_3(u)  \cdot  \varphi_3(u-h_2)\cdot\varphi_3(u-2h_2)
%\end{aligned}
%\end{cases}
%\end{equation}
%

For later use we also compute, using similar techniques,
\begin{equation}
x(z)\, |\overline{\blacksquare}+{\tiny\yng(1)_{\, 0}}\rangle=\sum_{\{z^{*}_{xe_0\bar{x}}\}}\frac{1}{z-z^{*}_{xe_0\bar{x}}} \, |\Phi^{xe_0\bar{x}}\rangle \ , 
\end{equation}
where 
\begin{equation}
|\Phi^{xe_0\bar{x}}\rangle:\qquad 
\begin{cases}
\begin{aligned}
\psi(u) &= \psi_0(u)\cdot \varphi_2(u-z^{*}_{xe_0\bar{x}})\cdot\varphi_3(u) \cdot \varphi^{-1}_2(-u-\sigma_3 \psi_0)  \\
\hat{\psi}(u) &=\hat{ \psi}_0(u)\cdot  \varphi^{-1}_2(-(u-z^{*}_{xe_0\bar{x}})-\sigma_3 \hat{\psi_0}) \cdot \varphi_2(u) \ . 
\end{aligned}
\end{cases}
\end{equation}
Again, apart from the solution $z^{*}_{xe_0\bar{x}}=h_2$, which leads to a descendant of the bi-adjoint representation $({\rm adj},{\rm adj})$, the only other solution appears at 
\begin{equation}\label{4.65}
z^{*}_{xe_0\bar{x}}=h_2-\sigma_3 \psi_0 \ , 
\end{equation}
and thus we find 
\begin{equation}\label{4.66}
%\boxed{
x(z)\, |\overline{\blacksquare}+{\tiny\yng(1)_{\, 0}}\rangle \sim  \frac{1}{(z+\sigma_3\psi_0-h_2)}\,  |({\tiny\yng(1)}, \widehat{{\tiny\yng(3)}}_{\, 2 })\rangle  + \frac{1}{(z-h_2)} \, | ({\rm adj},{\rm adj} +\hat{\tiny\yng(1)}_{\, \textrm{top}} ) \rangle   \ .
\end{equation}
The analysis also works similarly for the other $e$ descendant of $|\overline{\blacksquare}\rangle$, for which we find 
%\begin{equation}
%|\Phi_{xe_t\bar{x}}\rangle:\qquad 
%\begin{cases}
%\begin{aligned}
%\psi(u) &= \psi_0(u)\cdot \varphi_2(u-z^{*}_{xe\bar{x}})\cdot\varphi_3(u+\sigma_3\psi_0-h_2) \cdot \varphi^{-1}_2(-u-\sigma_3 \psi_0)  \\
%\hat{\psi}(u) &=\hat{ \psi}_0(u)\cdot  \varphi^{-1}_2(-(u-z^{*}_{xe\bar{x}})-\sigma_3 \hat{\psi_0}) \cdot \varphi_2(u)
%\end{aligned}
%\end{cases}
%\end{equation}
%There is only one solution
%\begin{equation}
%z^{*}_{xe_t\bar{x}}=-\sigma_3 \psi_0+2h_2
%\end{equation}
\begin{equation}
%\boxed{
x(z)\, |\overline{\blacksquare}+{\tiny\yng(1)_{\, \textrm{top}}}\rangle \sim  \frac{1}{(z+\sigma_3\psi_0-2h_2)}\, | \widehat{{\tiny\yng(4)}}_{\, 2 }\rangle + \frac{1}{z}\,  | ({\rm adj}+{\tiny\yng(1)_{\, \textrm{top}}},{\rm adj})   \ .
\end{equation}
In the conjugate case we find similarly\begin{equation}\label{4.70}
%\boxed{
\bar{x}(z)\, |{\blacksquare+\hat{\tiny\yng(1)}_{\, 0}}\rangle\sim  \frac{1}{(z+\sigma_3\hat{\psi}_0-h_2)}\,  |({\tiny\yng(3)}_{\, 2}, \widehat{{\tiny\yng(1)}})\rangle  + \frac{1}{(z-h_2)} \, | ({\rm adj} +{\tiny\yng(1)}_{\, \textrm{top}} ,{\rm adj} )\rangle \ , 
\end{equation}
and
\begin{equation}
%\boxed{
\bar{x}(z)\, |{\blacksquare}+{\hat{\tiny\yng(1)}_{\, \textrm{top}}}\rangle \sim  \frac{1}{(z+\sigma_3\hat{\psi}_0-2h_2)}\, | {{\tiny\yng(4)}}_{\, 2 }\rangle + \frac{1}{z}\,  | ({\rm adj},{\rm adj}+\hat{{\tiny\yng(1)}}_{\, \textrm{top}})   \ .
\end{equation}

\subsubsection{General formula}\label{sec:4.7}

As we have just seen, the action of $x$ can add a box $\blacksquare$ to $\lambda$ --- this is what we described above in Section~\ref{sec:4.5} --- but it can also remove one of the $\overline{\blacksquare}$ from $\hat{\rho}$. 

Consider a removable box $\overline{\blacksquare} \in \hat{\rho}$ with coordinates $(\hat{x}_1,\hat{x}_3)= (\hat{m},\hat{n})$. 
From the viewpoint of the unhatted algebra, this is part of the wall, so let us assume there are $\ell\geq 0$ (unhatted) boxes  on top of it. 
Before we apply the $x(u)$ action,  the $\overline{\blacksquare}$ at $(\hat{x}_1,\hat{x}_3)= (\hat{m},\hat{n})$ together with these $\ell$ boxes contribute to the charge function
\begin{equation}
\begin{cases}
\begin{aligned}
\psi(u): &\qquad   \varphi^{-1}_2(u-g(\overline{\blacksquare})) \prod_{j=0}^{\ell-1} \varphi_3 \bigl(u - g(\overline{\blacksquare})- j h_2   \bigr) \label{blackbar} \\
\hat{\psi}(u): &\qquad \varphi_2(u-\hat{g}(\overline{\blacksquare}))  \ ,
\end{aligned}
\end{cases}
\end{equation}
see eq.~(\ref{xbarforpsi}). Using eq.~(\ref{magic}) recursively as well as eq.~(\ref{varphi2m}), 
%\begin{equation}\label{varphi2m}
%\varphi_2(- u) = \varphi_2(u- h_2) \ , 
%\end{equation}
we can rewrite the $\psi(u)$ part of (\ref{blackbar}) as
\begin{equation}
\begin{aligned}
\psi(u): &\qquad \varphi_2^{-1}\bigl(u - (g(\overline{\blacksquare}) +\ell h_2 )\bigr) \ .
\end{aligned}
%\ , \qquad \hat{\psi}(u) = \varphi_2 \bigl(u - (mh_1+nh_3)\bigr) \ . 
\end{equation}

On the other hand, an $x(z)$ with pole at 
\begin{equation}\label{zast}
z_{\ast}= g(\overline{\blacksquare}) + \ell h_2  
\end{equation}
%where we have used $\hat{h}(\overline{\blacksquare})=g(\overline{\blacksquare})$ (eq. \ref{hbarblack}).
contributes to the charge function
\begin{equation}\label{xkilling}
\begin{cases}
\begin{aligned}
\psi(u): &\qquad   \varphi_2\bigl(u - (g(\overline{\blacksquare}) + \ell h_2 )\bigr) 
 \\
\hat{\psi}(u): &\qquad \varphi_2^{-1}\bigl(-u + (g(\overline{\blacksquare}) +\ell h_2 )-\sigma_3\hat{\psi}_0\bigr) \ .
\end{aligned}
\end{cases}
\end{equation}
Combining (\ref{blackbar}) and (\ref{xkilling}), we get
\begin{equation}
\begin{cases}
\begin{aligned}
\psi(u): &\qquad   1 
 \\
\hat{\psi}(u): &\qquad \prod_{j=0}^{\hat{m}+\hat{n}+\ell+3} \varphi_3\bigl(u - \hat{g}(\overline{\blacksquare}) - j h_2  \bigr) \ .
\end{aligned}
\end{cases}
\end{equation}
We see that applying $x(u)$ with the pole (\ref{zast}) removes the infinite row at $\overline{\blacksquare}$, and replaces it by $(\hat{m}+\hat{n}+\ell+3)$ hatted boxes, which appear on the right-hand-side at positions $(\hat{m},0,\hat{n}), (\hat{m},1,\hat{n}),\ldots, (\hat{m},\hat{m}+\hat{n}+\ell+2,\hat{n})$. 
The analysis for the $\bar{x}$ action is analogous.

%
%Similarly, the $\bar{x}$ action on $x_{(m,n)}$ with $l$ hatted boxes on top is of the form 
%\be\label{4.83}
%\bar{x}(z) |{\rm x}_{(m,n)}^{l}\rangle = \frac{1}{z-\bigl( (m+n+l+1) h_2 + (mh_1 + nh_3) - \sigma_3\hat{\psi}_0\bigr)} 
%|(m+n+l+3) {\rm boxes} \rangle \ , 
%\ee
%where the boxes are again at $(m,0,n), (m,1,n),\ldots, (m,m+n+l+2,n)$.

\section{Constraining the OPEs}\label{sec:5}

In the previous section we have understood the general structure of the action of $x$ and $\bar{x}$ on arbitrary twin-plane-partition configurations. 
As we shall explain in this section, this information is already sufficient to constrain the OPE relations of the algebra. 
In particular, we will be able to determine the functions $G(\Delta)$, $H(\Delta)$,  $\hat{G}(\Delta)$, and $\hat{H}(\Delta)$ separately (see the discussion in Section~\ref{sec:somerelations}). 

In this section, we only give first constraints based on the action of the algebra on suitably chosen low-lying states. 
In the next section, we will fix the remaining freedom, using the action on generic twin-plane-partitions.

%We can use these insights to constrain the OPEs of the relevant operators. In particular, this allows us to go beyond \cite{GLPZ} where  Let us begin with the case of $G(\Delta)$ and $H(\Delta)$.

%
%In \cite{GLPZ}, we wrote down the defining relations of the $\mathcal{N}=2$ affine Yangian at the free field point, using the explicit free field realizations of the generators. 
%Now we can constrain these defining relations at generic coupling, using the representation theory in terms of twin-plane partitions.
%
%Our guiding principle is that
%%, with respect to the the two bosonic affine algebras, denoted by $\mathcal{Y}$ and $\hat{\mathcal{Y}}$ respectively in the following, 
%%the fermionic generators sit in ``bi-minimal" representations. 
%a consistent OPE-type relation needs to satisfy two conditions
%\begin{enumerate}
%\item Have correct actions on any twin-plane-partition configurations, in particular, the resulting states should reproduce correct conformal dimension and higher spin charges.
%\item Reduces to the free field limit result of chapter 4. 
%\end{enumerate}
%In addition, all the relations should be mutually consistent. These three conditions turn out to be very constraining, especially the first one.
%
%\subsection{Limitation of free field result}
%
%
%
%\break

\subsection{Constraining $G$ and $H$}

In order to constrain the $G$ function, we apply the ansatz (\ref{g1})  
\begin{equation}\label{exG}
e(z)x(w)\sim G(z-w) x(w) e(z)
\end{equation}
on the vacuum $|\emptyset\rangle$, and compare the pole structures of the two expressions
\begin{equation} 
e(z) x(w) |\emptyset\rangle \qquad \textrm{and} \qquad x(w) e(z) |\emptyset\rangle \ .
\end{equation}  
The left-hand-side has been computed in (\ref{eonxfinal}) and is
%, and has the form 
\begin{equation}\label{exon0}
e(z) x(w) |\emptyset\rangle
=\frac{1}{w}\,e(z) |\blacksquare\rangle
=\frac{1}{w} \, \left(\frac{1}{z-h_1}\,  |\blacksquare+{\tiny\yng(1)_{\ 1}}\rangle +\frac{1}{z-h_3}\, |\blacksquare+{\tiny\yng(1)_{\ 3}}\rangle\right) \ . 
\end{equation}
On the other hand, we have determined the right-hand-side in Section~\ref{sec:4.3}, see in particular eq.~(\ref{xonefinal}), and we have 
\begin{equation}\label{xeon0}
x(w)e(z)  |\emptyset\rangle
=\frac{1}{z}\, x(w)\, |{\tiny\yng(1)}\rangle \ =\frac{1}{z}\, \frac{1}{w-h_2}\, |\blacksquare+\hat{{\tiny\yng(1)}}_{\, \textrm{top}}\rangle \ . 
\end{equation}
Combining the two equations, we conclude that 
\begin{equation}\label{exsimxe}
 \frac{1}{w} \, \left(\frac{1}{z-h_1} |\blacksquare+{\tiny\yng(1)_{\ 1}}\rangle +\frac{1}{z-h_3} |\blacksquare+{\tiny\yng(1)_{\ 3}}\rangle\right)\sim G(z-w) \, \frac{1}{z} \, \frac{1}{w-h_2} \, |\blacksquare+\hat{{\tiny\yng(1)}}_{\, \textrm{top}}\rangle \ .
\end{equation}
Since the two states on the l.h.s.\ and the one state on the r.h.s.\ are all independent, the function $G(\Delta)$ needs to be a rational function with at least two poles (at $h_1$ and $h_3$) and one zero (at $-h_2$),
\begin{equation}
G(\Delta) \sim \frac{(\Delta + h_2)}{(\Delta - h_1) (\Delta - h_3)} 
\end{equation}
in order to remove the poles on both sides --- recall that this identity is only expected to be true up to terms that are regular in $w$ or $z$. 
Given the structure of the free field answer, we expect $G(\Delta)$ to be homogeneous, and hence of the form 
\begin{equation}\label{Gfin}
G(\Delta) = \frac{(\Delta + h_2)(\Delta+a)}{(\Delta - h_1) (\Delta - h_3)} \ ,
\end{equation}
where $a=0$ in the free field limit. 
\smallskip

Similarly, we can constrain $H(\Delta)$ by applying the ansatz (\ref{g2}) 
\begin{equation}\label{fxansatz}
f(z)x(w)\sim H(z-w) x(w) f(z)
\end{equation}
on the excited state $|{\tiny\yng(1)}\rangle$. 
Using (\ref{xonefinal}), the l.h.s.\ gives 
\begin{equation}
f(z)x(w)\, |{\tiny\yng(1)}\rangle=\frac{1}{w-h_2} \, f(z)\,  |\blacksquare+\hat{{\tiny\yng(1)}}_{\, \textrm{top}}\rangle = 0 \ , 
\end{equation}
since the state $|\blacksquare+\hat{{\tiny\yng(1)}}_{\, \textrm{top}}\rangle$ does not have any box descendant on the left (and hence is killed by the unhatted $f_r$ modes). 
For the right-hand-side  of (\ref{fxansatz}), we have 
\begin{equation}
\begin{aligned}
x(w)f(z)|{\tiny\yng(1)}\rangle& \sim \frac{1}{z} \, x(w)|\emptyset\rangle\sim \frac{1}{z}\, \frac{1}{w} |\blacksquare\rangle \ . 
\end{aligned}
\end{equation}
%Since the poles have to cancel, this implies that 
For the poles on the two sides of (\ref{fxansatz}) to cancel, $H(\Delta)$ needs to contain the factor $\Delta$ in the numerator. 
From the discussion of \cite{GLPZ}, see also Section~\ref{sec:4.3}, it then follows that 
\begin{equation}\label{Hfin}
H(\Delta) = \frac{\Delta}{(\Delta+a)} \ , 
\end{equation}
such that the product reproduces $\varphi_2(\Delta)$, see eq.~(\ref{keyiden}). 
Note that this ansatz for $H(\Delta)$ leads to the correct commutation relations in the free field limit (where again $a=0$). 
 
We have explored the constraints that arise from the action on various other states, but we have not found any further constraints; in particular, we can not determine $a$.
We shall see in Section~\ref{sec:xaction} that $a=0$ is a preferred and natural choice; then
\begin{equation}\label{GHfin}
\boxed{G(\Delta) = \frac{(\Delta + h_2)\Delta}{(\Delta - h_1) (\Delta - h_3)}} \qquad \textrm{and} \qquad 
\boxed{ H(\Delta) = \frac{\Delta}{\Delta} \ . } 
\end{equation}
We emphasize that $H(\Delta)$ is not trivial: the OPE relation (\ref{fxansatz}) with this $H(\Delta)$ is a shorthand for 
\begin{equation}\label{almosttrivial}
(z-w)\, f(z) \, x(w) \sim (z-w) \, x(w)\, f(z) \ ,
\end{equation}
for which the presence of the factor $(z-w)$ is of significance --- recall that all these identities are only true up to regular terms in either $z$ or $w$, and indeed only with the factor $(z-w)$ does the OPE (\ref{almosttrivial}) reproduce the correct relations between the $f_j$ and $x_r$ modes in the free field limit.

\subsection{Constraining  $\hat{G}$ and $\hat{H}$}

Next we constrain the functions $\hat{G}(\Delta)$ and $\hat{H}(\Delta)$ in the OPE 
\begin{eqnarray}
\hat{e}(z) \, x(w)  &\sim  &\hat{G}(\Delta)\, x(w) \, \hat{e}(z)  \label{Ghatansatz}\\
\hat{f}(z) \, x(w) & \sim  &\hat{H}(\Delta)\, x(w) \, \hat{f}(z) \ .  \label{Hhatansatz}
\end{eqnarray}
Similar to (\ref{keyiden}), $\hat{G}(\Delta)$ and $\hat{H}(\Delta)$ are also related as \cite{GLPZ}
\begin{equation}\label{keyidenH}
\hat{G}(\Delta) \, \hat{H}(\Delta) = \varphi^{-1}_2(-\Delta-\sigma_3 \hat{\psi}_0) \ . 
\end{equation}
%From the symmetry between $(e,f)$ and $(\hat{e},\hat{f})$, it is tempting to further conjecture
%\begin{equation}\label{conjGHHH}
%\hat{G}(\Delta) = H^{-1}(-\Delta -\sigma_3 \hat{\psi}_0) \qquad \qquad \hat{H} (\Delta) = G^{-1}(-\Delta -\sigma_3 \hat{\psi}_0) \ , 
%\end{equation}
%To confirm this, we now constrain $\hat{G}(\Delta)$ and $\hat{H}(\Delta)$ using arguments similar to those used for $G$ and $H$ in the previous subsection.
%\smallskip
%
%%It turns out to be more convenient in this case to 
%The conjecture (\ref{conjGHHH}) suggests that we start with $\hat{H}(\Delta)$. 
Let us first study the constraints on $\hat{H}(\Delta)$. To this end, we apply the ansatz (\ref{Hhatansatz})
%\begin{equation}\label{Hhatansatz}
%\hat{f}(z)x(w)\sim \hat{H}(\Delta) \, x(w)\hat{f}(z)
%\end{equation}
on the state $|{\tiny\yng(1)}\rangle$, and compare the pole structures of the two sides:
\begin{equation} 
\hat{f}(z) x(w) |{\tiny\yng(1)}\rangle \qquad \textrm{and} \qquad x(w) \hat{f}(z) |{\tiny\yng(1)}\rangle \ .
\end{equation}
Using eq.~(\ref{xonefinal}), the left-hand-side leads to
\begin{equation}\label{fhatxone}
%\begin{aligned}
\hat{f}(z) x(w) |{\tiny\yng(1)}\rangle \sim \frac{1}{w-h_2}\,\hat{f}(z)\,  |\blacksquare+\hat{\tiny\yng(1)}_{\, \textrm{top}}\rangle
\sim \frac{1}{w-h_2}\, \frac{1}{(z+\sigma_3\hat{\psi}_0-h_2)}\, |\blacksquare\rangle \ , 
%\end{aligned}
\end{equation}
while for the right-hand-side we get 
\begin{equation}\label{xfhatone}
\begin{aligned}
 x(w)\hat{f}(z) |{\tiny\yng(1)}\rangle&=0 \ . 
 \end{aligned}
\end{equation}
Comparing (\ref{fhatxone}) and (\ref{xfhatone}), we conclude that $\hat{H}(\Delta)$ must have a pole at 
$\Delta=-\sigma_3\hat{\psi}_0$.
%, {\it i.e.}\ that it is of the form 
%\begin{equation}\label{Hhatans}
%\hat{H}(\Delta)=\frac{(\Delta + \hat{a})}{\Delta+\sigma_3\hat{\psi}_0} \ ,
%\end{equation}
%where, in the free field limit, $\hat{a}=-1$. 
%Again, we note that this is, in particular, compatible with the natural ansatz 
%\begin{equation}
%\hat{H}(\Delta)=H^{-1}(-\Delta-\sigma_3\hat{\psi}_0)=\frac{\Delta+\sigma_3\hat{\psi}_0-a}{\Delta+\sigma_3\hat{\psi}_0} \ . 
%\end{equation}
To constrain $\hat{H}(\Delta)$ further, we note that 
\be
x(w) \, |\blacksquare+ {\tiny\yng(1)}_j + \hat{\tiny\yng(1)}_{\, {\textrm{top}}}\rangle = 0  \qquad \textrm{with} \quad j=1,3  \ ,
\ee
%where $j=1$ or $j=3$. 
where we have used that, while there is a bud at $(1,0,0)$ or $(0,0,1)$, and hence the action of $x(w)$ might seem permissible, %possible, 
it is actually prevented by the presence of the hatted box on top of the wall (see the discussion at the end of Section~\ref{sec:4.5}.)
However, if we first apply $\hat{f}(z)$, the box $\hat{\tiny\yng(1)}_{\, {\textrm{top}}}$ can be removed, and we find 
\begin{align}
x(w)\, \hat{f}(z) \, |\blacksquare+ {\tiny\yng(1)}_j + \hat{\tiny\yng(1)}_{\, {\textrm{top}}}\rangle  &  = 
\frac{1}{(z +\sigma_3\hat{\psi}_0 - h_2 )} x(w)\,  |\blacksquare+ {\tiny\yng(1)}_j \rangle \nonumber \\
& \sim 
\frac{1}{(z +\sigma_3\hat{\psi}_0 - h_2  )} \, \frac{1}{(w - h_j - h_2)} |\blacksquare\blacksquare_j\rangle \ .
\end{align}
Thus it follows that the numerator of $\hat{H}(\Delta)$ must contain the factors 
$(\Delta + \sigma_3 \hat{\psi}_0 + h_j)$ for both $j=1$ and $j=3$. 
Together with the earlier constrain on its pole, this implies that $\hat{H}(\Delta)$ is of the form 
\be\label{Hhat}
%\boxed{
\hat{H}(\Delta)  
%=\varphi_2^{-1}(-\Delta-\sigma_3\hat{\psi}_0)
=\frac{(\Delta+\sigma_3\hat{\psi}_0+h_1)(\Delta+\sigma_3\hat{\psi}_0+h_3)}{(\Delta+ \sigma_3\hat{\psi}_0 )(\Delta+\sigma_3\hat{\psi}_0-b) } \ ,
\end{equation}
where $b=1$ in the free field limit. 
\smallskip

To constrain $\hat{G}(\Delta)$, we apply the ansatz (\ref{Ghatansatz})
%\begin{equation}\label{Ghatansatz}
%\hat{e}(z)\,x(w)\sim \hat{G}(z-w) x(w) \,\hat{e}(z)
%\end{equation}
on the state $|\emptyset\rangle$, and compare the pole structures of the two sides
\begin{equation} 
\hat{e}(z) x(w) |\emptyset\rangle \qquad \textrm{and} \qquad x(w) \hat{e}(z) |\emptyset\rangle \ . 
\end{equation}
It follows from (\ref{ehatonx}) that 
\begin{equation}\label{ehatxon0}
%\begin{aligned}
\hat{e}(z) x(w)\,  |\emptyset\rangle
=\frac{1}{w}\,\hat{e}(z)\, |\blacksquare\rangle
\sim \frac{1}{w}\left( \frac{1}{z} \, |\blacksquare+\hat{\tiny\yng(1)}_{\, 0}\rangle
+\frac{1}{(z+\sigma_3\hat{\psi}_0-h_2)} \, |\blacksquare+\hat{\tiny\yng(1)}_{\, \textrm{top}}\rangle \right) \ ,
%\end{aligned}
\end{equation}
while (\ref{xonehatfinal}) implies
\begin{equation}\label{xehaton0}
%\begin{aligned}
 x(w)\, \hat{e}(z) |\emptyset\rangle =\frac{1}{z}\, x(w)\, | \hat{{\tiny\yng(1)}}\rangle
\sim \frac{1}{z}\, \frac{1}{w} \, |\blacksquare+\hat{\tiny\yng(1)}_{\, 0}\rangle \ . 
%\end{aligned}
\end{equation}
To cancel the additional state on the left-hand-side, $\hat{G}(\Delta)$ needs to have a pole at $\Delta=h_2-\sigma_3\hat{\psi}_0$.
%\begin{equation}
%\hat{G}(\Delta)\sim \frac{\#}{\Delta+\sigma_3\hat{\psi}_0-h_2}
%\end{equation}
Then using the result of $\hat{H}(\Delta)$ in (\ref{Hhat}) and the relation (\ref{keyidenH}), we have
\begin{equation}\label{Ghat}
\hat{G}(\Delta)=\frac{(\Delta+\sigma_3\hat{\psi}_0-b)}{(\Delta+\sigma_3\hat{\psi}_0-h_2) } \ .
\end{equation}
%where we have used the relation from \cite{GLPZ} 
%\be
%\hat{G}(\Delta) \, \hat{H}(\Delta) = \varphi_2^{-1}(-\Delta-\sigma_3\hat{\psi}_0) 
%\ee
%to determine the factor in the numerator. 
%It remains to determine $b$. 
We have also  applied the ansatz (\ref{Ghatansatz}) and (\ref{Hhatansatz}) on various other states, but  have not been able to fix $b$. 
The analysis of Section~\ref{sec:xaction} below suggests that the most natural ansatz (that is also compatible with the free field limit) is $b=h_2$; under this assumption we then have 
%To summarize:
\begin{equation}\label{GhatHhatfin}
\boxed{
\hat{H}(\Delta)  
%=\varphi_2^{-1}(-\Delta-\sigma_3\hat{\psi}_0)
=\frac{(\Delta+\sigma_3\hat{\psi}_0+h_1)(\Delta+\sigma_3\hat{\psi}_0+h_3)}{(\Delta+ \sigma_3\hat{\psi}_0 )(\Delta+\sigma_3\hat{\psi}_0-h_2) } }  \quad \textrm{and}\quad 
\boxed{\hat{G}(\Delta)  = \frac{(\Delta+\sigma_3\hat{\psi}_0-h_2)}{(\Delta+\sigma_3\hat{\psi}_0-h_2)} } \ . 
\end{equation}
We should mention that $\hat{H}(\Delta) $ is compatible with the natural conjecture, see \cite{GLPZ}
\be
\hat{H}(\Delta) = G^{-1}(-\Delta -\sigma_3 \hat{\psi}_0) \ , 
\ee
provided that $a=0$ and $b=h_2$, giving further credence to this choice. However, the corresponding relation for $\hat{G}(\Delta)$, namely $\hat{G}(\Delta) = H^{-1}(-\Delta -\sigma_3 \hat{\psi}_0)$, does not hold. This is not in contradiction with eqs.~(\ref{keyiden}) and (\ref{keyidenH}) since both $H(\Delta)$ and $\hat{G}(\Delta)$ behave like the identity, i.e.\ their numerator is the same as their denominator.

\subsection{The $x$ $\bar{x}$ relation}

We proceed to use similar arguments to constrain the $x(z) \bar{x}(w)$ OPE. 
We start with the ansatz 
\begin{equation}\label{xbarxOPE}
x(z) \, \bar{x}(w) \sim - D(\Delta) \, \bar{x}(w) \, x(z) \ . 
\end{equation}
First, applying it on the vacuum $|\emptyset\rangle$, and using  (\ref{barxonxfinal}) and (\ref{xonxbar}), we conclude that $D(\Delta)$ has to contain the factor
\be\label{Dc1}
D(\Delta) \sim  \frac{(\Delta + h_2 - \sigma_3\hat{\psi}_0)}{(\Delta - h_2 + \sigma_3\psi_0)} 
\ee
in order to cancel the poles corresponding to the states $|\widehat{{\tiny\yng(3)}}_{\,2}\rangle $ and $|{\tiny\yng(3)}_{\,2}\rangle $, respectively.
%and determine $D(\Delta)$ from the condition that it should cancel the poles in (\ref{barxonxfinal}) and (\ref{xonxbar}). 
%This leads to the condition that 
We can also apply the ansatz (\ref{xbarxOPE}) 
%We have also analysed the OPE relation (\ref{xbarxOPE}) 
on $|{\tiny\yng(1)}\rangle$, leading to 
\begin{equation}
x(z)\, \bar{x}(w) \, |{\tiny\yng(1)}\rangle \sim \frac{1}{w} \Bigl[ 
 \frac{1}{(z+\sigma_3\psi_0-h_2)}\,  |({\tiny\yng(1)}, \widehat{{\tiny\yng(3)}}_{\, 2 })\rangle  + \frac{1}{(z-h_2)} \, | ({\rm adj},{\rm adj} +\hat{\tiny\yng(1)}_{\, \textrm{top}} ) \rangle \Bigr] 
\end{equation} 
and 
\begin{equation}
\bar{x}(w) \, x(z)\,  |{\tiny\yng(1)}\rangle \sim \frac{1}{(z-h_2)} \Bigl[ 
\frac{1}{(w+\sigma_3\hat{\psi}_0-2h_2)}\, | {{\tiny\yng(4)}}_{\, 2 }\rangle + \frac{1}{w}\,  | ({\rm adj},{\rm adj}+\hat{{\tiny\yng(1)}}_{\, \textrm{top}}) 
\Bigr] \ .
\end{equation} 
However, this does not give rise to any further constraint beyond (\ref{Dc1}). %, i.e.\ it only confirms (\ref{Dc1}) but does not provide new information.
%and hence just requires again 
%(\ref{Dc1}), but does not fix $d$ or $\hat{d}$. 

Recall that in the free field limit, the relations are  quadratic (see eq.\ (4.35) of \cite{GLPZ}),
%We also note from the free field limit, see \cite[eq.~(4.35)]{GLPZ}, that 
%he relation must be quadratic, i.e.\
and hence  $D(\Delta)$ should be of the form 
\begin{equation}\label{Dansatz}
D(\Delta) = \frac{(\Delta + h_2 - \sigma_3\hat{\psi}_0) \, (\Delta + \hat{d})}{(\Delta - h_2 + \sigma_3\psi_0)\,(\Delta+d)} \ , 
\end{equation}
where%, in order to reproduce the correct free field limit, 
\begin{equation}
d = 2\ , \qquad \hbox{and} \qquad \hat{d} = -1   
\end{equation}
in the free field limit. 
%We have also analysed the OPE relation (\ref{xbarxOPE}) on $|{\tiny\yng(1)}\rangle$, and this leads to 
%\begin{equation}
%x(z)\, \bar{x}(w) \, |{\tiny\yng(1)}\rangle \sim \frac{1}{w} \Bigl[ \frac{1}{(z+\sigma_3\psi_0-h_2)}\,  |({\tiny\yng(1)}, \widehat{{\tiny\yng(3)}}_{\, 2 })\rangle  + \frac{1}{(z-h_2)} \, | ({\rm adj},{\rm adj} +\hat{\tiny\yng(1)}_{\, \textrm{top}} ) \rangle \Bigr] 
%\end{equation} 
%and 
%\begin{equation}
%\bar{x}(w) \, x(z)\,  |{\tiny\yng(1)}\rangle \sim \frac{1}{(z-h_2)} \Bigl[ \frac{1}{(w+\sigma_3\hat{\psi}_0-2h_2)}\, | {{\tiny\yng(4)}}_{\, 2 }\rangle + \frac{1}{w}\,  | ({\rm adj},{\rm adj}+\hat{{\tiny\yng(1)}}_{\, \textrm{top}}) \Bigr] \ ,
%\end{equation} 
%and hence just requires again (\ref{Dc1}), but does not fix $d$ or $\hat{d}$. 
We shall make a proposal for $D(\Delta)$ in Section~\ref{sec:xxbar} below, 
after we have postulated an explicit action of $x$ and $\bar{x}$ on twin-plane-partitions.

\section{Action of fermionic creation operators}\label{sec:xaction}

In this section we postulate explicit actions of $x(w)$ and $\bar{x}(w)$ on twin-plane-partitions that are compatible with all these OPE relations. 
The action of $x(w)$ will be derived in detail, and that of $\bar{x}(w)$ will follow by symmetry. 
These actions then allow us to determine the OPE relations between the fermionic creation operators.

\subsection{Defining the action of $x(w)$}
We start with $x$.
As will become clear, the definition is most natural for $a=0$ and $b=h_2$, thus justifying the explicit choices (\ref{GHfin}) and (\ref{GhatHhatfin}). 
We shall also fix the form of $D(\Delta)$. 

Let us begin with a general ansatz, and then explain how the different factors can be determined by the OPE relations. 
First we recall that the action of $e(z)$ is defined via  
\begin{equation}\label{eactionbosonic}
e(z)|\Lambda\rangle =\sum_{{\tiny{\yng(1)}}\in \textrm{Add}(\Lambda)}\frac{E(\Lambda\rightarrow \Lambda+{\tiny{\yng(1)}})}{z-h({\tiny{\yng(1)}})}\, |\Lambda+{\tiny{\yng(1)}}\rangle \ ,
\end{equation}
where the sum is over all positions where a box ${\tiny{\yng(1)}}$ can be added, and we have introduced the short-hand notation 
\be\label{shorthand}
E(\Lambda\rightarrow \Lambda+{\tiny{\yng(1)}}) \equiv \Bigl[ -  \frac{1}{\sigma_3} {\rm Res}_{w = h({\tiny \yng(1)})} \psi_{\Lambda}(w) \Bigr]^{\frac{1}{2}} \ , 
\ee
see eq.~(\ref{ppart}). 
We now make a similar ansatz for the $x(w)$ action as 
\begin{equation}\label{xansatz}
x(w)|\Lambda\rangle =\sum_{\blacksquare\in \textrm{Add}(\lambda)}\frac{P_+(\Lambda\rightarrow \Lambda+\blacksquare)}{w-p_+(\blacksquare)}|\Lambda+\blacksquare\rangle
+ \sum_{\overline{\blacksquare}\in \textrm{Rem}(\hat{\rho})}\frac{P_-(\Lambda\rightarrow \Lambda-\overline{\blacksquare})}{w-p_-(\overline{\blacksquare})}|\Lambda-\overline{\blacksquare}\rangle  \ , 
\end{equation}
where we sum over all positions where a box $\blacksquare$ can be added to the Young diagram $\lambda$ or removed from the Young diagram $\hat{\rho}$, respectively. 
We also demand that the other requirements explained in Section~\ref{sec:4.5} and Section~\ref{sec:4.7} are satisfied. 
\smallskip

The positions of the poles have already been determined in Sections ~\ref{sec:4.5} and \ref{sec:4.7}: for $p_+(\blacksquare)$ we find from eq.~(\ref{pplus})
%Furthermore, $p_+(\blacksquare)$ is defined as (
%The result of  gives
\begin{equation}\label{p+f}
p_+(\blacksquare) \equiv h(\blacksquare) +\ell h_2 \ ,
\end{equation}
where $\ell$ is the number of additional boxes extending the minimal length bud; and eq.~(\ref{zast}) gives
%$p_-(\overline{\blacksquare})$ equals, see eq.~(\ref{zast})
\begin{equation}\label{p-f}
p_-(\overline{\blacksquare})\equiv\hat{h}(\overline{\blacksquare}) + (\ell+1) h_2 - \sigma_3 \psi_0 =g(\overline{\blacksquare})+\ell h_2  \ , 
\end{equation}
where $\ell$ is the number of boxes on top of the corresponding wall. 

In this section, we  determine the coefficients $P_+(\Lambda\rightarrow \Lambda+\blacksquare)$ and $P_-(\Lambda\rightarrow \Lambda-\overline{\blacksquare})$. 
%\smallskip
%\subsection{Determining $X_+(\Lambda\rightarrow \Lambda+\blacksquare)$ and $X_-(\Lambda\rightarrow \Lambda-\overline{\blacksquare})$ }
To this end we first apply the OPE relation 
\be\label{c1}
e(z) \, x(w) \sim G(z-w) \, x(w) \, e(z) 
\ee
on an arbitrary twin-plane-partition $\Lambda$. 
Since $e(z)$ can add a box in all permissible positions and  $x(w)$ can add a $\blacksquare$ or remove a $\overline{\blacksquare}$ in all permissible positions, there are two possible scenarios. 
Let us first consider the case where the action of $e$ and $x$ affect different positions of $\Lambda$ such that they may be performed in either order. 
Then it follows from (\ref{c1}) that 
\begin{equation}\label{case1}
\begin{aligned}
&\frac{E(\Lambda+\blacksquare\rightarrow \Lambda+\blacksquare+{\tiny{\yng(1)}})}{z-h({\tiny{\yng(1)}})} \cdot  \frac{P_+(\Lambda \rightarrow \Lambda+\blacksquare)}{w-p_+(\blacksquare)}\\
& \qquad  \qquad \sim G(z-w) \cdot \frac{P_+(\Lambda+{\tiny{\yng(1)}}\rightarrow \Lambda+{\tiny{\yng(1)}}+\blacksquare)}{w-p_+(\blacksquare)}  \cdot\frac{E(\Lambda \rightarrow \Lambda+{\tiny{\yng(1)}})}{z-h({\tiny{\yng(1)}})} 
 \end{aligned}
\end{equation}
as well as 
\begin{equation}\label{case2}
\begin{aligned}
&\frac{E(\Lambda-\overline{\blacksquare}\rightarrow \Lambda-\overline{\blacksquare}+{\tiny{\yng(1)}})}{z-h({\tiny{\yng(1)}})} \cdot  \frac{P_-(\Lambda \rightarrow \Lambda -\overline{\blacksquare})}{w-p_-(\overline{\blacksquare})}\\
& \qquad  \qquad \sim G(z-w) \cdot \frac{P_-(\Lambda+{\tiny{\yng(1)}}\rightarrow \Lambda+{\tiny{\yng(1)}}-\overline{\blacksquare})}{w-p_-(\overline{\blacksquare})}  \cdot\frac{E(\Lambda \rightarrow \Lambda+{\tiny{\yng(1)}})}{z-h({\tiny{\yng(1)}})} \ .
 \end{aligned}
\end{equation}
Next we use the explicit form of (\ref{shorthand}) to conclude that 
\begin{equation}
\frac{E(\Lambda+\blacksquare\rightarrow \Lambda+\blacksquare+{\tiny{\yng(1)}})}{E(\Lambda\rightarrow \Lambda+{\tiny{\yng(1)}})} =\sqrt{\frac{\textrm{Res}_{u\rightarrow h({\tiny{\yng(1)}})}\psi_{\Lambda+\blacksquare}(u)}{\textrm{Res}_{u\rightarrow h({\tiny{\yng(1)}})}\psi_{\Lambda}(u)}} =\sqrt{\varphi_2(h({\tiny{\yng(1)}}) - p_+(\blacksquare))} \ . 
\end{equation}
Thus, in order to satisfy (\ref{c1}), the most natural solution arises provided we take 
\be
G(\Delta) = \varphi_2(\Delta) \ , 
\ee
i.e.\ set $a=0$ in (\ref{Gfin}), leading to eq.~(\ref{GHfin}). 
Then the solution to (\ref{case1}) is simply 
\begin{equation}\label{ans1}
\frac{P_+(\Lambda+{\tiny{\yng(1)}}\rightarrow \Lambda+{\tiny{\yng(1)}}+\blacksquare)}{
P_+(\Lambda\rightarrow \Lambda+\blacksquare)}  =\sqrt{\varphi^{-1}_2(h({\tiny{\yng(1)}}) - p_+(\blacksquare))} \ .
\end{equation}
This relation can be naturally written in residue form, i.e.\  (\ref{ans1}) is satisfied provided that $P_+(\Lambda\rightarrow \Lambda+{\blacksquare})$ contains the factor 
\begin{equation}\label{fact1}
%X(\Lambda\rightarrow \Lambda+{\blacksquare}) \supset 
\Bigl\{ \textrm{Res}_{u=p_+(\blacksquare)}\prod_{{\tiny{\yng(1)}}\in \mathcal{E}} P_{\,{\tiny{\yng(1)}}}(u) \Bigr\}^{\frac{1}{2}} \quad \textrm{with}\quad \boxed{P_{\,{\tiny{\yng(1)}}}(u) = \varphi^{-1}_{2} (-u+h({\tiny{\yng(1)}}))} \ .
\end{equation}

The analysis for the second term in (\ref{xansatz}), i.e.\ eq.~(\ref{case2}), is similar, and we find that it is satisfied provided that $P_-(\Lambda\rightarrow \Lambda-\overline{\blacksquare})$ contains the factor 
\begin{equation}
\Bigl\{  \textrm{Res}_{u=p_-(\overline{\blacksquare})}\prod_{{\tiny{\yng(1)}}\in \mathcal{E}} P_{\,{\tiny{\yng(1)}}}(u) 
\Bigr\}^{\frac{1}{2}} \ ,  
\end{equation}
where $P_{\,{\tiny{\yng(1)}}}(u)$ is the same function as in (\ref{fact1}). 
We can also study the constraints arising from the OPE relation with $f(z)$, and they turn out to be compatible with this ansatz. 

Similarly, we can impose the constraints from the OPE relations with $\hat{e}(z)$ and $\hat{f}(z)$, and these are satisfied\footnote{Again, this is simplest provided that we choose $b=h_2$ in (\ref{Hhat}) and (\ref{Ghat}), leading to eq.~(\ref{GhatHhatfin}).} as long as 
$P_+(\Lambda\rightarrow \Lambda + {\blacksquare})$ and $P_-(\Lambda\rightarrow \Lambda - \overline{\blacksquare})$ contain the factors 
\begin{equation}
%X(\Lambda\rightarrow \Lambda+{\blacksquare}) \supset  
\Bigr\{ \textrm{Res}_{u=p_{+}({\blacksquare})}\prod_{\widehat{{\tiny{\yng(1)}}}\in \widehat{\mathcal{E}}} P_{\widehat{{\tiny{\yng(1)}}}}(u)   \Bigr\}^{\frac{1}{2}} \qquad \textrm{and}\qquad  
\Bigr\{ \textrm{Res}_{u=p_{-}(\overline{\blacksquare})}\prod_{\widehat{{\tiny{\yng(1)}}}\in \widehat{\mathcal{E}}} P_{\widehat{{\tiny{\yng(1)}}}}(u) \Bigr\}^{\frac{1}{2}} \ , 
\end{equation}
respectively, where 
\be
\boxed{P_{\widehat{{\tiny{\yng(1)}}}}(u)=\varphi^{-1}_{2} (u-\hat{h}(\widehat{\tiny{\yng(1)}})- \sigma_3 \hat{\psi}_0)} \ .
\ee
\smallskip

Finally, we expect factors corresponding to the boxes $\blacksquare$ of $\lambda$ and $\overline{\blacksquare}$ in $\hat{\rho}$; they can be fixed such that the action of $x(w)$ does not modify the overall coefficient function. 
This is achieved by setting 
\be\label{Xpm}
P_+(\Lambda\rightarrow \Lambda + {\blacksquare}) =  \Bigl\{ \textrm{Res}_{u=p_{+}({\blacksquare})} \, P_\Lambda(u) \Bigr\}^{\frac{1}{2}} \ , \quad 
P_-(\Lambda\rightarrow \Lambda - \overline{\blacksquare}) =  \Bigl\{ \textrm{Res}_{u=p_{-}(\overline{\blacksquare})} \, P_\Lambda(u) \Bigr\}^{\frac{1}{2}} \ , 
\ee
where 
\begin{equation}\label{XLambda}
P_\Lambda(u) = P_0(u)\, \Biggl\{
\prod_{\blacksquare \in \lambda}  P_{\,\blacksquare}(u) \,
\prod_{\overline{\blacksquare}\in \hat{\rho}} P_{\,\overline{\blacksquare}}(u) \, 
\prod_{ {\tiny{\yng(1)}} \in \mathcal{E}} P_{\,{\tiny{\yng(1)}}}(u) \, 
\prod_{\widehat{{\tiny{\yng(1)}}} \in \widehat{\mathcal{E}} } P_{\widehat{{\tiny{\yng(1)}}}}(u)  \Biggr\} \ .
\end{equation}
Here we take $P_0(u)$ to be the ``vacuum" factor\footnote{The precise form of $P_0(u)$ is not fixed by this analysis. However, we shall see below that this choice reproduces the correct free field limit.}
\be\label{P0}
P_0(u) = \frac{\psi_0(u)}{\hat{\psi}_0(u-\sigma_3 \hat{\psi}_0) }= 
\Bigl(1 + \frac{\sigma_3 \psi_0}{u}\Bigr) \, \Bigl( 1 - \frac{\sigma_3 \hat{\psi}_0}{u} \Bigr)   \ , 
\ee
and the remaining factors are 
\begin{equation}
\boxed{
\begin{aligned}
P_{\,\blacksquare}(u) & =\prod^{m+n-1}_{k=0}\varphi^{-1}_{2} \bigl(-u+g (\blacksquare)+ k  h_2\bigr) \\
P_{\,\overline{\blacksquare}\,}(u) & =\prod^{\hat{m}+\hat{n}+2}_{k=0}
\varphi^{-1}_{2} \bigl(u-(\hat{g}(\overline{\blacksquare}) +k h_2)- \sigma_3 \hat{\psi}_0\bigr) 
%\prod^{2}_{k=1}\varphi^{-1}_{2} (u-(h(\overline{\blacksquare})+k h_2)- \sigma_3 \hat{\psi}_0) \ . 
\end{aligned}} 
\end{equation}

\subsection{Properties}

The coefficient function $P_\Lambda(u)$ has a number of rather special properties. 
In particular, the value of $P_\Lambda(u)$ is invariant under 
the action of $x(w)$. 

\subsubsection{Moving individual boxes}

Let us start by showing that $P_\Lambda(u)$ stays invariant as boxes are moved from the left to the right under the $x$-creation action. 
For example, if the bud at $(x_1,x_3)=(m,n)$ is longer and contains a box ${\tiny{\yng(1)}}$ at position $(m,m+n+\ell,n)$ with $\ell\geq 0$, then this contributes to $P_{\,{\tiny{\yng(1)}}}(u)$ the factor
\be\label{6.7a}
\varphi_2^{-1} ( -u + h({\tiny{\yng(1)}}) )\qquad \textrm{with}\quad h({\tiny{\yng(1)}})= h(\blacksquare) + \ell h_2   \ .
% = \varphi_2^{-1} (- u - nh_1 - m h_3 + \ell h_2) \ . 
\ee
After adding $\blacksquare$ at $(x_1,x_3)=(m,n)$, the extra box is shifted to the right and appears on top of the corresponding wall, with position $(-n,\ell,-m)$ thus contributing to $P_{\widehat{{\tiny{\yng(1)}}}}(u)$, the factor
\begin{equation}
\varphi^{-1}_{2} (u-\hat{h}(\widehat{\tiny{\yng(1)}})- \sigma_3 \hat{\psi}_0) \qquad \textrm{with}\quad \hat{h}(\widehat{\tiny{\yng(1)}})= \hat{g}(\blacksquare) + \ell h_2  \ . 
\end{equation}
%\be
%\varphi_2^{-1} ( u - h(\blacksquare) + \sigma_3 \hat{\psi}_0 - (\ell+1) h_2 - \sigma_3 \hat{\psi}_0 ) = 
%\varphi_2^{-1} ( u -  h(\blacksquare) - \ell h_2 - h_2) \ , 
%\ee
This is identical to (\ref{6.7a}) because of (\ref{hblack}) and  (\ref{varphi2m}). 

The same phenomenon also happens for the $x$-annihilation action. 
A box ${\tiny{\yng(1)}}$ on top of the wall corresponding to $\overline{\blacksquare}$ at $(\hat{x}_1,\hat{x}_3)=(\hat{m},\hat{n})$ with position $(-\hat{n},\ell,-\hat{m})$  contributes to $P_{\,{{\tiny{\yng(1)}}}}(u)$ the factor
\be\label{aa1}
\varphi_2^{-1} ( -u + h({\tiny{\yng(1)}}) )\qquad \textrm{with}\quad h({\tiny{\yng(1)}})=   g(\overline{\blacksquare}) +\ell h_2 \ .
\ee
%\be\label{aa1}
%\prod_{r=1}^{\ell} \varphi_2^{-1}(-u+rh_2 - \sigma_3\psi_0  + \hat{h}(\overline{\blacksquare})  ) 
%\ee
After the removal of this $\overline{\blacksquare}$, this box appears at positions $(\hat{m},\hat{m}+\hat{n}+3+\ell,\hat{n})$  on the hatted side, and  contributes to $P_{\,\widehat{{\tiny{\yng(1)}}}}(u)$ the factor
\begin{equation}
\varphi^{-1}_{2} (u-\hat{h}(\widehat{\tiny{\yng(1)}})- \sigma_3 \hat{\psi}_0) \qquad \textrm{with}\quad \hat{h}(\widehat{\tiny{\yng(1)}})= \hat{h}(\overline{\blacksquare})   + (\ell+3) h_2  \ . 
\end{equation}
Again, this is identical to (\ref{aa1}) because of (\ref{hbarblack}) and  (\ref{varphi2m}). 
%\be
%\prod_{r=\hat{m}+\hat{n}+3}^{\hat{m}+\hat{n}+2+\ell} \varphi_2^{-1}(u-\sigma_3\hat{\psi}_0 - \hat{g}(\overline{\blacksquare})  - r h_2) 
%\ee
%to $P_{\widehat{{\tiny{\yng(1)}}}}(u)$. 
%We can rewrite the second product as 
%\be
%\prod_{r'=1}^{l} \varphi_2^{-1} ( u + \sigma_3 {\psi}_0 - \hat{h}(\overline{\blacksquare}) - r' h_2  -h_2) 
%= \prod_{r'=1}^{l} \varphi_2^{-1} (-u - \sigma_3\psi_0 + \hat{h}(\overline{\blacksquare})  + r'h_2) \ , 
%\ee
%where we have used $\sigma_3 \psi_0 + \sigma_3 \hat{\psi}_0 = - h_2$, see eq.~(\ref{3.30}),  as well as (\ref{varphi2m}). 
%This therefore then agrees with (\ref{aa1}).

\subsubsection{Invariance of $P_{\Lambda}(u)$ under $x(w)$}

For both the creation and the annihilation action, the contribution to $P_{\,{{\tiny{\yng(1)}}}}(u) \cdot P_{\widehat{{\tiny{\yng(1)}}}}(u)$ however changes. 
In the case of the creation-action of $x$ this is because the boxes that made up the bud\footnote{Because of the analysis of the previous subsection, it is sufficient to consider the case where the bud has minimal length.} before the $x$-action are part of ${\cal E}$, while they do not contribute to 
${\cal E}$ after the $x$-action. 
The difference in this case is 
\be\label{stub}
\prod_{r=0}^{{m}+{n}-1} \varphi_2^{-1}(-u + g(\blacksquare) + r h_2) \ . 
\ee
This is then precisely compensated by the additional contribution of the $P_{\,\blacksquare}(u)$ factor. 

Similarly, for the case of the $x$-annihilation action, the difference is due to the fact that even for $\ell=0$, $\hat{m}+\hat{n}+3$ hatted boxes appear after the $\overline{\blacksquare}$-wall has been removed; they contribute to $P_{\widehat{{\tiny{\yng(1)}}}}(u)$
\be
\prod_{r=0}^{\hat{m}+\hat{n}+2} \varphi_2^{-1}(u-\hat{g}(\overline{\blacksquare}) - r h_2 - \sigma_3 \hat{\psi}_0) \ . 
\ee
This is then precisely compensated by the fact that now one fewer $\overline{\blacksquare}$ term contributes to the $P_{\,\overline{\blacksquare}}(u)$ factor.

As a consequence, the generic $x x$ OPE is trivial, since acting with the first $x$ operator does not affect the coefficient for the second one and vice versa. 
In particular, this is compatible with the OPE relation 
\begin{equation}\label{xxOPE}
\boxed{
x(z)\, x(w) \sim - x(w) \, x(z) } \ , 
\end{equation}
which is also what the ${\cal W}_{\infty}$ perspective suggests, see the discussion in Section~\ref{sec:interactf}.\footnote{ 
Note that since the coefficient functions (\ref{Xpm}) are square roots, the sign appearing in (\ref{xxOPE}) is a bit delicate. 
We believe that these signs come out correctly (upon choosing a suitable prescription for %how to define the 
the signs of the coefficient functions), but have not carefully studied this.}

\subsubsection{Structure of poles}\label{sec:xpoles}

We %should also note 
now check that the coefficient functions defined above give the correct poles. 
For example, for the $x$-addition formula, the minimal length bud contributes (\ref{stub}) to $P_{\,{{\tiny{\yng(1)}}}}(u)$, which we can rewrite as 
\begin{equation}
\prod_{r=1}^{m+n} \varphi_2^{-1} (u - g(\blacksquare) - r h_2 ) \ . 
\end{equation}
In particular, this has a pole at, cf.\ eq.~(\ref{p+f}),
\begin{equation}
u^\ast =  g(\blacksquare) + (m+n) h_2   = h(\blacksquare) = p_+(\blacksquare)\ , 
\end{equation}
as desired. 
When the bud has $\ell$ extra boxes, the pole is shifted to $ h(\blacksquare)+\ell h_2$,  cf.\ eq.~(\ref{pplus}).

We should mention that the condition that the bud has the correct minimal length is enforced by the contribution of the neighbouring $\blacksquare$-boxes to $P_{\,\blacksquare}(u)$. To see this, we first note that if we can add a $\blacksquare$ at $(m,n)$, there must be either a $\blacksquare$ already at $(m-1,n)$ or at $(m,n-1)$. For definiteness let us assume that there is already a $\blacksquare$ present at $(m-1,n)$ ---  the analysis for the case $(m,n-1)$ is essentially identical. Then  it contributes to $P_{\,\blacksquare}(u)$ 
\be
\prod_{r=0}^{m+n-2}  \varphi_2^{-1} ( -u + (m-1) h_1 + n h_3 + r h_2) = 
\prod_{r=1}^{m+n-1} \varphi_2^{-1} \Bigl( u - \bigl( (m-1)h_1 + n h_3 + r h_2\bigr) \Bigr) \ . 
\ee
This has then zeros at 
\be
u^\ast = g(\blacksquare) + r h_2 \ , \qquad r = 1,\ldots, m+n-1 \ , 
\ee
thus killing the poles from any bud that is too short. 
\smallskip

Similarly, for the $x$-removal term, when there is no extra (unhatted) box sitting on top of the wall corresponding to a box $\overline{\blacksquare}$, 
%for $\ell=0$, 
the relevant pole comes from the contribution of $\overline{\blacksquare}$ to $P_{\,\overline{\blacksquare}}(u)$, whose $k=\hat{m}+\hat{n}+2$ term equals 
\be
\varphi^{-1}_{2} \bigl(u-\hat{h}(\overline{\blacksquare})- 2 h_2 - \sigma_3 \hat{\psi}_0\bigr)  = 
\varphi^{-1}_{2} \bigl( u-g(\overline{\blacksquare})\bigr) \ ,
\ee
which has a pole at $u^{\ast}=g(\overline{\blacksquare})$.
When there are $\ell\geq 1$ (unhatted) boxes on top of the wall corresponding to $\overline{\blacksquare}$, the $\ell$'th box contributes to $P_{\,{\tiny{\yng(1)}}}(u)$
\be
\varphi_2^{-1} \bigl( -u + g(\overline{\blacksquare}) + (\ell-1) h_2 \bigr)   = 
\varphi_2^{-1} \Bigl(u - \bigl(g(\overline{\blacksquare})+\ell h_2\bigr) \Bigr) \ . 
\ee
This has a pole at 
\be
u^\ast =  g(\overline{\blacksquare})+ \ell h_2= p_-(\overline{\blacksquare})\ , 
\ee
see eq.~(\ref{p-f}). 

\smallskip

We should mention though that in general the coefficient function $P_{\Lambda}(u)$ also has ``spurious" poles that do not correspond to any allowed addition (or removal) term. 
For example, if $\Lambda$ consists just of a single box in ${\cal E}$, then 
\begin{equation}
\begin{aligned}
P_\Lambda(u) =P_0(u) P_{\,{\tiny{\yng(1)}}}(u) &=\frac{\psi_0(u)}{\hat{\psi}_0(u-\sigma_3\hat{\psi}_0)} \varphi^{-1}_{2} (-u+h({\tiny{\yng(1)}})) \\
&= \frac{(u+\sigma_3\psi_0)(u-\sigma_3\hat{\psi}_0)(u+h_1)(u+h_3)}{u^3(u-h_2)} \ . 
\end{aligned}
\end{equation}
While the pole at $u=h_2$ corresponds to the usual $\blacksquare$ addition term, $u=0$ does not seem to have any obvious meaning. Similar ``spurious" poles also appear for other configurations.

\subsubsection{Special cases}

In deriving our coefficient functions above, we have assumed that the action of the two operators affect different positions of $\Lambda$, such that they may be performed in either order (see the comment after eq.~(\ref{c1})). 
We can also ask what happens if this is not the case. 

For the $e(z) x(w)$ OPE, two special cases may occur: first it can happen that a new position for the single box becomes available after the $\blacksquare$-row has been added; then we have a term from $e(z) x(w)$ that does not appear in $x(w) e(z)$. 
It is not difficult to see that the relevant poles of this term are given by
\begin{equation}
\frac{1}{(z-p_+(\blacksquare) - h_j)} \, \frac{1}{(w-p_+(\blacksquare))} \ , \qquad \hbox{where $j=1,3$} \ , 
\end{equation}
since the position of this single box differs from the place where the infinite $\blacksquare$-row has been added by either $h_1$ or $h_3$. 
This term is cancelled (in the usual way) by the denominator of the $G(z-w)=\varphi_2(z-w)$ function in (\ref{exG}). 

The other special case arises if the single box is added to the bud, and then becomes invisible after the $\blacksquare$ row has been added. 
(This configuration only contributes to $x(w) e(z)$ then.) 
In this case the poles are given by\footnote{Here $p_+(\blacksquare)$ is defined for $\Lambda$, before the individual box has been added.}
\begin{equation}
\frac{1}{(w-p_+(\blacksquare) - h_2)} \, \frac{1}{(z-p_+(\blacksquare))} \ , 
\end{equation}
which is indeed cancelled by the numerator of $G(z-w)=\varphi_2(z-w)$ in (\ref{exG}).
Thus we conclude that the above action respects the $e(z) x(w)$ OPE relations. 
\smallskip

For the case of the $f(z) x(w)$ OPE the special case arises if $f(z)$ removes the last box of a (sufficiently long) bud; this box cannot be removed any more by $f(z)$ after $x(w)$ has been added. 
In this case, the two poles appear at the same position, and the contribution is cancelled by the numerator of $H(\Delta)=\Delta/ \Delta$ in (\ref{fxansatz}). Thus also the $f(z) x(w)$ OPE is respected. 
\smallskip

Similarly, for the case of the $\hat{f}(z) x(w)$ OPE, the special cases arise for the situation where the bud is initially too short for the action of $x(w)$, but $\hat{f}(z)$ removes the hatted box that makes then the action of $x(w)$ possible, see the discussion at the very end of Section~\ref{sec:4.5}. 
In this case, the pole for the $x(w)$ is $p_+({\blacksquare})=h({\blacksquare}) + \ell h_2$, both before and after the hatted box was removed, since the removal of the hatted box affects a neighboring wall and hence does not modify $p_+({\blacksquare})$. 
The pole for the $\hat{f}(z)$ action is (depending on which of the two neighbouring walls the hatted box sits on, we either have $k=1$ or $k=3$)
\be
z^\ast = h({\blacksquare}) + (\ell+1) h_2 - \sigma_3 \hat{\psi}_0 + h_k = p_+({\blacksquare})-\sigma_3\hat{\psi}_0 - h_j \ ,
\ee
where  $j=1$ for $k=3$ and $j=3$ for $k=1$. (To check that this is indeed the correct pole, see eq.~(\ref{hhatbottop}).)
Therefore the pole structure for the $x(w)\hat{f}(z)$ action is given by 
\be\label{yetan}
\frac{1}{\bigl(w-p_+({\blacksquare})\bigr)} \, \frac{1}{\bigl(z- (p_+({\blacksquare})-\sigma_3\hat{\psi}_0 - h_j)\bigr)} \ , 
\ee
and hence the singular contribution of (\ref{yetan}) is cancelled by the numerator of $\hat{H}(\Delta)$, see eq.~(\ref{Hhat}). 
\smallskip

Finally, for the $\hat{e}(z) x(w)$ OPE, the special case arises when the hatted box is added at a position that only becomes available after the addition of the ``new wall". 
This only happens when the hatted box is added on top of the ``new" wall, i.e.\  with pole
\be
\frac{1}{\bigl(z-(p_+({\blacksquare})+h_2-\sigma_3\hat{\psi}_0)\bigr)} \, \frac{1}{\bigl(w-p_+({\blacksquare})\bigr)} \,  \ . 
\ee
This is then cancelled by the denominator of $\hat{G}(\Delta)$, see eq.~(\ref{Ghat}). 
\smallskip

In the above we have always assumed that $x(w)$ acts via creating a $\blacksquare$, i.e.\ via the first term in 
(\ref{xansatz}). 
There are similar considerations that apply when one considers the second term. 
For example, for the $\hat{e}(z) \, x(w)$ OPE, we can add a hatted box at the end of the $\hat{m}+\hat{n}+3+\ell$ hatted boxes that were created after $x$ removed the corresponding $\overline{\blacksquare}$. 
(This position is unavailable before $x(w)$ has acted.) 
The pole for the $\hat{e}(z)$ action is
\be
z^\ast = \hat{h}(\overline{\blacksquare}) + (\ell+3) h_2 =
p_{-}(\overline{\blacksquare}) + h_2 - \sigma_3 \hat{\psi}_0   \ , 
\ee
as follows from (\ref{3.30}). 
The total pole structure for the $\hat{e}(z) x(w)$ action is therefore 
\be
\frac{1}{\bigl(z- ( p_{-}(\overline{\blacksquare}) + h_2 - \sigma_3 \hat{\psi}_0 ) \bigr)} \, \frac{1}{\bigl(w - p_{-}(\overline{\blacksquare})\bigr)}  \ .
\ee
This is again cancelled by the denominator of $\hat{G}(\Delta)$ in (\ref{Ghat}). 
The other cases work similarly. 

\subsection{Action of $\bar{x}(w)$}

For $\bar{x}(w)$ the formula works very similarly: instead of (\ref{xansatz}) we write 
\begin{equation}\label{barxansatz}
\bar{x}(w)|\Lambda\rangle =\sum_{\overline{\blacksquare}\in \textrm{Add}(\hat{\rho})} \frac{\bar{P}_+(\Lambda\rightarrow \Lambda+\overline{\blacksquare})}{w-\bar{p}_+(\overline{\blacksquare})}|\Lambda+\overline{\blacksquare}\rangle
+ \sum_{\blacksquare\in \textrm{Rem}(\lambda)} \frac{\bar{P}_-(\Lambda\rightarrow \Lambda-{\blacksquare})}{w-\bar{p}_-({\blacksquare})}|\Lambda-{\blacksquare}\rangle  \ , 
\end{equation}
where the coefficient functions 
%$\bar{X}_+(\Lambda\rightarrow \Lambda+\overline{\blacksquare})$ and $\bar{X}_-(\Lambda\rightarrow \Lambda-{\blacksquare})$ 
are now given by 
\be\nonumber
\bar{P}_+(\Lambda\rightarrow \Lambda + \overline{\blacksquare}) =  \Bigl\{ \textrm{Res}_{u=\bar{p}_{+}(\overline{\blacksquare})} \, \bar{P}_\Lambda(u) \Bigr\}^{\frac{1}{2}} \ , \quad 
\bar{P}_-(\Lambda\rightarrow \Lambda - {\blacksquare}) =  \Bigl\{ \textrm{Res}_{u=\bar{p}_{-}({\blacksquare})} \, \bar{P}_\Lambda(u) \Bigr\}^{\frac{1}{2}} \ , 
\ee
with 
\begin{equation}\label{barXLambda}
\bar{P}_\Lambda(u) = \bar{P}_0(u)\, \Biggl\{
\prod_{\blacksquare \in \lambda}  \bar{P}_{\,\blacksquare}(u) \,
\prod_{\overline{\blacksquare}\in \hat{\rho}} \bar{P}_{\,\overline{\blacksquare}}(u) \, 
\prod_{ {\tiny{\yng(1)}} \in \mathcal{E}} \bar{P}_{\,{\tiny{\yng(1)}}}(u) \, 
\prod_{\widehat{{\tiny{\yng(1)}}} \in \widehat{\mathcal{E}} } \bar{P}_{\widehat{{\tiny{\yng(1)}}}}(u)  \Biggr\} \ ,
\end{equation}
and
\begin{equation}
\boxed{
\begin{aligned}
\bar{P}_{\,{\tiny{\yng(1)}}}(u) & = \varphi_2^{-1}(u-h({\tiny{\yng(1)}}) - \sigma_3 \psi_0) \\
\bar{P}_{\widehat{{\tiny{\yng(1)}}}}(u)  & = \varphi_2^{-1}(-u+\hat{h}(\widehat{\tiny{\yng(1)}})) 
\end{aligned}
}
\end{equation}
as well as 
\begin{equation}
\boxed{
\begin{aligned}
\bar{P}_{\,\blacksquare}(u) & =\prod^{m+n+2}_{k=0}\varphi^{-1}_{2} \bigl(u-(g(\blacksquare) +  k h_2) - \sigma_3 \psi_0\bigr) \\
\bar{P}_{\,\overline{\blacksquare}}(u) & =\prod^{\hat{m}+\hat{n}-1}_{k=0}
\varphi^{-1}_{2} \bigl(- u+ \hat{g}(\overline{\blacksquare}) + k h_2\bigr) 
%\prod^{2}_{k=1}\varphi^{-1}_{2} (u-(h(\overline{\blacksquare})+k h_2)- \sigma_3 \hat{\psi}_0) \ . 
\end{aligned}}
\end{equation}
\smallskip

\noindent For the $\bar{x}$ action, the relevant vacuum factor is determined by symmetry to be 
\be\label{barP0}
\bar{P}_0(u) = \frac{\hat{\psi}_0(u)}{{\psi}_0(u-\sigma_3 {\psi}_0) } = 
\Bigl(1 + \frac{\sigma_3 \hat{\psi}_0}{u}\Bigr) \, \Bigl( 1 - \frac{\sigma_3 {\psi}_0}{u} \Bigr)   \ , 
\ee
and the poles appear at positions 
\begin{align}
\bar{p}_+(\overline{\blacksquare}) & =  \hat{h}(\overline{\blacksquare}) + \ell h_2 \\ 
\bar{p}_-({\blacksquare}) & = \hat{g}(\blacksquare) + \ell h_2  \ ,
\end{align}
where in the first line $\ell$ denotes the number of (hatted) boxes extending the bud of length $\hat{m}+\hat{n}$, while in the second line $\ell$ denotes the number of (hatted) boxes on top of the wall corresponding to $\blacksquare$.

%
%\begin{equation}
%\bar{X}_{\widehat{{\tiny{\yng(1)}}}}(u)=\varphi^{-1}_{2} (-u+h(\widehat{{\tiny{\yng(1)}}}))
%\end{equation}
%\begin{equation}
%\bar{X}_{\blacksquare}(u)=\prod^{m+n}_{k=1}\varphi^{-1}_{2} (u-(h(\blacksquare)-k h_2)- \sigma_3 \psi_0)\prod^{2}_{k=1}\varphi^{-1}_{2} (u-(h(\blacksquare)+k h_2)- \sigma_3 \psi_0)
%\end{equation}
%\begin{equation}
%\bar{X}_{\overline{\blacksquare}}(u)=\prod^{m+n}_{k=1}\varphi^{-1}_{2} (-u+(h(\overline{\blacksquare}) - k h_2))
%\end{equation}

The analysis of the previous section then goes through essentially unaltered. 
In particular, the action of $\bar{x}$ does not change $\bar{P}_\Lambda(u)$, and hence we also have the OPE, cf.\ eq.~(\ref{xxOPE})
\be
\boxed{
\bar{x}(z)\, \bar{x}(w) \sim - \bar{x}(w) \, \bar{x}(z) } \ .
\ee

\subsection{The $x \bar{x}$ OPE}\label{sec:xxbar}

Given that we now have a well-defined action of both $\bar{x}(z)$ and $x(w)$, we can evaluate the OPE of $\bar{x}(z) x(w)$ on this representation. 
More specifically, we consider the coefficient of the two states
\be
x(z) \, \bar{x}(w) \, |\Lambda\rangle  = \frac{A}{ \bigl(z - {p}_+({\blacksquare})\bigr) \, \bigl(w - \bar{p}_+(\overline{\blacksquare})\bigr) } 
\, |\Lambda + \overline{\blacksquare} + \blacksquare \rangle \ , 
\ee 
and 
\be
\bar{x}(w) \, x(z) \, |\Lambda\rangle  = \frac{B}{\bigl(w - \bar{p}_+(\overline{\blacksquare})\bigr) \, \bigl(z - {p}_+({\blacksquare})\bigr) } 
\, |\Lambda + \blacksquare + \overline{\blacksquare} \rangle \ . 
\ee
We find\footnote{Recall that the action of $x$, say, does not just add a box $\blacksquare$, but also removes the corresponding bud.}
\be\label{Adef}
A^2 = C\, \prod_{k=0}^{2} \varphi_2^{-1} \bigl(  {p}_+({\blacksquare}) - \bar{p}_+(\overline{\blacksquare}) - k h_2 - \sigma_3\hat{\psi}_0 \bigr) \ ,
\ee
and 
\be\label{Bdef}
B^2 = C \, \prod_{k=0}^{2} \varphi_2^{-1} \bigl( \bar{p}_+(\overline{\blacksquare}) - {p}_+({\blacksquare}) - k h_2 - \sigma_3\psi_0 \bigr) \ , 
\ee
where $C$ is an overall factor that  comes from the various contributions which are unaffected by the addition of these boxes.
Using (\ref{varphi2m}) as well as $\sigma_3 \psi_0 + \sigma_3 \hat{\psi}_0 = - h_2$ one finds, quite remarkably, that $A^2 = B^2$. 
In particular, this suggests that the $D(\Delta)$ function of eq.~(\ref{Dansatz}) takes the simple form 
\be\label{Dfinal}
\boxed{
D(\Delta) = \frac{(\Delta + h_2 - \sigma_3\hat{\psi}_0) \, (\Delta - h_2 + \sigma_3\psi_0)}{(\Delta - h_2 + \sigma_3\psi_0)\, (\Delta + h_2 - \sigma_3\hat{\psi}_0) }  } \ .
\ee
Note that this is compatible with the free field limit. 
We should remind the reader that, just as for the case of the $H(\Delta)$ function, the corresponding OPE relation is not trivial, see the discussion around eq.~(\ref{almosttrivial}). Indeed, the non-trivial factors in (\ref{Dfinal}) are required to take care of the special cases (from which the ansatz for eq.~(\ref{Dansatz}) was derived).

\section{Action of fermionic annihilation operators}\label{sec:annih}

In the previous section we have defined the action of the fermionic creation operators $x$ and $\bar{x}$ on arbitrary twin-plane-partition configurations. 
To complete our construction it therefore only remains to define the corresponding annihilation operators $y(w)$ and $\bar{y}(w)$. 
We first discuss $y(w)$, and then explain how $\bar{y}(w)$ can be defined similarly.

\subsection{Action of $y(u)$}

We begin by reviewing the defining relations for $y(w)$. The analogue of (\ref{psiFBx0}) is 
\begin{equation}\label{psiFBy0} 
%\begin{cases}
\begin{aligned}
\psi(z) \, y(w)  &\sim  \varphi_2^{-1}(\Delta) \, y(w) \,\psi(z)   \\
\hat{\psi}(z) \, y(w)  &\sim  \varphi_2(-\Delta-\sigma_3\hat{\psi}_0) \, y(w) \,\hat{\psi}(z) \ , 
\end{aligned}
%\end{cases}
\end{equation}
see also Fig.~\ref{figOPEeverybody2}, from which it follows that the analogues of the charge contribution formulae (\ref{charge1}) and (\ref{charge2}) are 
\be
\phi[y](u)\equiv \varphi_2^{-1}(u) \ , \qquad \hat{\phi}[y](u)\equiv \varphi_2(-u-\sigma_3\hat{\psi}_0) \ . 
\ee
In particular, $y(w)$ therefore ``undoes" the action of $x(z)$. So, for example, $y(w)$ can remove a box $\blacksquare$ with coordinates $(x_1,x_3)=(m,n)$ from $\Lambda$; if the corresponding wall has $\ell$ (hatted) boxes on top, it will do so with a pole at 
\be
w^\ast = p_+(\blacksquare) = h(\blacksquare) +\ell h_2 \ , 
\ee
see eq.~(\ref{p+f}), and replace the infinite $\blacksquare$-row by $(m+n+\ell)$ boxes lined up along the $x_2$ direction at $(x_1,x_3)=(m,n)$, see the discussion in Section~\ref{sec:4.5}.  It can also add a $\overline{\blacksquare}$ row at $(\hat{x}_1,\hat{x}_3)=(\hat{m},\hat{n})$ to $\Lambda$, provided there is already a bud of length $(\hat{m}+\hat{n}+\ell +3)$ with $\ell\geq 0$. 
Then the pole will appear at position
\be
w^{\ast} = p_-(\overline{\blacksquare})= g(\overline{\blacksquare})+\ell h_2  \ , 
\ee
cf.\ eq.~(\ref{p-f}), and it will lead to $\ell$ boxes on top of the wall corresponding to $\overline{\blacksquare}$, see the discussion in Section~\ref{sec:4.7}. 

Using these rules we can then constrain the OPEs of $e(z) y(w)$ and $f(z) y(w)$. 
For example, evaluating $f(z) y(w)$ and $y(w)f(z)$ on $|\blacksquare+\hat{{\tiny\yng(1)}}_{\, \textrm{top}}\rangle$ and on $|\blacksquare+{\tiny\yng(1)_{\, j}}\rangle$ with $j=1,3$, we conclude that 
\be
f(z)\, y(w) \sim \frac{ (\Delta  - h_1) (\Delta - h_3)}{(\Delta + h_2)} \, y(w)\, f(z) \ ,
\ee
where we have used (\ref{xonefinal}). 
Similarly, evaluating $e(z) y(w)$ and $y(w)e(z)$ on $|\blacksquare\rangle$ we learn that 
\be
e(z)\, y(w) \sim \frac{1}{\Delta} \, y(w)\, e(z) \ ,
\ee
thus suggesting that the correct OPEs are 
\be
\boxed{f(z)\, y(w) \sim \varphi_2^{-1}(\Delta) \, y(w)\, f(z)}  \ , \qquad
\boxed{e(z)\, y(w) \sim \frac{\Delta}{\Delta} \, y(w)\, e(z) } \ . 
\ee
This agrees with the prediction of \cite{GLPZ}, see also Fig.~\ref{figOPEbosonic-xy} above. By similar arguments we also conclude that, see eq.~(\ref{GhatHhatfin}) and Fig.~\ref{figOPEbosonicfull} 
\be
\boxed{\hat{f}(z)\, y(w) \sim \frac{(\Delta+\sigma_3 \hat{\psi}_0 - h_2) }{(\Delta +\sigma_3 \hat{\psi}_0 - h_2) }   \, y(w)\, \hat{f}(z)}  \ , \quad
\boxed{\hat{e}(z)\, y(w) \sim \varphi_2(-\Delta-\sigma_3\hat{\psi}_0) \, y(w)\, \hat{e}(z) } \ . 
\ee
With these relations at hand, we can then construct the $y(w)$ action on an arbitrary twin-plane-partition, paralleling what we did for $x(w)$ above, and we find 
\begin{equation}\label{yansatz}
y(w)|\Lambda\rangle = \!\!\! \sum_{\blacksquare\in \textrm{Rem}(\lambda)}
\frac{ \Bigl[ \textrm{Res}_{u=p_{+}({\blacksquare})} \, P_\Lambda(u) \Bigr]^{\frac{1}{2}}
}{w-p_+(\blacksquare)}
\, |\Lambda-\blacksquare\rangle
+ \!\!\! \sum_{\overline{\blacksquare}\in \textrm{Add}(\hat{\rho})}\frac{\Bigl[ \textrm{Res}_{u=p_{-}(\overline{\blacksquare})} \, P_\Lambda(u) \Bigr]^{\frac{1}{2}}
}{w-p_-(\overline{\blacksquare})}|\Lambda+\overline{\blacksquare}\rangle  \ , 
\end{equation}
where $P_\Lambda(u)$ is the same function as defined before, see eq.~(\ref{XLambda}). 
We should mention that this implies, in particular, that (\ref{yansatz}) has the same poles as for the case of the $x$ action, in agreement with what we need. 
We also conclude by the same token that these generators satisfy the OPE relation 
\begin{equation}
y(z) \, y(w) \sim - y(w) \, y(z) \ . 
\end{equation}

\subsection{Action of $\bar{y}(u)$}

The analysis for $\bar{y}(u)$ is completely analogous, so we shall be brief. 
The relevant OPE relations are, see Figs.~\ref{figOPEeverybody2} and \ref{figOPEbosonic-xy}
\begin{equation}\label{psiFBbary0} 
%\begin{cases}
\begin{aligned}
\psi(z) \, \bar{y}(w)  &\sim  \varphi_2(-\Delta -\sigma_3 {\psi}_0 ) \,\, \bar{y}(w) \,\psi(z)   \\
\hat{\psi}(z) \, \bar{y}(w)  &\sim  \varphi_2^{-1}(\Delta) \,\, \bar{y}(w) \,\hat{\psi}(z) \\[4pt]
\hat{e}(z) \, \bar{y}(w) & \sim \frac{\Delta}{\Delta} \,\, \bar{y}(w)\, \hat{e}(z) \\
\hat{f}(z) \, \bar{y}(w) & \sim \varphi_2^{-1}(\Delta) \,\, \bar{y}(w)\, \hat{f}(z) \ ,
\end{aligned}
%\end{cases}
\end{equation}
as well as, see Fig.~\ref{figOPEbosonicfull} 
\be
\begin{aligned}
e(z)\, \bar{y}(w) & \sim  \varphi_2(-\Delta - \sigma_3\psi_0) \,\, \bar{y}(w) \, e(z) \\
f(z)\, \bar{y}(w) & \sim  \frac{(\Delta + \sigma_3 \psi_0 - h_2)}{(\Delta + \sigma_3 \psi_0 - h_2)} \,\,
 \bar{y}(w) \, f(z) \ . 
\end{aligned}
\ee
These relations are respected by the action 
\begin{equation}\label{ybaransatz}
\bar{y}(w)|\Lambda\rangle = \!\!\!\sum_{\overline{\blacksquare}\in \textrm{Rem}(\hat{\rho})}
\frac{ \Bigl[ \textrm{Res}_{u=\bar{p}_+(\overline{\blacksquare})} \, \bar{P}_\Lambda(u) \Bigr]^{\frac{1}{2}}
}{w-\bar{p}_+(\overline{\blacksquare})}
\, |\Lambda-\overline{\blacksquare}\rangle
+ \!\!\!\sum_{\blacksquare\in \textrm{Add}(\lambda)} 
\frac{\Bigl[ \textrm{Res}_{u=\bar{p}_{-}({\blacksquare})} \, \bar{P}_\Lambda(u) \Bigr]^{\frac{1}{2}}
}{w-\bar{p}_-({\blacksquare})}|\Lambda+{\blacksquare}\rangle  \ , 
\end{equation}
where $\bar{P}_\Lambda$ is again the same function as before, see eq.~(\ref{barXLambda}). 
By the same arguments as for the $\bar{x}$ case we then conclude that 
\be
\begin{aligned}
y(z) \, \bar{y}(w) & \sim - \, D(\Delta)\, \, \bar{y}(w) \, y(z) \\
\bar{y}(z) \, \bar{y}(w) & \sim - \, \bar{y}(w) \, \bar{y}(z) \ . 
\end{aligned}
\ee

\subsection{OPEs between fermionic creation and annihilation operators}

%\subsubsection{OPEs}
This leaves us with determining the OPEs of the $x$ generators with the $y$ generators. 
They can be deduced from the representation using the same techniques as for the corresponding bosonic identity --- the last equation in (\ref{bosonicdef}) ---  and we find that on the state $\Lambda$ 
\be
x(z)\, y(w) + y(w) \, x(z) \ \sim \ \frac{P_\Lambda(z) - P_\Lambda(w)}{z-w} \ ,
\ee
where $P_\Lambda(u)$ is defined in eq.~(\ref{XLambda}). We also have a similar identity for the barred fields, 
\be
\bar{x}(z)\, \bar{y}(w) + \bar{y}(w) \, \bar{x}(z) \ \sim \ \frac{\bar{P}_\Lambda(z) - \bar{P}_\Lambda(w)}{z-w} \ ,
\ee
where $\bar{P}_\Lambda$ is defined in eq.~(\ref{barXLambda}). 
We can think of $P_\Lambda(u)$ as the eigenvalue of a field $P(u)$, and similarly for $\bar{P}_\Lambda(u)$, i.e.
\begin{equation}
P(u) \, |\Lambda\rangle = P_\Lambda(u) \, |\Lambda\rangle \qquad \textrm{and} \qquad
\bar{P}(u) \, |\Lambda\rangle = \bar{P}_\Lambda(u) \, |\Lambda\rangle \  . 
\end{equation}

\subsubsection{Expressing $P(u)$ and $\bar{P}(u)$ in terms of $\psi(u)$ and $\hat{\psi}(u)$ }

Each twin-plane-partition is an eigenstate of all four operators $\psi(u)$, $\hat{\psi}(u)$, $P(u)$, and $\bar{P}(u)$. However, since each twin-plane-partition is uniquely characterized by the pair of charge functions $(\psi_{\Lambda}(u),\hat{\psi}_{\Lambda}(u))$, the eigenfunctions $(P_{\Lambda}(u),\bar{P}_{\Lambda}(u))$ must be expressible in terms of $(\psi_{\Lambda}(u),\hat{\psi}_{\Lambda}(u))$. 
\smallskip

In order to derive this relation we note that each twin-plane-partition configuration can be generated by the action of its four building blocks: ${\tiny\yng(1)}$, $\widehat{{\tiny\yng(1)}}$ , $\blacksquare$, and $\overline{\blacksquare}$. Each of these building blocks has a definite eigenvalue with respect to $\psi(u)$, $\hat{\psi}(u)$, $P(u)$, and $\bar{P}(u)$, see Table~\ref{tab2} in Appendix~\ref{app:eigen}.
%For all four building blocks, we have
%\begin{equation}
%\begin{aligned}
%\frac{X_{\Lambda}(u)}{X_{\Lambda}(u+h_2)} & = \psi_{\Lambda}(u) \, \hat{\psi}_{\Lambda}(u+h_2-\sigma_3 \hat{\psi}_0) \\
%\frac{\bar{X}_{\Lambda}(u)}{\bar{X}_{\Lambda}(u+h_2)} & =  {\psi}_{\Lambda}(u+h_2-\sigma_3 {\psi}_0) \,\hat{\psi}(u) 
%\end{aligned}
%\end{equation}
%for $\Lambda= {\tiny\yng(1)}$, $\widehat{{\tiny\yng(1)}}$, $\blacksquare$, or $\overline{\blacksquare}$, where we have striped off the vacuum factor. 
%Restoring the vacuum factor, 
It is then easy to check that, for an arbitrary twin-plane-partition configuration $\Lambda$
\begin{equation}
\begin{aligned}
\frac{P_{\Lambda}(u)\, P_0(u+h_2) }{P_{\Lambda}(u+h_2)\, P_0(u) } & = \frac{\psi_{\Lambda}(u) \, \hat{\psi}_{\Lambda}(u+h_2-\sigma_3 \hat{\psi}_0) }{\psi_0(u)\, \hat{\psi}_{0}(u+h_2-\sigma_3 \hat{\psi}_0)} 
%\cdot \frac{P_0(u)}{P_0(u+h_2)}
\\
\frac{\bar{P}_{\Lambda}(u)\, \bar{P}_0(u+h_2)}{\bar{P}_{\Lambda}(u+h_2)\, \bar{P}_0(u)} & = \frac{ {\psi}_{\Lambda}(u+h_2-\sigma_3 {\psi}_0) \,\hat{\psi}_{\Lambda}(u)}{{\psi}_{0}(u+h_2-\sigma_3 {\psi}_0) \,\hat{\psi}_0(u)} \ ,
%\cdot\frac{\bar{P}_0(u)}{\bar{P}_0(u+h_2)}
\end{aligned}
\end{equation}
where the vacuum factors $P_0(u)$ and $\bar{P}_0(u)$ are given in (\ref{P0}) and (\ref{barP0}), respectively.
%with 
%\begin{equation}
%\begin{aligned}
%X_0(u)&=(1+\sigma_3 \frac{\psi_0}{u})(1-\sigma_3 \frac{\hat{\psi}_0}{u}) \\
%\bar{X}_0(u)&=(1-\sigma_3 \frac{\hat{\psi}_0}{u})(1+\sigma_3 \frac{\hat{\psi}_0}{u}) \\
%\end{aligned}
%\end{equation}
%
%
%
Since these equations are true on arbitrary twin-plane partitions, we can write them as operator identities
\begin{equation}\label{conv1}
\begin{aligned}
\frac{P(u)\, P_0(u+h_2)}{P(u+h_2)\, P_0(u) } & = \frac{\psi(u) \, \hat{\psi}(u+h_2-\sigma_3 \hat{\psi}_0) }{\psi_0(u)\, \hat{\psi}_{0}(u+h_2-\sigma_3 \hat{\psi}_0)}
% \cdot \frac{X_0(u)}{X_0(u+h_2)}
\\
\frac{\bar{P}(u)\,\bar{P}_0(u+h_2)}{\bar{P}(u+h_2)\, \bar{P}_0(u)} & = \frac{ {\psi}(u+h_2-\sigma_3 {\psi}_0) \,\hat{\psi}(u)}{{\psi}_{0}(u+h_2-\sigma_3 {\psi}_0) \,\hat{\psi}_0(u)} \ .
%\cdot\frac{\bar{X}_0(u)}{\bar{X}_0(u+h_2)}
\end{aligned}
\end{equation}
Similarly, we find 
\begin{equation}\label{conv2}
\begin{aligned}
\frac{P(u-2h_2)\, \bar{P}_0(u+\sigma_3\psi_0)}{\bar{P}(u+\sigma_3\psi_0)\, P_0(u-2h_2)} & = 
\frac{\psi(u) \, \psi(u-h_2) \, \psi(u-2h_2)}{\psi_0(u) \, \psi_0(u-h_2) \, \psi_0(u-2h_2)} \\[2pt]
\frac{\bar{P}(u-2h_2)\, P_0(u+\sigma_3 \hat{\psi}_0)}{{P}(u+\sigma_3\hat{\psi}_0)\, \bar{P}_0(u-2h_2)} & = 
\frac{\hat{\psi}(u) \, \hat{\psi}(u-h_2)\, \hat{\psi}(u-2h_2)}{\hat{\psi}_0(u) \, \hat{\psi}_0(u-h_2)\, \hat{\psi}_0(u-2h_2)} \ .
\end{aligned}
\end{equation}
\smallskip

We can use these relations to solve the modes of $P(u)$ and $\bar{P}(u)$, which we define via 
\begin{equation}\label{XP}
P(u)   =  1 + \sigma_3 \sum_{r=0}^{\infty} \frac{ P_r}{u^{r+1}}   \ \qquad \textrm{and} \qquad
\bar{P}(u)   =  1 + \sigma_3 \sum_{r=0}^{\infty} \frac{ \bar{P}_r}{u^{r+1}} \  
\end{equation} 
recursively in terms of $\psi_r$ and $\hat{\psi}_s$, and vice versa; for the first few cases the explicit results are given in Appendix~\ref{app:exact}, see eqs.~(\ref{Pexp}). 
Note that $P(u)$ and $\bar{P}(u)$ map to each other under the interchange of $\psi_r$ and $\hat{\psi}_r$,
\begin{equation}\label{PPbar}
P(u) \longleftrightarrow \bar{P}(u) \qquad\quad \textrm{under} \qquad\quad \psi(u) \longleftrightarrow \hat{\psi}(u) \ ,
\end{equation}
as follows from (\ref{conv1}).
% and (\ref{Pbarexp}).

We have checked that, in the free field limit, the result (\ref{Pexp}) reproduces exactly the expressions of $P_r$ and $\bar{P}_r$ with $r\geq 1$, which can be deduced from the equations in Appendix~A of \cite{GLPZ}. 
We have spelled out the first few cases explicitly in Appendix~\ref{sec:PbarP}, see eqs.~(\ref{P1}) and  (\ref{Pbar1}). 
The fact that these fairly complicated expressions match is a strong consistency check of our analysis.
\smallskip

It remains to determine the OPE of $x(z) \, \bar{y}(w)$. This can be done using effectively  the same approach as in Section~\ref{sec:xxbar}, except that now the factors of $\varphi_2$ that appear in (\ref{Adef}) have the inverse power since $\bar{y}(w)$ removes a box $\overline{\blacksquare}$, whereas $\bar{x}(w)$ adds one. 
Using the identity between (\ref{Adef}) and (\ref{Bdef}) we thus conclude that the OPE takes the form 
\be
x(z)\, \bar{y}(w) \ \sim \ - \Bigl[ \prod_{k=0}^{2} \, \varphi_2 (\Delta - k h_2 - \sigma_3\hat{\psi}_0)\Bigr] \, \bar{y}(w)\, x(z) \ . 
\ee
Finally, the remaining OPE of $\bar{x}(z)\, y(w)$ is determined by symmetry to be 
\be
\bar{x}(z)\, y(w) \ \sim \ - \Bigl[ \prod_{k=0}^{2} \, \varphi_2 (\Delta - k h_2 - \sigma_3\psi_0)\Bigr] \, y(w)\, \bar{x}(z) \ . 
\ee

\section{Conclusion and discussion}\label{sec:sum}

In this paper we have completed the program begun in \cite{GLPZ} and identified the defining relations of the ${\cal N}=2$ generalization of the affine Yangian of $\mathfrak{gl}_1$. 
Our main guiding principle was to use the fact that this affine Yangian must act on twin-plane-partitions, i.e.\ pairs of plane partitions that are ``glued" together along one common direction. 
We have shown that this ansatz, together with the (known) representation theory of the two bosonic affine Yangians, is strong enough to fix the structure of the full algebra almost completely. 
Together with some natural assumptions (as well as the constraint that the answer must have the correct free field limit), this allowed us to write down a set of defining relations: in addition to the familiar bosonic relations, they are 
\be
\begin{aligned}
\psi(z)\, x(w) & \sim \varphi_2(\Delta)\, x(w) \psi(z) \\ 
e(z)\, x(w) & \sim \frac{(\Delta + h_2)\Delta}{(\Delta - h_1) (\Delta - h_3)} \, x(w) \, e(z) \\ 
f(z)\, x(w) & \sim \frac{\Delta}{\Delta} \, x(w)\, f(z) \\
\hat{e}(z) \, x(w) & \sim \frac{(\Delta+\sigma_3\hat{\psi}-h_2)}{(\Delta+\sigma_3\hat{\psi}_0-h_2)}\, x(w)\, \hat{e}(z) \\ 
\hat{f}(z) \, x(w) & \sim \frac{(\Delta+\sigma_3\hat{\psi}_0+h_1)(\Delta+\sigma_3\hat{\psi}_0+h_3)}{(\Delta+ \sigma_3\hat{\psi}_0 )(\Delta+\sigma_3\hat{\psi}_0-h_2) }  \, x(w) \, \hat{f}(z)  \ ,
\end{aligned}
\ee
as well as similar relations with $x(w)$ replaced by $\bar{x}(w)$, $y(w)$ and $\bar{y}(w)$. %, see Fig.~\ref{figOPEeverybody3}. 
%\begin{figure}[h!]
%	\centering
%	\includegraphics[trim=2cm 12cm 0cm 4cm, width=.7\textwidth]{"OPEeverybody3"}
%	\caption{The OPE relations of the bosons with the fermions.	}
%	\label{figOPEeverybody3}
%	
%\end{figure}
The OPE relations of the fermionic generators among themselves are 
\be
\begin{aligned} 
x(z) \, x(w)  & \sim - x(w) \, x(z) \\
x(z) \, \bar{x}(w) & \sim - \frac{(\Delta + h_2 - \sigma_3\hat{\psi}_0) \, (\Delta - h_2 + \sigma_3\psi_0)}{(\Delta - h_2 + \sigma_3\psi_0)\, (\Delta + h_2 - \sigma_3\hat{\psi}_0) } \, \bar{x}(w) \, x(z) \\
x(z)\, y(w) + y(w) \, x(z) & \sim \ \frac{P(z) - P(w)}{z-w}  \ , 
\end{aligned}
\ee
as well as similar relations for the other cases. 
Here the eigenvalue of $P(u)$ on $\Lambda$, $P_\Lambda(u)$, is given explicitly in (\ref{XLambda}), and there are identities that express $P(u)$ and $\bar{P}(u)$ in terms of $\psi(u)$ and $\hat{\psi}(u)$, see eqs.~(\ref{conv1}) and (\ref{conv2}). 
This allows us, at least recursively, to write down the algebra in closed form.  

We have defined the algebra by constructing a non-trivial (faithful) representation on twin-plane-partitions. 
In particular, this therefore also establishes that the above set of relations is consistent.
\smallskip

There are a number of open problems that are interesting directions for future research. 
We have tried to see whether there are any non-trivial Serre relations we may have to impose in addition, but we have not found any such relations that are compatible with our OPE relations --- recall that in the bosonic case, the Serre relations are closely related to the corresponding OPE relations, see the discussion in eq.~(\ref{Serre1}). 
Similarly, there may be some initial relations, analogous to (\ref{initial}), one may have to impose, but beyond those already spelled out in \cite{GLPZ},  see section~\ref{sec:N=2initial} above, we have not seen any scope for them. We have mainly focused on the vacuum representation (corresponding to trivial asymptotics along the $x_1$, $x_3$, $\hat{x}_1$ and $\hat{x}_3$ directions); it would also be interesting to study in more detail the non-trivial representations. In particular, this may shed light on the ``spurious poles" we mentioned in Section~\ref{sec:xpoles}. It should also allow us to make more direct contact with the ${\cal N}=2$ ${\cal W}_\infty$ algebra and its representation theory, see \cite{Candu:2012tr}. In particular, it may help in identifying the 
detailed dictionary of our supersymmetric Yangian to the ${\cal N}=2$ ${\cal W}_\infty$ algebra, generalizing the analysis of \cite{Gaberdiel:2017dbk} to the supersymmetric case. Finally, one may expect some relation to the integrable structure of \cite{OhlssonSax:2011ms,Sax:2012jv,Borsato:2015mma}, see \cite{Sfondrini:2014via} for a review. We hope to return to some of these issues in the near future.

\section*{Acknowledgements}

We thank Hong Zhang for initial collaboration on this project. The work of MRG is partially supported by  the NCCR SwissMAP, funded by the Swiss National Science Foundation. WL is grateful for support from the ``Thousand talents grant" and from the ``Max Planck Partnergruppe". The work of CP is supported by the US Department of Energy under contract DE-SC0010010 Task A. MRG and WL thank ICTS Bangalore, MRG thanks the Yukawa Institute in Kyoto, WL thanks APCTP Korea, and CP thanks the Institute for Advanced Study in Princeton for hospitality at various stages of this project.

\appendix

\section{Relating the modes of $P$ and $\bar{P}$ to $\psi_s$ and $\hat{\psi}_s$}\label{app:exact}
\subsection{Solving $P_r$ and $\bar{P}_r$ recursively}

The modes $P_r$ and $\bar{P}_r$, defined in (\ref{XP}), can be solved recursively in terms of $\psi_s$ and $\hat{\psi}_s$ from (\ref{conv1}). 
We give the first few $P_r$ explicitly:
\begin{equation}\label{Pexp}
\begin{aligned}
P_0=&(\psi_0-\hat{\psi}_0)+\frac{1}{h_2}\cdot[\psi_1+\hat{\psi}_1]\\
P_1=&\Bigl(\frac{\psi_1-\hat{\psi}_1}{2}-\sigma_3\psi_0\hat{\psi}_0\Bigr)+\frac{1}{h_2}\cdot\Bigl[\frac{\psi_2+\hat{\psi_2}}{2}+\sigma_3\Bigl(\frac{\psi_0\psi_1-\hat{\psi}_0\hat{\psi}_1}{2}+\psi_0\hat{\psi}_1-\hat{\psi}_0\psi_1\Bigr)\Bigr]\\&+\frac{1}{h_2^2}\cdot\frac{\sigma_3(\psi_1+\hat{\psi}_1)^2}{2}\\
P_2=&h_2\cdot\frac{\psi_1+\hat{\psi}_1}{6}+\Bigl(\frac{\psi_2-\hat{\psi}_2}{2}-\frac{\sigma_3(\psi_0\hat{\psi}_1+\hat{\psi}_0\psi_1)}{2}\Bigr)\\
&+\frac{1}{h_2}\cdot \Bigl[\frac{\psi_3+\hat{\psi}_3}{3}+\sigma_3\Bigl(\frac{\psi_1^2-2\hat{\psi}_1^2}{3}+\frac{\psi_0\psi_2+\hat{\psi}_0\hat{\psi}_2}{6}+\frac{{\psi}_0\hat{\psi}_2-\hat{\psi}_0\psi_2}{2}\Bigr)\\&\qquad\qquad-\sigma_3^2\frac{\psi_0^2\psi_1+\hat{\psi}_0^2\hat{\psi}_1+3\psi_0\hat{\psi}_0(\psi_1+\hat{\psi_1})}{6} \Bigr]\\
&+\frac{1}{h_2^2}\cdot\frac{\sigma_3(\psi_1+\hat{\psi}_1)}{2} \bigl( (\psi_2+\hat{\psi}_2)+\sigma_3 (\psi_0\hat{\psi}_1-\hat{\psi}_0\psi_1)\bigr)+\frac{1}{h_2^3}\cdot \frac{\sigma_3^2(\psi_1+\hat{\psi_1})^3}{6} \ . 
\end{aligned}
\end{equation}
The expressions for $\bar{P}_r$ can be obtained from those for $P_r$ by interchanging $\psi_j$ and $\hat{\psi}_j$, see eq.~(\ref{PPbar}).
%\begin{equation}
%\bar{P}_r
%\end{equation}

\subsection{Free field limit}\label{sec:PbarP}

The free field relations for $P_r$ and $\bar{P}_r$ were already given in Appendix~A of \cite{GLPZ}, and they take the explicit form 
\bal\label{Pdef1}
P_r & =  {  \sum_{m\in \mathbb{Z}+\frac{1}{2}} \sum_{i} (m-\tfrac{3}{2})(-m+\tfrac{1}{2})^{r-1} \, :\bar{\chi}^i_{-m} {\chi}^i_{m}:
+\sum_{m\in \mathbb{Z}}  \sum_{i}  \bigl(-m-1 \bigr)^{r-1} \, : \bar{\alpha}^i_{-m} \alpha^i_{m}:}    \\
\bar P_r & =  {  \sum_{m\in \mathbb{Z}+\frac{1}{2}} \sum_{i} (m+\tfrac{3}{2})(-m-\tfrac{1}{2})^{r-1} \, :\bar{\chi}^i_{-m}{\chi}^i_{m} :
+\sum_{m\in \mathbb{Z}}  \sum_{i}  \bigl(-m+1 \bigr)^{r-1} \, : \bar{\alpha}^i_{-m} \alpha^i_{m}:}  \ . \label{Pbardef1}
\eal
We can express them recursively in terms of the $\psi_r$ and $\hat{\psi}_r$ generators which were also defined there:
\begin{eqnarray}\label{psidef}
\psi_r & = & {\displaystyle  \sum_{m\in\mathbb{Z}+\frac{1}{2}} \sum_{i} \Bigl( (-m-\tfrac{1}{2})^r - (-m+\tfrac{1}{2})^r \Bigr)\, :\bar{\chi}^i_{-m}\chi^i_m :}   \\ 
\hat{\psi}_r & = &  {\displaystyle   \sum_{m\in\mathbb{Z}}  \sum_{i} \Bigl( (  m + 1) (-m)^{r-2} + \bigl(-m+1 \bigr)^{r-1} \Bigr)\, 	:   \bar\alpha^i_{-m} \alpha^i_{m}:  } \ . 
\end{eqnarray}
Explicitly, one finds then for the $P_r$ modes 
\be\label{P1}
\begin{aligned}
P_1 &=  \frac{1}{2} \bigl( \psi_2 + \hat{\psi}_2) + \frac{3}{2} \psi_1 + N \\
P_2 & = \frac{2}{3} \psi_1 + \psi_2 +\frac{1}{3} \psi_3  -\frac{2}{3} \hat{\psi}_2   + \frac{1}{3} \hat{\psi}_3 \\
P_3 & = \frac{1}{6} \psi_1 + \frac{3}{4} \psi_2 + \frac{5}{6} \psi_3 + \frac{1}{4} \psi_4 
+ \frac{5}{6} \hat{\psi}_2 - \frac{11}{12} \hat{\psi}_3 +\frac{1}{4} \hat{\psi}_4  \\ 
P_4 & = - \frac{1}{30} \psi_1 + \frac{1}{4} \psi_2 + \frac{5}{6} \psi_3 + \frac{3}{4} \psi_4 +\frac{1}{5} \psi_5 
-\frac{29}{30} \hat{\psi}_2 + \frac{107}{60} \hat{\psi}_3  - \frac{21}{20} \hat{\psi}_4 + \frac{1}{5} \hat{\psi}_5 \\ 
P_5 & =  -\frac{1}{30} \psi_1-\frac{1}{12} \psi_2+ \frac{1}{3} \psi_3+ \frac{11}{12}\psi_4+\frac{7}{10}\psi_5+\frac{1}{6}\psi_6  \\
& \qquad 
+ \frac{31}{30} \hat{\psi}_2 - \frac{173}{60} \hat{\psi}_3  + \frac{167}{60} \hat{\psi}_4 - \frac{17}{15} \hat{\psi}_5  + \frac{1}{6} \hat{\psi}_6 \\
P_6 & =  \frac{1}{42}\psi_1 - \frac{1}{12}\psi_2 - \frac{1}{6}\psi_3 + \frac{5}{12}\psi_4 +\psi_5 + \frac{2}{3}\psi_6 + \frac{1}{7}\psi_7  \\
& \qquad -\frac{43}{42} \hat{\psi}_2 + \frac{341}{84} \hat{\psi}_3 - \frac{485}{84} \hat{\psi}_4 + \frac{80}{21} \hat{\psi}_5  - \frac{25}{21} \hat{\psi}_6 + \frac{1}{7} \hat{\psi}_7 \ ,
\end{aligned}
\ee
while for $\bar{P}_r$ we get 
\be\label{Pbar1}
\begin{aligned}
\bar{P}_1 & =  \frac{1}{2} \bigl( \psi_2 + \hat{\psi}_2) - \frac{3}{2} \psi_1 + N \\ 
\bar{P}_2 & = \frac{2}{3} \psi_1 {-} \psi_2 +\frac{1}{3} \psi_3  
+ { \frac{1}{3}} \hat{\psi}_2   + \frac{1}{3} \hat{\psi}_3 \\
\bar{P}_3 & = {-} \frac{1}{6} \psi_1 + \frac{3}{4} \psi_2 {-}\frac{5}{6} \psi_3 + \frac{1}{4} \psi_4 
+ { \frac{1}{6}} \hat{\psi}_2  + {\frac{5}{12}} \hat{\psi}_3 +\frac{1}{4} \hat{\psi}_4   \\
\bar{P}_4 & = - \frac{1}{30} \psi_1 {-} \frac{1}{4} \psi_2 + \frac{5}{6} \psi_3 {-} \frac{3}{4} \psi_4 +\frac{1}{5} \psi_5 
+ { \frac{1}{30}} \hat{\psi}_2 + {\frac{17}{60}} \hat{\psi}_3  + { \frac{9}{20}} \hat{\psi}_4 + \frac{1}{5} \hat{\psi}_5 \\ 
\bar{P}_5 & =  {+} \frac{1}{30} \psi_1-\frac{1}{12} \psi_2{-} \frac{1}{3} \psi_3+ \frac{11}{12}\psi_4{-}\frac{7}{10}\psi_5+\frac{1}{6}\psi_6  \\
& \qquad 
- {  \frac{1}{30}} \hat{\psi}_2 + { \frac{1}{20}} \hat{\psi}_3  + {\frac{23}{60}} \hat{\psi}_4 + { \frac{7}{15}} \hat{\psi}_5  + \frac{1}{6} \hat{\psi}_6 \\
\bar{P}_6 & =  \frac{1}{42}\psi_1 {+} \frac{1}{12}\psi_2 - \frac{1}{6}\psi_3 {-} \frac{5}{12}\psi_4 +\psi_5 {-} \frac{2}{3}\psi_6 + \frac{1}{7}\psi_7  \\
& \qquad - { \frac{1}{42}} \hat{\psi}_2 - { \frac{3}{28}} \hat{\psi}_3 + { \frac{5}{84} }\hat{\psi}_4 + {\frac{10}{21}} \hat{\psi}_5  + { \frac{10}{21}} \hat{\psi}_6 + \frac{1}{7} \hat{\psi}_7 \ . 
\end{aligned}
\ee
We have checked that these are the correct free field limits of $P_r$ and $\bar{P}_r$ as determined from (\ref{conv1}). (For brevity we have only written the expressions for $P_r$
in (\ref{Pexp}) for $r\leq 2$; however, we have checked the free field limit also for $r\leq 6$.) 
Note that in order for $P_1$ and $\bar{P}_1$ to be well-defined in the free field limit, we need to take 
\begin{equation}
\lambda=\frac{N}{N+k} \rightarrow 0 
\end{equation}
while keeping $N$ fixed. In particular, this implies that $\lambda N \rightarrow 0$. 
We should also mention that in this free field limit, the symmetry (\ref{PPbar}) is obscured since 
%$\psi_0=N$, while $\hat{\psi}_0= k \rightarrow \infty$, and 
the limit is taken such that $\sigma_3 \psi_0=0$ and $\sigma_3 \hat{\psi}_0 = -1$.

\section{OPEs between $(P(u),\bar{P}(u))$ and bosonic operators}\label{app:eigen}

The eigenvalues of the four building blocks ${\tiny\yng(1)}$, $\widehat{{\tiny\yng(1)}}$ , $\blacksquare$, and $\overline{\blacksquare}$ with respect to $\psi(u)$, $\hat{\psi}(u)$, $P(u)$, and $\bar{P}(u)$ are summarized  in Table~\ref{tab2} below. 
Note that it follows from these eigenvalues that we have the OPE relations 
\be\label{B1}
\begin{aligned}
P(z) \, e(w) & \sim \varphi_2^{-1}(-\Delta)\, e(w) \, P(z)  \\
P(z) \, f(w) & \sim \varphi_2 (-\Delta)\, f(w) \, P(z)  \\
\bar{P}(z) \, e(w) & \sim \varphi_2^{-1}(\Delta-\sigma_3 \psi_0 )\, e(w) \, \bar{P}(z)  \\
\bar{P}(z) \, f(w) & \sim \varphi_2 (\Delta -\sigma_3 \psi_0)\, f(w) \, \bar{P}(z) \ , 
\end{aligned}
\ee
and similarly for the hatted generators, 
\be\label{B2}
\begin{aligned}
P(z) \, \hat{e}(w) & \sim \varphi_2^{-1}(\Delta-\sigma_3\hat{\psi}_0)\, \hat{e}(w) \, P(z)  \\
P(z) \, \hat{f}(w) & \sim \varphi_2 (\Delta-\sigma_3\hat{\psi}_0)\, \hat{f}(w) \, P(z) \\
\bar{P}(z) \, \hat{e}(w) & \sim \varphi_2^{-1}(-\Delta)\, \hat{e}(w) \, \bar{P}(z)  \\
\bar{P}(z) \, \hat{f}(w) & \sim \varphi_2 (-\Delta)\, \hat{f}(w) \, \bar{P}(z)  \ . 
\end{aligned}
\ee
(On the other hand, there do not seem to be any simple OPE relations for $P$ or $\bar{P}$, and $x$, $\bar{x}$, $y$ or $\bar{y}$.)
One checks that the OPE relations (\ref{B1}) and (\ref{B2}) reduce to the commutation relations of Appendix~A of \cite{GLPZ} in the free field limit. 

\begin{sidewaystable}[h!]
\begin{center}
\begin{tabular}{ccccc}
%\hline \\[-5pt]
 & $\psi(u)$ 
 & $\hat{\psi}(u)$&$P(u)$
 & $\bar{P}(u)$ \vspace{4pt}\\ 
 \hline  \\[-10pt]
 \hbox{vac.} & $\psi_0(u)$ & $\hat{\psi}_0(u)$ & $P_0(u)$ &$\bar{P}_0(u)$ \vspace{4pt} \\
 \hline  \\[-10pt]
${\tiny\yng(1)}$ 
& $\varphi_3(u-h({\tiny\yng(1)}))$ 
& $1$
& $\varphi^{-1}_2(-u+h({\tiny\yng(1)}))$ 
& $\varphi^{-1}_2(u-h({\tiny\yng(1)})-\sigma_3 \psi_0)$ \vspace{4pt}\\
\hline  \\[-10pt]
$\widehat{{\tiny\yng(1)}}$ 
&  $1$ 
& $\varphi_3(u-\hat{h}(\widehat{\tiny\yng(1)}))$ 
& $\varphi^{-1}_2(u-\hat{h}(\widehat{\tiny\yng(1)})-\sigma_3 \hat{\psi}_0)$ 
& $\varphi^{-1}_2(-u+\hat{h}(\widehat{\tiny\yng(1)}))$  \vspace{4pt} \\
\hline  \\[-10pt]
$\blacksquare$ 
& $\varphi_2(u-g(\blacksquare))$ 
& $\varphi^{-1}_2(-u+h(\blacksquare)-\sigma_3 \hat{\psi}_0)$ 
& ${\displaystyle \prod^{m+n-1}_{k=0}\varphi^{-1}_2(-u+g(\blacksquare)+k h_2)}$
& ${\displaystyle \prod^{m+n+2}_{k=0}\varphi^{-1}_2(u-(g(\blacksquare)+k h_2) -\sigma_3 \psi_0)}$ \vspace{4pt} \\
\hline  \\[-10pt]
$\overline{\blacksquare}$ 
& $\varphi^{-1}_2(-u+\hat{h}(\overline{\blacksquare})-\sigma_3 \psi_0)$
& $\varphi_2(u-\hat{g}(\overline{\blacksquare}))$ 
&${\displaystyle \prod^{m+n+2}_{k=0}\varphi^{-1}_2(u-(\hat{g}(\overline{\blacksquare})+k h_2) -\sigma_3 \hat{ \psi}_0)}$
& ${\displaystyle \prod^{m+n-1}_{k=0}\varphi^{-1}_2(-u+\hat{g}(\overline{\blacksquare})+k h_2)}$ \vspace{4pt} \\ 
\hline
\end{tabular}
\end{center}
\caption{The eigenvalues of the different factors. Recall that $h(\blacksquare)=g(\blacksquare)+(x_1(\blacksquare)+x_3(\blacksquare))h_2$ and $\hat{h}(\overline{\blacksquare})=\hat{g}(\overline{\blacksquare})+(\hat{x}_1(\overline{\blacksquare})+\hat{x}_3(\overline{\blacksquare}))h_2$.}\label{tab2}
\end{sidewaystable}

\bibliographystyle{JHEP}

\end{document}